\documentclass[useAMS,usenatbib,usegraphicx,fleqn]{mn2e}

\usepackage{amsmath}
\usepackage{amssymb}

\newcommand{\be}{\begin{equation}}
\newcommand{\ee}{\end{equation}}

\newcommand{\ba}{\begin{eqnarray}}
\newcommand{\ea}{\end{eqnarray}}

\newcommand{\rmd}{\textrm{d}}

\usepackage{times}
\usepackage{aas_macros}
\usepackage{graphicx}
\usepackage{bm}
\usepackage{natbib}
\usepackage{mathrsfs}
\usepackage{color}
\usepackage[mediumspace,Gray,squaren]{SIunits}

\title[Thermal dust emission in distant galaxies]{Constraining thermal dust emission in distant galaxies with number counts and angular power spectra}
\author[G. E. Addison et al.]{G. E. Addison$^{1,2}$\thanks{E-mail: gaddison@phas.ubc.ca}, J. Dunkley$^{2}$ and J. R. Bond$^3$\\
$^{1}$Department of Physics and Astronomy, University of British Columbia, 6224 Agricultural Road, Vancouver, BC V6T 1Z1, Canada\\
$^{2}$Sub-department of Astrophysics, University of Oxford, Denys Wilkinson Building, Keble Road, Oxford OX1 3RH, UK\\
$^{3}$Canadian Institute for Theoretical Astrophysics, University of Toronto, Toronto, ON, Canada M5S 3H8}

\begin{document}

\date{Accepted xxx. Received xxx; in original form xxx}

\pagerange{\pageref{firstpage}--\pageref{lastpage}} \pubyear{2012}

\maketitle

\label{firstpage}

\begin{abstract}
We perform a joint fit to differential number counts from \emph{Spitzer}'s MIPS and \emph{Herschel}'s SPIRE instruments, and angular power spectra of cosmic infrared background (CIB) anisotropies from SPIRE, \emph{Planck}, the Atacama Cosmology Telescope, and the South Pole Telescope, which together span $220\lesssim\nu\textrm{ / GHz}\lesssim4300$ ($70\lesssim\lambda/\mu \textrm{m}\lesssim1400$). We simultaneously constrain the dust luminosity function, thermal dust spectral energy distribution (SED) and clustering properties of CIB sources, and the evolution of these quantities over cosmic time. We find that the data strongly require redshift evolution in the thermal dust SED. In our adopted parametrization, this evolution takes the form of an increase in graybody dust temperature at high redshift, but it may also be related to a temperature -- dust luminosity correlation or evolution in dust opacity. The counts and spectra together constrain the evolution of the thermal dust luminosity function up to $z\sim2.5-3$, complementing approaches relying on rest-frame mid-infrared observations of the rarest bright objects. We are able to fit the power spectra without requiring a complex halo model approach, and show that neglecting scale-dependent halo bias may be impairing analyses that do use this framework.

\end{abstract}


\section{Introduction}

A galaxy's star formation rate (SFR) and far-infrared (FIR) emission are known to be strongly connected \citep{kennicutt:1998}, due to absorption and thermal re-emission of starlight by dusty stellar birth clouds. A simple comparison of the energy in cosmic infrared background \citep[CIB;][]{puget/etal:1996} photons with infrared emission from nearby galaxies indicates that FIR emission (and thus SFR) was considerably higher in the past \citep[e.g.][and references therein]{hauser/dwek:2001}, and extensive measurements of infrared (IR) emission have been made in order to exploit this connection and understand how SFR and stellar formation environments have evolved over cosmic time. A prevailing picture is that luminous and ultra luminous infrared galaxies (LIRGs and ULIRGs; $10^{11}<L_{\rm IR}/L_{\odot}<10^{12}$ and $10^{12}<L_{\rm IR}/L_{\odot}<10^{13}$, respectively), which are rare in the local universe, become far more important to the global SFR at higher redshift \citep[e.g.][]{lefloc'h/etal:2005,perez-gonzales/etal:2005,daddi/etal:2005,caputi/etal:2007,magnelli/etal:2011,lapi/etal:2011}. This trend has predominantly been probed using IR luminosity functions (LFs) measured using instruments including the \emph{Infrared Astronomical Satellite} \citep[IRAS;][]{neugebauer/etal:1984}, the \emph{Spitzer space telescope} \citep{werner/etal:2004} and, more recently, the \emph{Herschel} Space Observatory \citep{pilbratt/etal:2010}.

Current constraints on dust-enshrouded star formation beyond $z\sim2-2.5$ are limited, for several reasons. Inferring the IR luminosity associated with star formation from rest-frame mid-infrared measurements (e.g. using \emph{Spitzer}), is subject to increasingly large uncertainties at high redshift, relating to, for instance, lack of information about high-redshift source spectral energy distributions (SEDs), and the extent to which high-redshift IR emission is associated with obscured active galactic nuclei \citep[AGN; e.g.][]{alexander/etal:2005,lutz/etal:2005,yan/etal:2005,lefloch/etal:2007,sajina/etal:2012}. Probing the thermal dust emission associated with star formation near its rest-frame peak at $\lambda\sim100$ $\mu$m at these high redshifts, using observations at lower frequencies, is also difficult. Source confusion, arising from the high number density of CIB sources per instrumental beam area \citep[e.g.][]{blain/etal:1998,dole/etal:2003}, greatly impairs efforts to resolve the CIB in the submillimeter (submm). Resolved objects in maps at 250, 350 and 500 $\mu$m from \emph{Herschel}'s Spectral and Photometric Imaging Receiver \citep[SPIRE;][]{griffin/etal:2010} account for less than a fifth of the total CIB intensity \citep{oliver/etal:2010}. Deeper constraints have been obtained with fits to the pixel flux histogram of confused maps \citep[the `$P(D)$' approach:][]{patanchon/etal:2009,glenn/etal:2010}, and stacking analyses \citep[e.g.][]{dole/etal:2006,marsden/etal:2009,pascale/etal:2009,bethermin/etal:2010,bethermin/etal:2012b}, however these studies are typically limited to small areas of the sky, and robustly quantifying systematic uncertainties in the deep counts is challenging. While submm-selected samples observed using, for example, the Shared Common-User Bolometric Array \citep[SCUBA;][]{holland/etal:1999} have given us important information about high-redshift dust emission \citep[e.g.][]{smail/etal:1997,hughes/etal:1998,eales/etal:1999,chapman/etal:2005,coppin/etal:2006,michalowski/etal:2010b}, our understanding is fundamentally limited if we have access to only the rare few brightest objects of the population.

Given the success of recent cosmic microwave background background (CMB) temperature anisotropy measurements \citep[e.g.][]{spergel/etal:2003,lueker/etal:2010,fowler/etal:2010,komatsu/etal:2011}, it is perhaps not surprising that the angular power spectrum has emerged as a statistic with which to study the properties of the numerous unresolved dusty sources \citep[e.g.][]{lagache/etal:2007,viero/etal:2009,hajian/etal:2012,amblard/etal:2011,planckcib:2011,viero/etal:2013}, especially given that extragalactic dust emission is a significant CMB temperature foreground on small angular scales \citep[e.g.][]{hall/etal:2010,fowler/etal:2010,dunkley/etal:2011,shirokoff/etal:2011}. The fluctuations in the CIB surface brightness are correlated across angular scales considerably larger than a beam, meaning the confusion phenomenon does not inhibit extracting information from the power spectrum. Dusty galaxies, like many other luminous populations, are understood to exhibit clustering behaviour because galaxies trace the matter density field, whose own fluctuations are strongly scale dependent \citep{peebles:1980}. Combining an understanding of the dark matter clustering with a prescription for how the galaxies and dark matter are related, either through a simple biasing prescription, or a more complex framework, such as the halo model \citep[e.g.][]{bond/etal:1991,seljak:2000,peacock/smith:2000,scoccimarro/etal:2001,cooray/sheth:2002}, allows us to constrain the environmental properties of galaxies (such as characteristic host halo mass scales) using measurements of the clustering. The galaxy correlation function is the statistic that has been widely for these purposes in analyses of, for example, local galaxies in the main Sloan Digital Sky Survey (SDSS) sample \citep[e.g.][]{zehavi/etal:2002,zehavi/etal:2005,zehavi/etal:2011}, massive galaxies at $z\sim0.5$ \citep{white/etal:2011} and luminous red galaxies \citep[LRGs; e.g.][]{blake/etal:2007,zheng/etal:2009}. Interpreting angular power spectra of unresolved CIB sources is more challenging, because of the difficulty in separating the contribution from highly biased intrinsically faint sources from less biased intrinsically brighter objects. Furthermore, existing analyses have shown that the simplest dusty galaxy clustering model, based on a linear biasing ansatz, is inadequate to explain small-scale clustering power \citep[hereafter A12]{planckcib:2011,addison/etal:2012}.

In this work, we perform a joint fit to deep number counts from SPIRE and the Multiband Imaging Photometer for Spitzer \citep[MIPS;][]{rieke/etal:2004}, as well as angular power spectra covering degree to arcminute scales from 250 $\mu$m to 1.4 mm (1200 to 217 GHz) from SPIRE, \emph{Planck}'s High Frequency Instrument \citep[HFI;][]{lamarre/etal:2010,planckHFIperformance:2011}, the Atacama Cosmology Telescope \citep[ACT;][]{fowler/etal:2007,swetz/etal:2011,dunner/etal:2013}, and the South Pole Telescope \citep[SPT;][]{carlstrom/etal:2011}. We simultaneously constrain a bolometric dust luminosity function, the dust SED, the CIB source clustering properties, and the evolution of the these quantities. We require that our model successfully reproduces not only the clustered power in the power spectrum, but also the Poisson, shot-noise, power. Thanks to this joint analysis, we are able to demonstrate the ability of the angular power spectrum to provide constraints on the SEDs of the numerous, faint, unresolved CIB sources.

Many recent dusty galaxy anisotropy analyses have interpreted angular power spectra using the halo model \citep{viero/etal:2009,amblard/etal:2011,planckcib:2011,shang/etal:2012,xia/etal:2012,debernardis/cooray:2012,viero/etal:2013}. In this approach, CIB sources inhabit collapsed dark matter haloes according to a prescription such as the halo occupation distribution \citep[H.O.D.; e.g.][]{berlind/weinberg:2002,kravtsov/etal:2004}. The clustered source contribution to the angular power spectrum arises from correlations between objects in the same halo (`one-halo' term, dominant on small angular scales) or different haloes (`two-halo' term, dominant on large angular scales). The abundance and bias of these haloes, typically inferred from large $N$-body simulations, is used to calculate the contribution of the clustered dusty sources to the angular power spectrum, rather than directly dealing with the dusty source bias. The analyses listed above make a number of assumptions, including:
\begin{enumerate}
\item the halo bias (relative to the linear theory dark matter power spectrum) may be taken as independent of scale when calculating the two-halo term, and
\item the spatial distribution of the dusty sources within their host halo follows the halo mass density profile, truncated at the virial radius.
\end{enumerate}
\cite{cooray/sheth:2002} suggested using the scale-independent, large-scale halo bias with respect to the linear matter power spectrum in the two-halo term as a crude way to prevent overestimating the two-halo power on scales where the non-linear matter power spectrum would begin to receive contributions from within a single halo (since galaxies in a single halo are supposed to be accounted for in the one-halo term). Several more recent analyses have found that, in fact, the halo bias with respect to the linear matter power increases with scale for $k\gtrsim0.1h$ Mpc$^{-1}$, both at low and high redshift \citep[e.g.][]{cole/etal:2005,tinker/etal:2005,yoo/etal:2009,fernandez/etal:2010,tinker/etal:2012,mandelbaum/etal:2013}. Wavenumber $k\simeq0.1h$ Mpc$^{-1}$ corresponds to multipole moment $\ell$ of several hundred at redshift $z\sim1-2$, where much of the CIB anisotropy signal in the far-infrared and longer wavelengths originates. The two-halo term dominates the one-halo on these angular scales, only becoming subdominant at $\ell>750$, even when only the linear matter power spectrum is used in the two-halo term \citep{amblard/etal:2011,planckcib:2011,shang/etal:2012,xia/etal:2012}. This suggests that, if the scale dependence of the halo bias is neglected, either power over a range of scales may be wrongly attributed to the one-halo term, or the inferred redshift distribution of the dusty sources will be altered, in either case likely biasing the inferred H.O.D. parameters and characteristic halo mass scales.

The angular scale dependence of the one-halo term is determined by the Fourier transform of the spatial distribution of sources within haloes (sometimes called $u_{\rm gal}$). Setting $u_{\rm gal}=u_{\rm DM}$, the Fourier space density profiles of a spherical NFW \citep{navarro/etal:1996} halo, truncated at the virial radius, is a convenient choice, but there is no evidence that it is an accurate reflection of the distribution of dusty sources. There is, indeed, considerable evidence that galaxies near the center of groups and clusters, where the NFW dark matter density profile peaks, are forming stars less actively than those closer to the outskirts (e.g. \citeauthor{kennicutt:1983} 1983, \citeauthor{hashimoto/etal:1998} 1998, \citeauthor{bai/etal:2006} 2006, \citeauthor{bai/etal:2007} 2007; also \citeauthor{boselli/gavazzi:2006} 2006 and references therein). These systems are relevant for the calculation of the one-halo power, which receives no contribution from smaller haloes hosting a single galaxy, and allowing the CIB sources to lie further from the center of their halo than a virial radius has a non-negligible effect on the one-halo power at scales of a few arcminutes \citep{viero/etal:2009,addison/etal:prep}.

It should be noted that the above issues are related to how the halo model is implemented rather than being limitations of the formalism itself. It is not clear, however, that quantifying the uncertainties induced in the halo model calculations by scale-dependent halo bias or alternative spatial distributions of dusty sources in haloes may be straightforwardly accomplished. It is also unclear that quantities such as the halo mass function, large-scale halo bias, and halo concentration, are sufficiently well known at $z\gtrsim2$ that their uncertainties may be safely neglected, given the excellent signal-to-noise of current and future SPIRE and \emph{Planck} measurements. For these reasons, we do not attempt a halo model analysis in this work, instead phenomenologically parametrizing the CIB source bias as a function of redshift and scale directly.

This work could be considered an extended version of the analysis of A12, constraining a more physically motivated model using higher quality spectra, as well as number counts. Compared to the CIB evolution model of \citeauthor{bethermin/etal:2011} (2011, hereafter B11), this work uses the angular power spectra and counts together, rather than just one-point statistics (counts and LFs). B11 utilise data from a wider range of frequencies. Our selection is restricted to the MIPS 70 $\mu$m data and lower frequencies; while our SED parametrization allows phenomenologically for the presence of hot dust components (see Section 2.2), our focus is on the thermal dust emission associated with star formation. We also do not consider a model for polycyclic aromatic hydrocarbon (PAH) or other spectral line emission, which begin to contribute to the CIB at higher frequencies \citep[e.g.][]{leger/puget:1984,lagache/etal:2004}. An expanded SED model, and data from, for example, IRAS, or the Wide-Field Infrared Survey Explorer \citep[WISE;][]{wright/etal:2010}, will be included in our model at a future date.

The layout of this paper is as follows: in Section 2 we describe the model parameters we are attempting to constrain with our fitting, in Section 3 we describe our treatment of the data and uncertainties, results are presented in Section 4, and a discussion and conclusions follow in Sections 5 and 6. Calculations are performed using a flat, $\Lambda$CDM cosmology, with $\Omega_m=0.2715$, $\Omega_b=0.0455$, $h=0.704$, $n_s=0.967$, and $\sigma_8=0.81$ \citep{komatsu/etal:2011}.

\section{Model}

In this section we present our model for the spectral and clustering properties of the dusty sources and show how it may be used to predict various statistics. Our model is divided into three parts -- the luminosity function, which describes the abundance of dusty sources as a function of redshift and integrated dust luminosity, the spectral energy distribution, which contains the frequency dependence of the dust emission, and the clustering properties (i.e. bias), which connects the spatial distribution of the sources to that of the underlying dark matter.

Existing studies have largely dealt with only one or two of these three components at a time. B11 constrained the luminosity function evolution assuming fixed source SEDs from earlier work \citep{lagache/etal:2003,lagache/etal:2004}. \cite{penin/etal:2012b} then calculated predictions for the angular power spectra using the constraints obtained by B11. Similarly, \cite{xia/etal:2012} fixed the spectral properties of the CIB sources using the results of \cite{granato/etal:2004} and \cite{lapi/etal:2011}, and constrained source clustering properties using the spectra. We favour a combined analysis because, while it is more complex, it is capable of obtaining more robust parameter constraints and performing a more thorough exploration of the parameter degeneracies. Both \cite{shang/etal:2012} and \cite{viero/etal:2013} attempted to constrain elements of the LF, SED and clustering properties simultaneously, however these analyses failed to successfully fit the \emph{Planck} and SPIRE data, respectively. This may be partly because of issues with their implementations of the halo model, discussed earlier, and also because of overly simplistic parametrizations of the LF and SED, and insufficient freedom in the evolution of these quantities (discussed further in Section 4).

Given that the parametrization we adopt is largely phenomenological, an important question is whether our model is capable of reproducing any data outside that directly used to constrain it, and we make comparisons with several other data sets in Section 5.

\subsection{Luminosity function}

For brevity we refer to the bolometric thermal dust luminosity, $L_{\rm dust}$, integrated over $8<\lambda/\mu {\rm m}<1000$, simply as $L$. The luminosity function, $\Phi$, is defined such that the comoving number density of sources with luminosity in the interval $[L,L+dL]$ is given by
\be
\Phi(L)\,d\log L=\frac{dN}{dV_c}.
\ee
Here, and throughout, `$\log$' refers to the base-10 logarithm. We adopt a double-exponential form for $\Phi$:
\be
\Phi(L,z)=\Phi_c(z)\left(\frac{L}{L_c(z)}\right)^{1-\alpha_{\rm LF}}\exp\left(-\frac{1}{2\sigma^2_{\rm LF}}\log^2\left[1+\frac{L}{L_c(z)}\right]\right),
\ee
with normalisation $\Phi_c$, characteristic luminosity $L_c$, power law index $\alpha_{\rm LF}$ and spread parameter $\sigma_{\rm LF}$. This parametrization was introduced by \cite{saunders/etal:1990}, and has been used in a range of CIB source analyses \citep[e.g.][B11]{pozzi/etal:2004,lefloc'h/etal:2005,caputi/etal:2007}. The number of faint sources diverges for $\alpha_{\rm LF}>1$ but we choose to allow this behaviour since, provided the total luminosity is convergent (i.e. $\alpha_{\rm LF}<2$), it is not necessarily prohibited by the data.

We parametrize the evolution of the characteristic dust luminosity, $L_c$, and normalisation, $\Phi_c$, using expansions about a pure power law in $1+z$,
\be
\begin{split}
\ln L_c(z)&=\ln L_0 + \epsilon_L\ln(1+z)+\zeta_L\ln^2(1+z)\\
L_c(z)&=L_0(1+z)^{\epsilon_L}\exp\left[\zeta_L\ln^2(1+z)\right],
\end{split}
\ee
and, similarly,
\be
\Phi_c(z)=\Phi_0(1+z)^{\epsilon_{\Phi}}\exp\left[\zeta_{\Phi}\ln^2(1+z)\right].
\ee
These expression reduce to a simple power law if $\epsilon$ is the only non-zero evolution parameter, however more complex evolution, featuring a turning point or plateau, is allowed if the higher order parameters are non-zero. This expansion ensures all derivatives are continuous, and, if a quantity peaks at a particular redshift, this redshift will emerge naturally (provided enough terms of the series can be constrained), without having to be separately fitted for or put in by hand. The baseline model contains only as many higher order evolution parameters for each quantity as improve the model likelihood; some alternative or extended parametrizations are considered in Section 4.

We do not find significant evidence for evolution of the shape parameters $\alpha_{\rm LF}$ and $\sigma_{\rm LF}$ with redshift (see discussion in Section 4.2).

\subsection{Spectral energy distribution}

The flux observed at frequency $\nu$ from a source at redshift $z$ with bolometric luminosity $L$ is given by
\be
S_{\nu}(L,z)=\frac{L_{\nu(1+z)}}{4\pi(1+z)\chi^2(z)},
\ee
where $\chi$ is the comoving distance to redshift $z$, and the luminosity as a function of frequency is related to the SED function $\phi_{\nu}$ via
\be
L_{\nu}=\frac{dL}{d\nu}=L\frac{\phi_{\nu}}{\int d\nu'\phi_{\nu'}},
\ee
where the integral in the denominator of the fraction is taken over 8 $\mu$m $<c/\nu<1000$ $\mu$m.

Our model is intended to phenomenologically capture the global properties of dust emission, such that all the parameters of the model are constrained from the data, while still providing a good fit. We consequently make several simplifying assumptions when modelling the dust SED. Most importantly, we consider only a single SED at each redshift, motivated by recent studies that have found relatively little scatter about such a `universal' SED shape for observed dusty galaxies \citep{elbaz/etal:2011,lapi/etal:2011}. 

A graybody function of the form $\phi_{\nu}\propto\nu^{\beta}B_{\nu}(T)$ has been used extensively to model FIR SEDs, where the emissivity spectral index $\beta\sim1-2$ is related to the emission properties of the dust grains \citep{hildebrand:1983}. Various modifications to a single-temperature model have been suggested to better fit observed dusty galaxy SEDs \citep[e.g.][and references therein]{blain/etal:2003,hayward/etal:2012} and we find that, indeed, the data reject a single-$T$, single-$\beta$ dust component. We adopt a six-parameter model to describe the dust SED and its evolution. There are two dust components, one at temperature $T_a$, which evolves with redshift as
\be
T_a(z) = \left\{
         \begin{array}
         {l@{\quad {\rm for}\quad}l}
 T_0 & z\leq z_T \\
 T_0\left(\frac{1+z}{1+z_T}\right)^{\epsilon_T} &  z>z_T,
 \end{array}
\right.  \ee 
that is, constant below $z=z_T$ and then rising as a power law in $1+z$, and the other at a fixed temperature, $T_b$. We would expect dust temperature to increase naturally with increasing redshift due to the increasing temperature of the CMB \citep[e.g.][]{blain:1999b}, but find that, for the range of SED parameter values preferred by the data, this effect is negligible over the relevant redshift range. A single spectral emissivity index, $\beta$, is common to both dust components, and the final parameter, $f_b$, describes the contribution of the `$b$' dust component to the total dust luminosity. The SED, summing the contributions from both dust components, is given by
\be
\phi_{\nu}\propto\frac{\nu^{\beta}B_{\nu}(T_a)}{\int d\nu'\nu'^{\beta}B_{\nu'}(T_a)}+f_b\frac{\nu^{\beta}B_{\nu}(T_b)}{\int d\nu'\nu'^{\beta}B_{\nu'}(T_b)},
\ee
so that $f_b=1$ corresponds to the `$a$' and `$b$' components making equal contributions to the total luminosity.

The form of the temperature evolution in equation (7) is motivated by the fact that the data used show a strong preference for an increase in temperature, but appear to disfavour a continuous rise from redshift zero. We also investigated the effect of adding a flattening of the $T_a-z$ relation at high redshift, but found no evidence for this behaviour. We show in Section 4.3 that, while somewhat arbitrary, the parametrization we adopt yields results consistent with more physical analyses of SED evolution from measurements of individual galaxies.

We considered a range of modifications to this model, such as allowing a separate emissivity index for each dust component, or evolution in the emissivity index. For the data considered, none of these modifications lead to sufficient improvements in the model likelihood to warrant their inclusion in the baseline model (see Section 4.3).

\cite{hall/etal:2010}, \cite{shang/etal:2012} and \cite{viero/etal:2013} follow \cite{blain:1999a} and replace the graybody Wien tail in the rest-frame mid-infrared (MIR) with a power law in frequency, $\phi_{\nu}\propto\nu^{-2}$, as a phenomenological way to account for the presence of hotter dust without explicitly parametrizing additional dust components. We choose to parametrize a second dust component directly in the baseline model, in order to facilitate, for instance, assessing the effect of allowing different emissivity indices between the two components, or evolution in their relative contribution to the total dust luminosity, however, we also investigated a range of SED models involving a power-law modification to the graybody Wien tail. We find that the two approaches yield similar results, provided evolution in dust temperature is allowed in each case (see Section 4.3). While an SED falling as a power law may provide a more realistic description of the emission from very small dust grains shortwards of the SED peak \citep[e.g.][]{desert/etal:1990}, there is sufficient flexibility in our baseline parametrization to mimic this behaviour for the data considered.

It is important to note that the extent to which physical dust properties are captured by our model is not clear. Varying degrees of correlation between luminosity and dust temperature, not included in our parametrization, are observed in different samples of dusty sources (e.g. \citealt{dunne/etal:2000}; \citealt{dunne/eales:2001}; \citealt{blain/etal:2003}, and references therein; \citealt{chapman/etal:2005}; \citealt{sajina/etal:2006}; \citealt{greve/etal:2012}). There is, however, considerable scatter in this relation, both within and between different samples, and the extent to which it applies to the distant faint sources that contribute significantly to the angular power spectra, is unknown. Similarly, we have not considered the effect of dust self-absorption (the graybody form we adopt is only valid in the optically thin limit), which may be significant, particularly at high redshift \citep[e.g.][]{benford/etal:1999,blain/etal:2003,draine:2006,hayward/etal:2012}. The fact that we see good consistency between our model predictions and various data over a range of frequency suggests our treatment is, however, at least phenomenologically reasonable, and it seems likely that the effects of the $L-T$ correlation or dust opacity, and their evolution, may be partially absorbed into our SED parameters (see Section 4.3 for further discussion).

\subsection{CIB source clustering}

We write the three-dimensional power spectrum of dusty galaxies as
\be
P_{\rm gal}(k,z)=\langle b_{\rm gal}(k,z)\rangle^2P_{\rm DM}(k,z),
\ee
where the bias, $\langle b_{\rm gal}\rangle$, is averaged over all sources at a given redshift. We model the scale and redshift dependence of the bias as
\be
\langle b_{\rm gal}(k,z)\rangle=b_0(1+z)^{\epsilon_{\rm bias}}\left[1+A_{\rm bias}\left(\frac{k}{k_0}\right)\right]\exp\left(-\frac{k^2}{k_c^2}\right),
\ee
where the pivot scale, $k_0$, is chosen to be 1 Mpc$^{-1}$, and $b_0$, $\epsilon_{\rm bias}$, $A_{\rm bias}$, and $k_c$ are free parameters. In our chosen model, $k_c$ provides a cut-off scale at small scales, similar to that arising from $u_{\rm gal}$ in the halo model. We choose an expansion in powers of $k$ to mimic scale-dependent halo biasing, finding that the current data do not require a quadratic term in $k$, and that its inclusion does not impact our results.

A range of galaxy biasing treatments exist in the literature; we repeated our analysis using both the `$Q$-model' \citep{cole/etal:2005}, with $P_{\rm gal}=\frac{1+Qk^2}{1+Ak}P_{\rm DM}$, and a pure power law, $P_{\rm gal}\propto k^{-\gamma}$ (such that on small scales the clustered angular power spectrum $C^C_{\ell}\propto\ell^{-\gamma}$), motivated by the remarkable success of this simple form in describing the clustering of various galaxy populations \cite[e.g.][and references therein]{watson/etal:2011}. We find that neither of these forms provides a better fit than the baseline model. The power law form is disfavoured at a significance level of over $3\sigma$, indicating, for the first time, a preference for a deviation from power-law clustering of unresolved CIB sources, which arises from the wide range of angular scale probed by SPIRE and \emph{Planck}. We find $\gamma=1.33\pm0.03$, somewhat higher than, though not inconsistent with, the value of $1.25\pm0.06$ from A12, obtained using spectra spanning a similar range of frequency.

Importantly, adopting these alternative $P_{\rm gal}$ descriptions leads to minimal changes in the LF and SED parameters; these other parts of our model are largely decoupled from the source biasing treatment due to including both counts and spectra in our fit (see Section 4.1).

For simplicity, we take the dark matter power spectrum in equation (10) as the linear dark matter power spectrum, but find that treating the bias as being relative to the nonlinear matter power spectrum instead has, again, very little effect on the LF and SED parameters. We calculate the linear matter power spectrum using the CAMB\footnote{http://camb.info/} distribution \citep{lewis/etal:2000}.

Having described the dusty source properties we aim to constrain, we now turn to how model predictions necessary to perform a fit to real data may be calculated. A summary of the model parameters is given, along with the marginalised parameter constraints, in Section 4.

\subsection{Angular power spectrum}

The angular power spectrum of CIB anisotropies from correlating a pair of maps at frequency $\nu$ is written as the sum of clustered and Poissonian components \citep[e.g.][]{peebles:1980,bond:1996}:
\be
C_{\ell,\nu}=C^C_{\ell,\nu}+C^P_{\ell,\nu},
\ee
where $\ell$ is the multipole moment. We take the Poisson contribution, which can be thought of arising from the correlation of the image of a source in one map with the image of the same source in the other map, to be independent of angular scale. On scales larger than $18''$ (the beam FWHM of the 250 $\mu$m SPIRE detector), all but fairly nearby galaxies will appear as point sources, and we assume that any local, extended sources, which may contribute scale-dependent Poisson power, since some internal structure is resolved, are masked from maps prior to calculating spectra (for instance by cross-matching with existing catalogues).

In the flat-sky limit \citep{limber:1953,kaiser:1992}, the clustered power is written as a single line-of-sight integral \citep[e.g.][]{bond/etal:1991c,kashlinsky/odenwald:2000,knox/etal:2001,tegmark/etal:2004}. Since the number of faint sources is not required to be finite in our model, it proves helpful to work in terms of the comoving emissivity density, $j_{\nu}$ \citep{haiman/knox:2000,knox/etal:2001}. If source bias and luminosity are uncorrelated (see below), the clustered power can be written as
\be
C^C_{\ell,\nu}=\int \frac{dz}{\chi^2}\,\frac{d\chi}{dz}\frac{1}{(1+z)^2}\langle b_{\rm gal}(k,z)\rangle^2\,\langle j_{\nu}(z)\rangle^2_{\rm cut} P_{\rm DM}(k,z),
\ee
where wavenumber $k=(\ell+1/2)/\chi(z)$, and $\langle j_{\nu}\rangle_{\rm cut}$ is the mean emissivity density once bright sources (with flux $S>S_{\rm cut}$) have been masked from the map, given, in terms of the quantities defined above, by
\be
\langle j_{\nu}(z)\rangle_{\rm cut}=(1+z)\chi^2\int_0^{S_{\rm cut}}dS_{\nu}\frac{dL}{dS_{\nu}}\frac{\Phi(L,z)}{L\ln 10}S_{\nu},
\ee
where here $L$ is considered a function of $S_{\nu}$, by inverting equations (5) and (6). The fact that $S_{\nu}$ and $L$ are related by a simple constant at each redshift in our model simplifies this calculation.

\cite{bethermin/etal:2012a} recently found evidence for dependence of star formation rate (and thus bolometric IR luminosity) on halo mass through abundance matching, and \cite{wang/etal:2013} also found that the SFR-halo mass relation is not flat but has a peak at some characteristic mass scale \citep[see also][and references therein]{behroozi/etal:2013b}. Since the halo bias increases with mass \citep[e.g.][]{sheth/tormen:1999,tinker/etal:2010}, we may therefore expect a correlation between dust luminosity and bias. To account for this, we considered allowing $b_{\rm gal}$ to depend on $L$ as well as $z$ and $k$ via the replacement
\be
\begin{split}
\langle b_{\rm gal}(k,z)\rangle&\langle j_{\nu}(z)\rangle_{\rm cut}\to\\
&(1+z)\chi^2\int_0^{S_{\rm cut}}dS_{\nu}\frac{dL}{dS_{\nu}}\frac{\Phi(L,z)}{L\ln 10}S_{\nu}\,b'_{\rm gal}(k,z,L).
\end{split}
\ee
We investigated a range of treatments for the dependence of $b'_{\rm gal}$ on $L$, and found that, for the data considered, allowing a bias--luminosity correlation serves primarily to degrade constraints on $b_0$, and does not have a significant effect on either the goodness-of-fit or the LF and SED parameters. We therefore do not include such a correlation in the baseline model; more discussion is provided in Section 4.4.

\subsection{Poisson anisotropy power}

Existing studies have typically treated the Poisson power either by adding an additional free parameter to each spectrum \citep[e.g.][]{hajian/etal:2012,amblard/etal:2011,xia/etal:2012}, or using the predictions of the B11 CIB model \citep{planckcib:2011,shang/etal:2012}. We instead require that our model for the dusty source LF and SEDs successfully reproduce the Poisson, as well as the clustered component of the angular power spectrum. In terms of the quantities introduced earlier, the Poisson contribution to the anisotropy power is given by
\be
C^P_{\nu}=\int dz\frac{d\chi}{dz}\chi^2\int_0^{S_{\rm cut}}dS_{\nu}\frac{dL}{dS_{\nu}}\frac{\Phi(L,z)}{L\ln 10}S^2_{\nu}.
\ee
Since the Poisson power is weighted more towards the brighter sources, one may expect that fitting to deep number counts already constrains the Poisson power. We find that, in fact, this is not the case, and that the Poisson tail in the SPIRE, ACT, and SPT spectra carries additional constraining power (see Section 5.5). The fact that there is useful signal in the shot-noise is one of the main results of this work.

Removing the contribution from sources above $S_{\rm cut}$ in the integrals in equations (13) and (15) does not exactly mimic the removal of bright sources in confused maps, which typically involves masking a region of roughly a beam size per source, and may therefore also remove flux from from objects in the same group or cluster. In Section 5.5 we show that any bias introduced by this issue is likely to be small. Constructing a simulated CIB sky map from our theoretical model, and utilising the actual masking scheme that was used on the real data, will be used as a more rigorous test in future work.

\subsection{Differential number counts}

The differential number counts are modelled using the quantities introduced above as (see, e.g. discussion in B11)
\be
\frac{dN}{dS_{\nu}}=\int dz\frac{d\chi}{dz}\chi^2\frac{dL}{dS_{\nu}}\frac{\Phi(L,z)}{L\ln 10},
\ee
where, as in equation (14), $L$ is taken as a function of $S_{\nu}$ and $z$. The Euclidean-nomalised counts (multiplied by a factor of $S^{2.5}$) are integrated over the finite width of each flux bin to compare with the measured counts.

\section{Data treatment and model fitting}

\begin{table*}
  \centering
  \caption{Angular power spectra used for constraining the baseline model}
  \begin{tabular}{lcccccl}
\hline
Instrument&$\nu_0$&Bandpowers&Angular scale&$S_{\rm cut}^{\dagger}$&$\sigma_{\rm cal}^{\ddagger}$&Reference\\
&(GHz)&&&(mJy)&(per cent)&\\
\\
\hline
\emph{Herschel}-SPIRE&1200&23&$2000<\ell<63300$&50&15&\cite{amblard/etal:2011}\\
&857&21&$2000<\ell<41200$&50&15\\
&600&19&$2100<\ell<26300$&50&15\\
\\
\emph{Planck}-HFI&857&9&$80<\ell<2240$&710&9$^*$&\cite{planckcib:2011}\\
&545&9&$80<\ell<2240$&540&9$^*$\\
&353&9&$80<\ell<2240$&325&4$^*$\\
&217&9&$80<\ell<2240$&245&4$^*$\\
\\
ACT&219.6&13&$2391<\ell<9900$&20$^{\dagger}$&7&\cite{das/etal:2011}\\
\\
SPT&219.6&15&$2000<\ell<9400$&6.4$^{\dagger}$&4.8$^{\ddagger}$&\cite{reichardt/etal:2012}\\
\hline
\end{tabular}
\begin{center}
$^{\dagger}$ level above which bright sources are masked in the map prior to calculating spectra -- $S_{\rm cut}$ is given at 150~GHz for ACT and SPT (see text)\\
$^{\ddagger}$ absolute map calibration uncertainty: given in units of power for the SPT spectrum, otherwise in units of flux or temperature\\
$^*$ we take the \emph{Planck} calibration uncertainties to be 9 or 4 per cent, rather than the nominal 7 or 2 per cent, to account for uncertainty in the bandpass filter responses (Section 3.6.1)\\
\end{center}
\end{table*}

\subsection{Data sets used}

We constrain the model parameters described in Section 2 using a joint fit to the following data sets:
\begin{enumerate}
\item differential number counts from \emph{Spitzer}-MIPS at 70 and 160 $\mu$m \citep[4300 and 1900 GHz;][]{bethermin/etal:2010}
\item differential number counts from \emph{Herschel}-SPIRE at 250, 350 and 500 $\mu$m \citep[1200, 857 and 600 GHz;][hereafter B12]{bethermin/etal:2012b}
\item angular power spectra bandpowers from SPIRE at 250, 350 and 500 $\mu$m \citep[1200, 857 and 600 GHz;][]{amblard/etal:2011}
\item angular power spectra bandpowers from \emph{Planck}-HFI at 857, 545, 353 and 217 GHz \citep{planckcib:2011}
\item angular power spectrum bandpowers from ACT at 218 GHz \citep{das/etal:2011}
\item angular power spectrum bandpowers from SPT at 220 GHz \citep{reichardt/etal:2012}
\end{enumerate}

The MIPS number counts are described in Tables 3 to 6 of \cite{bethermin/etal:2010}, and the SPIRE number counts in Tables 2, 3 and 4 of B12. The baseline model does not include the SPIRE counts obtained by stacking from the GOODS field (see Section 3.9); we also exclude the highest flux bin in the MIPS 160 $\mu$m stacked counts since the lowest flux bin from the MIPS 160 $\mu$m resolved counts covers the same flux range with a smaller uncertainty. 

We summarise key properties of the power spectra in Table 2. Note that we do not use the full range of SPIRE or ACT bandpowers in the baseline fit; we use only data from $\ell\gtrsim2000$ for SPIRE, and $\ell\gtrsim2400$ for ACT, because of uncertainty over Galactic cirrus contamination and spectrum-to-spectrum correlations in the SPIRE data, and because data from lower $\ell$ in ACT do not contribute significant constraining power once the primary CMB power spectrum is subtracted (see Sections 3.8.2 and 3.9, below). Bright sources in the ACT and SPT maps were masked based on flux at the 150 GHz bands of these instruments, not 220 GHz. Our model allows us to account for this, since, for given model parameters, we can predict both the 150 and 220 GHz flux of all sources, and so calculate the equivalent $S_{\rm cut}$ at 220 GHz and substitute these values into equations (13) and (15).

We do not consider counts from lower wavelengths than the 70 $\mu$m MIPS band because, as stated earlier, our focus is on thermal dust emission rather than other contributions -- such as emission associated with AGN, which may be non-negligible in, for instance, \emph{Spitzer} 24 $\mu$m counts  \citep[e.g.][]{bethermin/etal:2012c}, or PAH features -- to the SED. Counts from the \emph{Herschel} Photodetector Array Camera and Spectrometer \citep{berta/etal:2011} are not included as they are in good agreement with the MIPS counts and span a smaller range in flux. We do not use counts from lower frequencies than 600 GHz because of possible enhancement of the intrinsic counts by strong gravitational lensing (see Section 3.10). Our choice of spectra was limited to $\nu\geq220$ GHz due to the presence of the thermal Sunyaev Zel'dovich effect \citep[tSZ;][]{sunyaev/zeldovich:1970}, possible tSZ--CIB cross-correlations \citep[e.g.][]{shirokoff/etal:2011,reichardt/etal:2012,zahn/etal:2012,addison/etal:prep}, and the increased importance of the primary CMB anisotropies, in data at lower frequencies.

Towards the end of this study, \cite{viero/etal:2013} presented new SPIRE power spectrum results. We will consider these data in future work; for the purposes of this paper we note that the SPIRE bandpowers from \cite{amblard/etal:2011} and \cite{viero/etal:2013} are, for $\ell>2000$, in agreement within a multiplicative shift of 10 per cent \citep[Figure 7 of][]{viero/etal:2013}, which is small compared to the flux calibration uncertainty of the maps used by \cite{amblard/etal:2011} of 15 per cent (corresponding to a $>30$ per cent uncertainty in units of power).

\subsection{Model likelihood and data covariance}

We explore the parameter space using a Markov Chain Monte Carlo analysis \citep[MCMC;][]{metropolis/etal:1953} using the optimal sampling step size from \cite{dunkley/etal:2005} -- see also \cite{gelman/roberts/gilks:1996}. The results presented later in this paper were obtained from chains of approximately $6\times10^7$ steps. Running longer chains did not lead to significant shifts in the parameter covariance or marginalised constraints. Evaluating the full posterior probability at each step can be achieved in under a millisecond on a modern 16-core processor, meaning even these long chains take no more than a day to run. The long chain length required to achieve convergence is a consequence of the complex parameter degeneracies rather than solely the number of parameters (which -- including model parameters and the additional parameters described below -- is 40 for the baseline model).

The posterior probability is proportional to the product of the likelihood, ${\cal L}$, and the prior probability. We work with the negative log-likelihood, given by
\be
-2\ln {\cal L}=\sum_{i=1}^N[{\bf M}_i(\theta)-{\bf D}_i]^T\cdot {\bf C}^{-1}_i\cdot[{\bf M}_i(\theta)-{\bf D}_i],
\ee
where $i$ labels the different data sets, the elements of the data vectors ${\bf D}_i$ are either the binned spectrum bandpowers (see below) or the binned number counts, as appropriate, ${\bf M}_i(\theta)$ are the corresponding vectors of model predictions, which depend on the model parameter values at each step, denoted by $\theta$, and ${\bf C}_i$ are the covariance matrices, which contains the data uncertainties.

Note that we include off-diagonal data covariance elements for many of the data sets. These are discussed further in Sections 3.7 and 3.9.

At each step of the MCMC chain, the likelihood is multiplied by the contribution from parameter priors described later in this section. In particular, Gaussian priors are adopted on the photometric calibration parameters, $f_{\rm cal}$, which are discussed in Section 3.5.

The remainder of this section contains discussion of various issues relating to fitting models to the counts and angular power spectra. We draw the reader's attention to our treatment of the tension between the SPIRE and \emph{Planck} spectra discussed by \cite{planckcib:2011} and \cite{viero/etal:2013}. These papers suggest that the discrepancy may arise from incorrect calculation of the SPIRE effective beam areas, or uncertainty in the \emph{Planck} beam shape, respectively, and we allow for both these possibilities as described in Section 3.7, below.

\subsection{Binning of spectra in $\ell$}
The measured power spectra were calculated using bins in $\ell$ rather than individual $\ell$ values. The resulting bandpower values $C_b$ are related to the unbinned $C_{\ell}$s by
\be
C_b=\frac{\sum_{\ell}W_{b,\ell}C_{\ell}}{\sum_{\ell}W_{b,\ell}},
\ee
where the sum is over all $\ell$ values in bin $b$, and $W_{b,\ell}$ is a weight function, which is equal to unity for the SPIRE spectra, and $\ell$ for the \emph{Planck} spectra \citep[see Section 4.1 of][]{planckcib:2011}. The weighting of the ACT and SPT spectra is slightly more complex; see \cite{das/etal:2011} and \cite{reichardt/etal:2012} for more details. We use the same binning and weighting scheme used for the measured bandpowers when calculating the model predictions. 

\subsection{Correlated uncertainties for stacked counts}

The deepest MIPS and SPIRE counts used in our fit were obtained by stacking on 24 $\mu$m \emph{Spitzer} sources. This process is subject to systematic uncertainties that may be expected to be correlated across bins in flux and also from band to band \citep[e.g.][]{viero/etal:2012}. This is supported by the fact that the scatter in the stacked counts is smaller than the quoted uncertainties. We assume a 50 per cent covariance between the uncertainties in the stacked MIPS counts at 70 $\mu$m and 160 $\mu$m, and a 50 per cent covariance between the uncertainties in the stacked SPIRE counts between all flux bins and across all three bands within each redshift bin. While our results are largely unaffected by changes in the assumed covariance by several tens of per cent, more detailed analysis of the covariance in future deep number count extraction is important for ensuring robust constraints can be obtained.

\subsection{Absolute flux calibration uncertainties}

The data considered in this work were obtained using instruments that do not measure absolute flux and must therefore be calibrated using ancillary measurements (see \citeauthor{stansberry/etal:2007} 2007, \citeauthor{gordon/etal:2007} 2007, \citeauthor{swinyard/etal:2010} 2010, \citeauthor{planckHFI:2011} 2011, \citeauthor{hajian/etal:2011} 2011, and \citeauthor{lueker/etal:2010} 2010, for details of the MIPS 160 $\mu$m, MIPS 70 $\mu$m, SPIRE, HFI, ACT, and SPT data, respectively). Each power spectrum was calculated from maps with the best-guess flux calibration correction applied, however the uncertainty in this calibration is not negligible compared to the statistical uncertainties in the bandpowers, and it is therefore necessary to correctly account for the calibration uncertainty during model fitting.

Let the calibration of the maps for the quoted bandpower values equal unity, and let $f_{\rm cal}$ be the factor, in flux units, that the map must be multiplied by in order to achieve the correct calibration. We fit for $f_{\rm cal}$ as nine extra free parameters (one for each spectrum), as in A12. At each step of the MCMC chain, the bandpowers and errors of each spectrum are multiplied by the corresponding $f_{\rm cal}^2$ before the likelihood is calculated. Furthermore, in order to penalise models requiring $f_{\rm cal}$ to be considerably different from one, we impose Gaussian priors on each $f_{\rm cal}$, with mean of unity and standard deviation equal to the quoted calibration uncertainty for that band. We also account for an additional effect of the calibration uncertainty on the power spectrum, namely that bright sources are not masked to the nominal flux value, $S_{\rm cut}$, but actually to $f_{\rm cal}S_{\rm cut}$, by substituting this modified value into equations (13) and (15).

Absolute flux calibration uncertainty also affects the differential number counts; the lower and upper limits of each flux bin, $S_{\rm min}$ and $S_{\rm max}$, must be replaced by $f_{\rm cal}S_{\rm min}$ and $f_{\rm cal}S_{\rm max}$, respectively, at each MCMC step, while $S^{2.5}\frac{dN}{dS}\to f_{\rm cal}^{1.5}S^{2.5}\frac{dN}{dS}$. In the time between the analyses of \cite{amblard/etal:2011} and B12, the SPIRE flux calibration uncertainty was improved from the 15 per cent quoted above to 7 per cent in each band, with a band-to-band relative calibration uncertainty of 2 per cent (values taken from the SPIRE Observers' Manual\footnote{http://herschel.esac.esa.int/Docs-SPIRE/html/spire\_om.html\#x1-880005.2.1}, hereafter SOM). We account for this by fitting for separate calibration parameters for each band for the SPIRE spectra and the SPIRE counts. Two additional calibration parameters are included for the MIPS 70 and 160 $\mu$m counts.

\subsection{Accounting for bandpass filter profiles}

The CIB intensity and anisotropy depend strongly on frequency over the wavelength range considered \citep[e.g.][]{fixsen/etal:1998,planckcib:2011}, with the anisotropy power at a given angular scale increasing by up to an order of magnitude across the microwave \emph{Planck} filters (see Figure 3 of A12). Accounting for the bandpass filter transmission profile when fitting models to the measured data is therefore important.

\subsubsection{Planck spectra}

We convert the \emph{Planck} bandpowers from temperature to flux density units using the values given in the notes accompanying Table 4 of \cite{planckcib:2011}. This conversion assumes a source SED that varies as $S_{\nu}\propto1/\nu$ (i.e. $\nu S_{\nu}=$ const.). Converting the spectra to the `real' flux units for a source with observed SED $S_{\nu}$ requires multiplying both the bandpowers and errors by a factor $f_{\rm SED}^2$, where \citep[e.g.][]{planckHFI:2011}
\be
f_{\rm SED}=\frac{\int d\nu\,\tau(\nu)\frac{\nu_0}{\nu}}{\int d\nu\,\tau(\nu) \frac{S_{\nu}}{S_{\nu_0}}},
\ee
with bandpass filter transmission profile $\tau(\nu)$ and nominal band frequency $\nu_0$ (these values are listed in Table 1). \cite{planckcib:2011} report $1/f_{\rm SED}$ values of 1.00, 1.06, 1.08, and 1.08 at 857, 545, 353 and 217 GHz, respectively, calculated using the CIB SED measured from FIRAS data by \cite{gispert/etal:2000}. We find that the CIB SED predicted by our best-fit model is very similar in shape to that measured by FIRAS (see Section 5.1), and do not make further corrections to the \emph{Planck} bandpowers. Uncertainties in the \emph{Planck} $f_{\rm SED}$ values are estimated at 2 per cent \citep{planckHFI:2011}; to allow for this we take the \emph{Planck} calibration uncertainties to be 9 per cent for the 857 and 545 GHz bands, and 4 per cent for the 353 and 217 GHz bands, rather than the nominal 7 and 2 per cent.

\subsubsection{SPIRE spectra and counts}

The SPIRE maps were made assuming point source emission and $S_{\nu}\propto1/\nu$ (see the SOM for further information). For the SPIRE spectra, we calculate $f_{\rm SED}$ values using Table 5.3 of the SOM for extended source emission and input SEDs of the form $S_{\nu}\propto\nu^{\alpha}$ with effective spectral indices, $\alpha$, of 0.0, 1.0, and 2.0, for the 1200, 857 and 600 GHz bands, respectively. These values were selected based on the frequency dependence of \cite{gispert/etal:2000} CIB SED mentioned above; the best-fit CIB SED from our model yields very similar values. A separate factor, denoted $K_{4E}/K_{4P}$ in the SOM, is also applied to account for the detector response to extended as opposed to point-like emission. These factors together yield effective $f_{\rm SED}^2$ correction factors of 0.989$^2$, 0.991$^2$, and 0.995$^2$ for the 1200, 857 and 600 GHz SPIRE spectra, respectively.

For the SPIRE counts, the SED correction is applied as $S\to f_{\rm SED}S$, and $S^{2.5}dN/dS\to f_{\rm SED}^{1.5}S^{2.5}dN/dS$, where here the conversion factors are for point-like emission, and we do not include the $K_{4E}/K_{4P}$ correction, yielding $f_{\rm SED}$ values of 0.989, 0.974, and 0.940 at 1200, 857 and 600 GHz, respectively.

\subsubsection{ACT and SPT spectra}

The ACT and SPT bandpowers are reported in units of CMB temperature, which can be converted to flux density (i.e. $\mu$K$^2$ to Jy$^2$ sr$^{-1}$) by multiplying by $(\partial B_{\nu}/\partial T|_{T=T_{\rm CMB}})^2$, where
\be
\frac{\partial B_{\nu}}{\partial T}=\frac{2k_{\rm B}\nu^2}{c^2}\frac{x^2e^x}{\left(e^x-1\right)^2},
\ee
$x=h_{\rm p}\nu/k_{\rm B}T_{\rm CMB}$, and $T_{\rm CMB}$ is the present-day CMB temperature. We adopt an effective frequency, $\nu_0$, of 219.6 GHz for performing this units conversion and evaluating the model predictions for the ACT spectra; this value accounts for the ACT bandpass filter and assumes an SED rising as $S_{\nu}\propto\nu^{3.5}$ \citep{swetz/etal:2011}. We assume that small differences in SED or uncertainties in the effective frequency can be absorbed into the ACT spectrum calibration uncertainty.

\cite{reichardt/etal:2012} report the same effective frequency as ACT, 219.6 GHz, for the response of the SPT 220 GHz channel to sources with $S_{\nu}\propto\nu^{3.5}$, and we use this value for calculating the units conversion and model predictions for the SPT bandpowers.

\subsubsection{MIPS counts}

\cite{bethermin/etal:2010} reported MIPS counts assuming, again, a source SED that varies as $S_{\nu}\propto1/\nu$. We calculate the MIPS $f_{\rm SED}$ values by integrating the \cite{gispert/etal:2000} CIB SED across the MIPS bandpass filters (equation 19), finding 0.977 and 1.012 at 70 and 160 $\mu$m, respectively. We assume, as above, that small uncertainties in these values can be absorbed into the flux calibration factors.

\subsection{Instrumental beam treatment}

Given the observed discrepancy between SPIRE and \emph{Planck} spectra \citep{planckcib:2011,viero/etal:2013}, we opt to allow additional freedom in the SPIRE and \emph{Planck} beam treatment compared to the \cite{amblard/etal:2011} and \cite{planckcib:2011} analyses.

\cite{planckcib:2011} found from a preliminary SPIRE / \emph{Planck} comparison that the discrepancy between the SPIRE and \emph{Planck} spectra could be resolved if the SPIRE beam areas were underestimated by around 5 and 10 per cent at 857 and 600 GHz, respectively. We therefore introduce an additional beam area correction parameter, $f_{\rm beam}$, for each SPIRE band, with a Gaussian prior centred at unity with a 1$\sigma$ uncertainty of 10 per cent. At each MCMC step the SPIRE bandpowers and errors are multiplied by the corresponding $f^2_{\rm beam}$ before the likelihood is calculated.

For the \emph{Planck} spectra, we use the beam error estimates given in Table 4 of \cite{planckcib:2011}, however we multiply the \emph{Planck} beam errors by five before taking the quadrature sum of the statistical and beam uncertainties, motivated by the possible SPIRE and \emph{Planck} spectrum shape discrepancy discussed by \cite{viero/etal:2013}. We also assume a 100 per cent correlation in the beam uncertainty across bandpowers in each \emph{Planck} spectrum, appropriate for a roughly Gaussian beam whose width is uncertain.

Allowing a larger SPIRE beam area correction, or increasing the \emph{Planck} beam shape uncertainties further, has no impact on our results. If the nominal treatments presented in \cite{amblard/etal:2011} and \cite{planckcib:2011} are used, the mismatch between the SPIRE and \emph{Planck} spectra is largely absorbed by the SPIRE calibration parameters, with shifts in the CIB model parameter constraints by no more than 0.5$\sigma$; our choice of beam treatment does not have a significant impact on our results (see Section 4.5).

We use the bandpower covariance matrix provided by \cite{reichardt/etal:2012}, which includes beam uncertainties, when calculating the SPT spectrum likelihood contribution.

\cite{amblard/etal:2011} and \cite{das/etal:2011} found that bin-to-bin correlations in the SPIRE  and ACT spectra, due either to correlated beam shape uncertainties, or the source mask, are small, and we also therefore neglect them in this work.

\subsection{Non-CIB contributions to spectra}

All the spectra used in this work were calculated from regions of the sky that are known to be relatively free from Galactic dust contamination, however diffuse emission (cirrus) is present; we also consider several additional non-CIB contributions to the microwave ACT and SPT spectra.

\subsubsection{Planck spectra}

The cirrus was cleaned from the \emph{Planck} maps using cross-correlations with high-resolution Galactic HI maps \citep[Section 2.5 of][]{planckcib:2011}, and the 143 GHz \emph{Planck} channel was used to remove the primary CMB contribution. We do not consider any modifications to these methods in this analysis.

\subsubsection{SPIRE spectra}

\cite{amblard/etal:2011} remove the cirrus in the SPIRE power spectrum using an extrapolation from lower wavelengths, however \cite{planckcib:2011} found that this treatment overestimated the cirrus contamination. We therefore adopt a similar treatment to \cite{debernardis/cooray:2012}, and fit for the cirrus contamination in the SPIRE power spectra along with the CIB model parameters. We further reduce the possible impact of incorrect cirrus modelling by only including SPIRE bandpowers at $\ell>2000$ in our fit. We assume that the cirrus power spectrum has the form $C_{\ell}^{\rm cirr}=A_{\rm cirr}(\ell/2000)^{-n_{\rm cirr}}$ \citep[e.g.][]{gautier/etal:1992,miville-deschenes/etal:2007}. We fit for the amplitude $A_{\rm cirr}$ at 1200 GHz, and assume a graybody frequency dependence of the cirrus flux with emissivity index $\beta=1.5$ and effective temperature $T=20$ K \citep[motivated by the findings of e.g.][]{bracco/etal:2011} to scale the cirrus to 857 and 600 GHz. \cite{planckcib:2011} found that the cirrus at these lower frequencies is negligible in the field from which the SPIRE spectra were calculated, and so we do not expect our results to be sensitive to the exact frequency dependence adopted. We adopt a Gaussian prior on $n_{\rm cirr}$, centred at 2.89, with a 1$\sigma$ spread of 0.44, using the constraint obtained by \cite{lagache/etal:2007}, but doubling the width of the uncertainty to account for possible variation of the cirrus index with frequency. We find that, in fact, even the cirrus in the 1200 GHz SPIRE power spectrum is not significantly detected in our fit, and that the exact treatment of the SPIRE cirrus has negligible impact on the CIB constraints.

\subsubsection{ACT spectrum}

We include in our model a lensed CMB component with a shape in $\ell$ calculated assuming our fiducial cosmology using CAMB, but fit for the an overall amplitude, $A_{\rm CMB}$, as a free parameter, using a Gaussian prior with a 1$\sigma$ uncertainty of 10 per cent. This freedom is more than sufficient to mimic changing, for example, $\sigma_8$ by $\pm0.025$ (roughly the 1$\sigma$ constraint obtained by \citealt{keisler/etal:2011}) on the angular scales on which the CMB makes a non-negligible contribution to the total spectrum. A more detailed treatment of the CMB power spectrum has no impact on our results, at least for a standard cosmological model.

No subtraction of Galactic dust is performed for the ACT spectrum \citep[see][]{das/etal:2011}. We include the amplitude of the subdominant ACT radio Poisson component as a nuisance parameter with a Gaussian prior of $4.1\pm0.8$ $\mu$K$^2$ (or $0.48\pm0.09$ Jy$^2$ sr$^{-1}$), adopting the mean value obtained by \cite{dunkley/etal:2011}, but conservatively doubling the measured uncertainty.

While the ACT 218 GHz band is virtually at the thermal Sunyaev Zel'dovich effect null, the blackbody kinematic Sunyaev Zel'dovich effect \citep[kSZ;][]{sunyaev/zeldovich:1980} will be present. We allow for a non-zero kSZ component, fitting for the power at $\ell=3000$ in units of $\ell(\ell+1)C_{\ell}/2\pi$, $A_{\rm kSZ}$, using a fixed shape in $\ell$ obtained from recent hydrodynamical simulations \citep{battaglia/etal:2010,battaglia/etal:2012}, and imposing a uniform prior of $0<A_{\rm kSZ}/\mu\textrm{K}^2<8$, based on constraints obtained by \cite{dunkley/etal:2011} and \cite{reichardt/etal:2012}. In the latter work it was found that a larger kSZ amplitude was possible only for very high degrees of correlation between the tSZ and CIB (over 50 per cent in units of power at $\ell=3000$), for which there appears little physical motivation \citep[see discussion in][]{addison/etal:prep}. The choice of the kSZ amplitude prior does not have a significant effect on any of the CIB model parameters.

In principle, power spectrum measurements at $\sim$220 GHz could allow meaningful kSZ constraints to be obtained in the absence of the tSZ and tSZ--CIB contributions. We find that, for the data considered, the dominant dusty source contribution is not sufficiently well-constrained for the kSZ constraints to be useful (see Section 5.3).

\subsubsection{SPT spectrum}

The lensed CMB and kSZ power are included when modelling the SPT spectrum in the same way as for ACT. Radio source Poisson power is minimal at 220 GHz with the SPT source mask; we include a radio Poisson component with amplitude fixed to the mean of the prior adopted by \cite{reichardt/etal:2012}, which corresponds to 0.16 Jy$^2$ sr$^{-1}$ in flux density units. \cite{reichardt/etal:2012} also find that Galactic cirrus contributes only at the per cent level to the total 220 GHz spectrum, and we adopt the model described in their Section 5.6, without adding any additional free parameters. Allowing more freedom in the SPT radio source or cirrus levels has a negligible impact on our results.

\subsection{Cosmic and sample variance} 

The deepest B12 number counts were obtained from an area of only half a square degree (GOODS-N field). It is apparent from Figure 10 of B12 that the counts, binned by redshift, are not always consistent between GOODS and the larger COSMOS field within the quoted error bars. There is good consistency between the counts in the lowest GOODS flux bin and those found from the $P(D)$ analysis of \cite{glenn/etal:2010}, but, since both analyses use the same field, this does not help inform our treatment of possible cosmic variance uncertainty. We choose to include only the counts from the larger COSMOS field counts in our fit. While we cannot rule out the possibility that the faint counts in the smaller field are, in fact, more representative of the global population, we find no current motivation for favouring these data; we also note that \cite{bethermin/etal:2012c} recently found that the GOODS counts systematically fell below their CIB galaxy model predictions while the COSMOS counts were fairly well-reproduced. We consider the effect of including the GOODS counts in Section 4.

On large scales, uncertainties in the \emph{Planck} and SPIRE bandpowers are dominated by the limited sampling of Fourier modes, due to finite sky coverage. For binned bandpower, $C_b$, with bin $b$ spanning $\ell_{\rm min}\le\ell\le\ell_{\rm max}$, this cosmic variance contribution is given analytically by \citep[e.g.][]{fowler/etal:2010,das/etal:2011}
\be
\sigma^2_{b,\nu}=\frac{2}{\ell_{\rm max}^2-\ell_{\rm min}^2}C^2_{b,\nu}\frac{1}{f_{\rm sky}},
\ee
where $f_{\rm sky}$ is the fraction of sky covered by the map used to calculate the spectrum. For spectra calculated from the same patch of sky, there will be a correlation in this cosmic variance uncertainty between spectra at frequencies $\nu_1$ and $\nu_2$, given, for bandpower $b$ (assuming the same binning of the two spectra), by
\be
\sigma^2_{b,\nu_1\nu_2}=\frac{2}{\ell_{\rm max}^2-\ell_{\rm min}^2}C^2_{b,\nu_1\nu_2}\frac{1}{f_{\rm sky}},
\ee
where $C^2_{b,\nu_1\nu_2}$ is the binned cross-spectrum. The effect of the spectrum-to-spectrum correlation in the SPIRE spectra may be expected to be larger than for \emph{Planck}, because around ten times less sky was used. Due to this fact, and the uncertain cirrus contamination on large scales in the SPIRE spectra (Section 3.8.2), we do not include any SPIRE bandpowers from $\ell<2000$, preferring the large-scale power to be constrained by \emph{Planck} alone. Removal of the large-scale SPIRE data does not significantly degrade parameter constraints.

We find that we can adequately account for the \emph{Planck} and remaining SPIRE spectrum-to-spectrum correlation with a simplified iterative approach (similar to that described in A12), without requiring additional calculations at each MCMC step. We re-run the MCMC chain several times. At the end of each chain, we use the best-fit model parameters to calculate the ratio $C^2_{b,\nu_1\nu_2}/(C_{b,\nu_1}C_{b,\nu_2})$ for each bandpower and for each cross-spectrum (e.g. $857\times545$, $857\times353$, $857\times217$, $545\times353$, $545\times217$ and $353\times217$ GHz for \emph{Planck}, where the multiplication symbol here denotes cross-correlation). This ratio is then multiplied by the corresponding measured auto-spectra bandpowers to calculate an estimate for the cross-spectrum bandpowers, which are used to calculate the off-diagonal covariance elements, $\sigma^2_{b,\nu_1\nu_2}$, for use in the subsequent chain. We find that good convergence is achieved after three chains, and that there is good agreement between the SPIRE cross-correlation ratios obtained by this method and those measured by \cite{viero/etal:2013}, discussed further in Section 5.6.

We note that there is a partial overlap between the Lockman Hole region used for the \cite{amblard/etal:2011} SPIRE analysis and one of the regions used to calculate the \emph{Planck} spectra. Since five separate, larger, regions are also used in the \emph{Planck} analysis, we neglect the effect of any correlation in the cosmic variance uncertainty \emph{between} the \emph{Planck} and SPIRE spectra. There is also partial overlap between the ACT and SPT coverage, but as we are fitting to these spectra on angular scales where noise dominates the error budget, we similarly neglect any correlation between the ACT and SPT spectra.

\subsection{Effect of strong gravitational lensing}

The observed bright-end CIB source counts in the submm and at longer wavelengths are expected to differ from the intrinsic counts due to the effect of strong gravitational lensing by foreground groups and clusters \citep[e.g.][]{blain:1996,perrotta/etal:2002,negrello/etal:2007,lima/etal:2010,negrello/etal:2010}. While, for any given source, this effect is not frequency dependent, the fact that lower frequencies receive a larger relative contribution from sources at higher redshift, where the lensing cross-section is higher, means that we may expect the enhancement of the intrinsic counts to increase with decreasing frequency. 

A full treatment for the effect of lensing requires modelling the distribution of lens systems and lensing cross-sections \citep[e.g.][]{lima/etal:2010b,hezaveh/holder:2011}. Furthermore, \cite{mead/etal:2010} find that baryonic physics, including active galactic nuclei feedback, may double the strong lensing cross-section for massive clusters, while \cite{hezaveh/etal:2012} show that compact sources are preferentially more strongly lensed, introducing a source-size dependence. Accounting for the lensing effect on CIB counts thus requires extensive modelling beyond that used to describe the intrinsic $\rmd N/\rmd S$.

In this work, we conservatively choose not to constrain our model with counts from lower frequencies than the SPIRE 600 GHz band. Even at this frequency, significant enhancement of the counts is possible for $S\gtrsim100$ mJy (e.g. \citealt{negrello/etal:2007}, B11, \citealt{wardlow/etal:2013}, \citealt{bethermin/etal:2012c}). The uncertainties in the B12 counts at the bright end are large due to the limited sky coverage, suggesting that the bias introduced in our parameters by ignoring the effects of the lensing is likely to be small. We consider the effect of excluding either the brightest SPIRE counts ($S\gtrsim60$ mJy), or the high-redshift ($z\geq2$) SPIRE counts completely, and find that, indeed, there is no systematic change in the parameter constraints. We ignore the effect of strongly-lensed sources on the angular power spectra because the lensing preserves surface brightness, and it is fluctuations in surface brightness that the spectrum measures. In addition, the brightest sources are masked from the maps before the spectra are calculated.

\section{Results}

\subsection{Goodness of fit and parameter constraints}

Globally, the model provides a good fit to the data, with $\chi^2/{\rm d.o.f.}=239/248=0.96$. The degrees of freedom are calculated by summing the number of bandpowers (36 \emph{Planck}, 63 SPIRE, 13 ACT and 15 SPT) and counts (105 SPIRE and 39 \emph{Spitzer} flux bins), and subtracting the number of free parameters (18 CIB model parameters, plus the SPIRE cirrus amplitude and index, and kSZ, primary CMB and ACT radio source amplitudes).

The SPIRE, \emph{Planck}, ACT and SPT spectra are individually well-fit by the model, with $\chi^2/\textrm{d.o.f.}$ of $54/63$, $35/36$, $11/13$, and $14/15$, respectively. Likewise for the SPIRE and MIPS counts, with $\chi^2/\textrm{d.o.f.}$ of $80/105$ and $34/39$. Note that the difference in the sum of the $\chi^2$ values for the individual data sets and the total value reported in the previous paragraph is due to the contribution from parameter priors, particularly flux calibration parameters (Sections 3.5 and 4.5). If the analysis is performed without allowing the SPIRE beam area correction or increased \emph{Planck} beam shape uncertainty discussed in Section 3.7 the model remains a good fit, with $\chi^2/{\rm d.o.f.}=1.04$.

The one-dimensional marginalised parameter constraints, and CIB parameter correlation matrix, are shown in Tables  2 and 3, respectively. The parameters are grouped into three sections -- luminosity function, SED and bias. While there are strong correlations within each section, a combined fit to the counts and spectra greatly reduces the correlation between parameters in different sections. In particular, the parameters describing the CIB source bias parameters are virtually decoupled from the LF and SED parameters. This is a clear advantage of combining a range of data sets. Constraints on the SPIRE cirrus scale dependence and primary CMB and ACT radio source amplitudes are not shown as they are driven by the priors given in Section 3.

\begin{table*}
  \centering
  \caption{Marginalised parameter constraints}
  \begin{tabular}{lclc}
\hline
Parameter&Marginalised&Description&Equation\\
&constraint\\
\hline
$\alpha_{\rm LF}$&$0.79_{-0.42}^{+0.40}$&power-law index describing faint-end slope of LF ($\Phi\propto L^{1-\alpha}$ for low L)&2\\
$\sigma_{\rm LF}$&$0.34\pm0.03$&LF spread&2\\
$\log L_0/L_{\odot}$&$9.8\pm0.3$&characteristic bolometric dust luminosity at redshift zero&3\\
$\epsilon_L$&$5.2\pm0.5$&first order characteristic luminosity redshift evolution&3\\
$\zeta_L$&$-0.98^{+0.27}_{-0.26}$&second order characteristic luminosity redshift evolution&3\\
$\log \Phi_0$ / gal dex$^{-1}$ Mpc$^{-3}$&$-1.9\pm0.2$&LF normalisation at redshift zero&4\\
$\epsilon_{\Phi}$&$-2.5\pm0.8$&first order LF normalisation redshift evolution&4\\
$\zeta_{\Phi}$&$-0.3\pm0.5$&second order LF normalisation redshift evolution&4\\
\\
$T_0$ / K&$24.1\pm1.4$&temperature of dust component `$a$' at $z=0$&7\\
$z_T$&$1.27^{+0.14}_{-0.15}$&redshift for turn-on of component `$a$' temperature evolution&7\\
$\epsilon_T$&$0.75\pm0.10$&component `$a$' temperature evolution index&7\\
$T_b$ / K&$52\pm7$&temperature of dust component `$b$' (does not vary with redshift)&8\\
$\beta$&$1.78\pm0.11$&dust spectral emissivity index&8\\
$f_b$&$0.46\pm0.10$&contribution of dust component `$b$' to total luminosity&8\\
\\
$b_0$&$0.84^{+0.19}_{-0.18}$&large-scale dusty source bias at redshift zero&10\\
$\epsilon_{\rm bias}$&$1.43\pm0.21$&bias redshift evolution parameter&10\\
$A_{\rm bias}$&$1.61\pm0.19$&source bias scale-dependence parameter&10\\
$k_c$ / Mpc$^{-1}$&$4.9^{+0.7}_{-0.8}$&small-scale bias truncation scale&10\\
\\
$A_{\rm cirr}$ / Jy$^2$ sr$^{-1}$&$(1.3\pm1.0)\times10^3$&Galactic cirrus power at $\ell=2000$ in SPIRE 1200 GHz spectrum\\
\\
$A_{\rm kSZ}$ / $\mu$K$^2$ CMB&$5.3^{+2.2}_{-2.4}$&amplitude of the kinematic Sunyaev Zel'dovich angular power spectrum at $\ell=3000$\\
\hline
\end{tabular}
\end{table*}

\begin{table*}
	\centering
	\caption{CIB source parameter correlation matrix}
	\begin{tabular}{lcccccccccccccccccccc}
&$\alpha_{\rm LF}$&$\sigma_{\rm LF}$&$\log L_0$&$\epsilon_L$&$\zeta_L$&$\log \Phi_0$&$\epsilon_{\Phi}$&$\zeta_{\Phi}$&&$T_0$&$z_T$&$\epsilon_T$&$T_b$&$\beta$&$f_b$&&$b_0$&$\epsilon_{\rm bias}$&$A_{\rm bias}$&$k_c$\\
\hline
$\alpha_{\rm LF}$&100&-76&89&4&-7&-6&-3&2&&-18&9&1&-30&10&39&&-33&4&1&2\\ 
$\sigma_{\rm LF}$&-76&100&-82&-5&0&22&4&1&&18&-16&-3&24&-5&-31&&27&-10&0&-3\\ 
$\log L_0$&89&-82&100&-32&28&-39&29&-29&&-1&18&24&-13&-1&44&&-32&5&2&-2\\ 
$\epsilon_L$&4&-5&-32&100&-96&72&-94&91&&-24&-32&-56&-16&14&-19&&10&-4&-5&12\\ 
$\zeta_L$&-7&0&28&-96&100&-65&91&-94&&22&40&52&14&-20&11&&-13&13&6&-12\\ 
$\log \Phi_0$&-6&22&-39&72&-65&100&-77&71&&-1&-21&-32&8&2&-34&&11&0&-1&9\\ 
$\epsilon_{\Phi}$&-3&4&29&-94&91&-77&100&-97&&10&28&43&0&-7&19&&-12&5&7&-13\\ 
$\zeta_{\Phi}$&2&1&-29&91&-94&71&-97&100&&-14&-32&-44&-4&16&-14&&17&-15&-6&13\\ 
\\ 
$T_0$&-18&18&-1&-24&22&-1&10&-14&&100&-17&29&92&-82&-50&&9&1&-8&-1\\ 
$z_T$&9&-16&18&-32&40&-21&28&-32&&-17&100&68&-23&3&27&&1&-9&13&-3\\ 
$\epsilon_T$&1&-3&24&-56&52&-32&43&-44&&29&68&100&21&-8&13&&3&-11&9&-4\\ 
$T_b$&-30&24&-13&-16&14&8&0&-4&&92&-23&21&100&-69&-52&&14&1&-8&1\\ 
$\beta$&10&-5&-1&14&-20&2&-7&16&&-82&3&-8&-69&100&40&&-11&-2&7&1\\ 
$f_b$&39&-31&44&-19&11&-34&19&-14&&-50&27&13&-52&40&100&&-17&-2&1&0\\ 
\\ 
$b_0$&-33&27&-32&10&-13&11&-12&17&&9&1&3&14&-11&-17&&100&-85&-45&17\\ 
$\epsilon_{\rm bias}$&4&-10&5&-4&13&0&5&-15&&1&-9&-11&1&-2&-2&&-85&100&21&-4\\ 
$A_{\rm bias}$&1&0&2&-5&6&-1&7&-6&&-8&13&9&-8&7&1&&-45&21&100&-60\\ 
$k_c$&2&-3&-2&12&-12&9&-13&13&&-1&-3&-4&1&1&0&&17&-4&-60&100\\ 
\hline
\end{tabular}
\end{table*}

The measured angular power spectra and best-fit model are shown in Figure 1. The bandpowers shown include the corrections for source SEDs and bandpass filter transmission described in Section 3.6, as well as the best-fit calibration parameters. Figures 2 and 3 show differential number counts from SPIRE and MIPS, respectively, with the binned counts likewise corrected for SED, filter and best-fit calibration. When comparing our model with other data sets, or predictions from other models, it is not sufficient to consider only the model curves plotted in Figures 1 to 3, which do not reflect the variation allowed by the calibration uncertainties.

\begin{figure*}
	\centering
	\includegraphics[width=150mm]{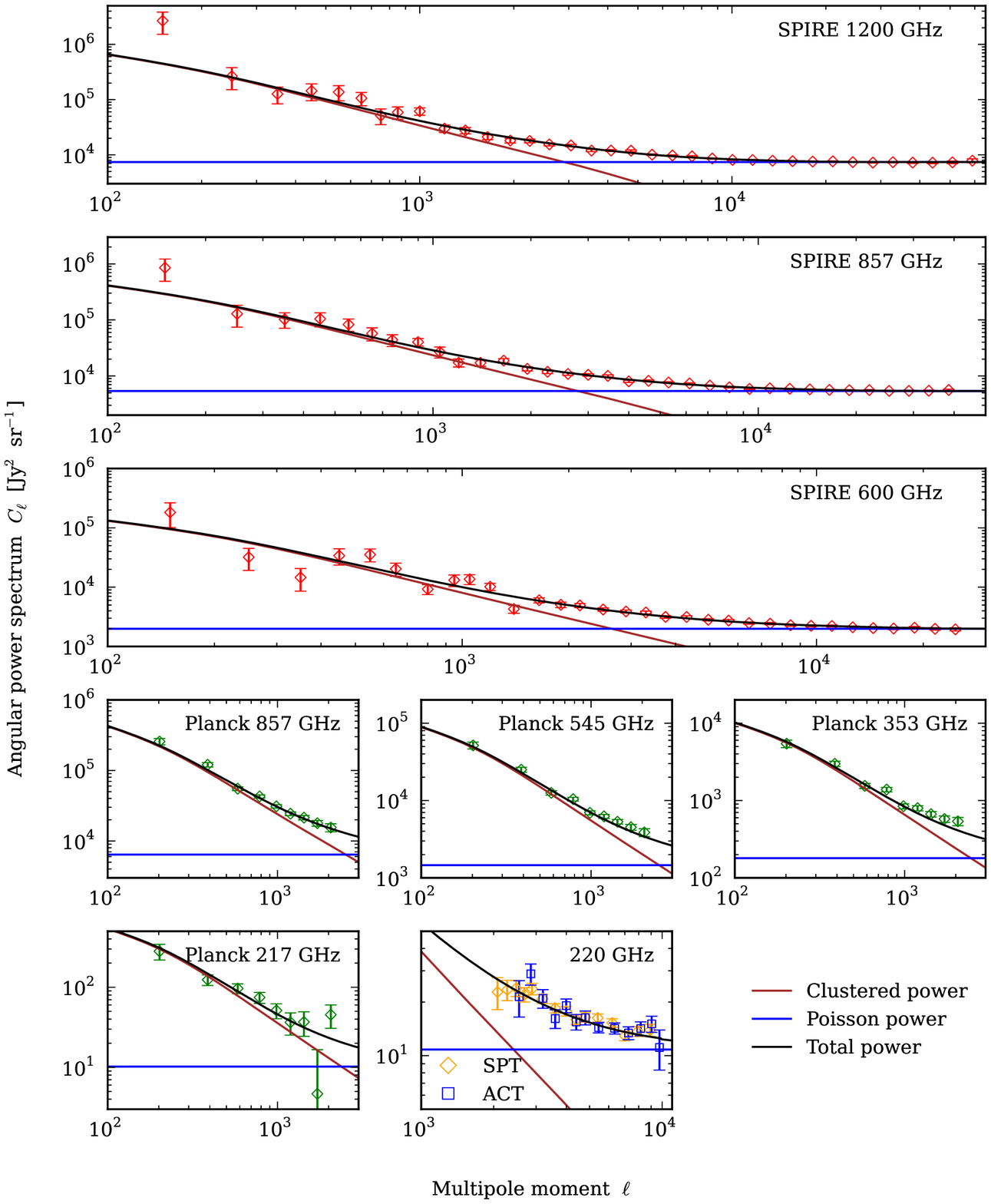}
	\caption{Angular power spectra of unresolved CIB sources from \emph{Herschel}-SPIRE \citep{amblard/etal:2011}, \emph{Planck} \citep{planckcib:2011}, ACT \citep{das/etal:2011}, and SPT \citep{reichardt/etal:2012}, with our best-fit model overplotted. The power spectra consist of a scale-independent Poisson term, and a clustered component that falls roughly as a power law. The bandpower uncertainties do not include uncertainties in map calibration (see Section 4.5). The best-fit Galactic cirrus power in the SPIRE spectra is approximately zero and not shown. SPIRE bandpowers from $\ell<2000$ were not included in the fit (Section 3), but are shown here for comparison. The \emph{Planck} bandpower uncertainties plotted include only the nominal beam uncertainties; additional freedom was allowed in the baseline model, as described in Section 3.7. The kSZ effect, Galactic cirrus and unresolved radio sources contribute to the ACT and SPT spectra but are highly subdominant to the dusty CIB source contribution. The best-fit primary lensed CMB has been subtracted from the ACT and SPT bandpowers (Section 3.7.3).}
\end{figure*}

\begin{figure*}
	\centering
	\includegraphics[width=150mm]{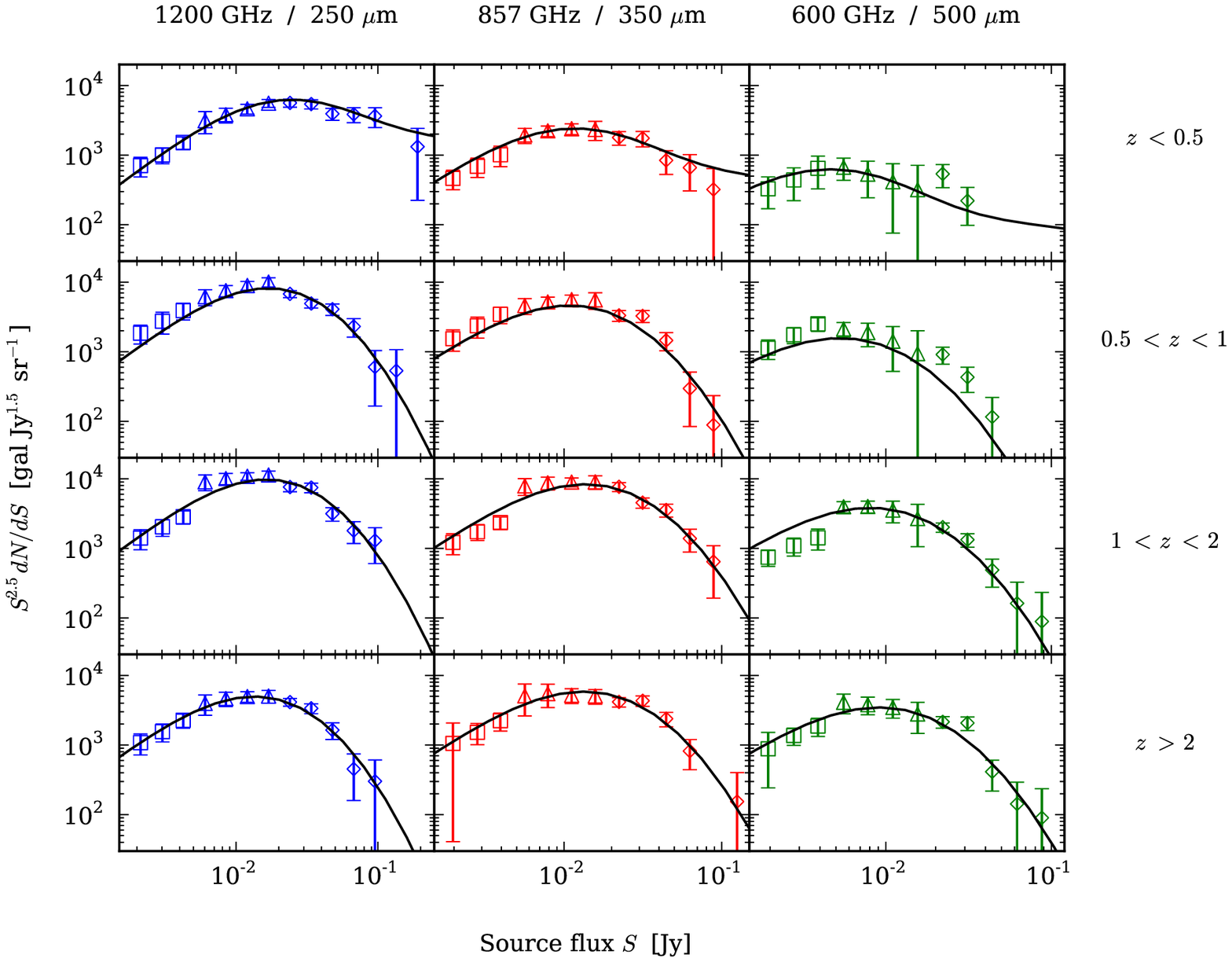}
	\caption{Far-infrared differential number counts from \emph{Herschel}-SPIRE \citep{bethermin/etal:2012b}, obtained from a stacking analysis of the GOODS-N (squares) and COSMOS (triangles) fields, and by counting resolved sources in COSMOS (diamonds), with our best-fit model overplotted. The best-fit calibration values have been applied to the measured counts, as have corrections for source SED and the SPIRE bandpass filter transmission, as described in Section 3. The GOODS counts were not included in the baseline model (Section 3.9).}
\end{figure*}

The remainder of this section deals, in turn, with the three parts of our model. We make comparisons with existing work and discuss a range of model extensions or modifications.

\subsection{Luminosity function}

While our constraint on the faint-end LF slope of $\alpha_{\rm LF}=0.79^{+0.40}_{-0.42}$ is consistent with the value of $\sim1.2$ typically obtained in the literature \citep[e.g.][B11]{saunders/etal:1990}, considerably lower values are also permitted. B11 found $\alpha_{\rm LF}=1.223\pm0.044$; their constraint is tighter than ours only because of including very local constraints from IRAS and does not correspond to a tighter constraint on the behaviour of faint high-redshift sources. Given the mild preference shown for lower values of $\alpha_{\rm LF}$, we consider allowing $\alpha_{\rm LF}$ to evolve as a power law in $1+z$, but with $\alpha_{\rm LF}$ fixed to 1.2 at $z=0$, and find that the data do indeed prefer a decrease in $\alpha_{\rm LF}$ with increasing $z$, but only at the 1.3$\sigma$ level of significance.

A range of values for $\sigma_{\rm LF}$, which determines the bright-end behaviour, has been reported in the literature. Our constraint of $0.34\pm0.03$ agrees fairly well with the $0.20^{+0.11}_{-0.07}$ obtained by \cite{caputi/etal:2007} from the 8 $\mu$m rest-frame luminosity function of star-forming galaxies at $z=1$, for instance, but appears low compared to the $0.406\pm0.019$ obtained by B11. We note, however, that $\sigma_{\rm LF}$ is somewhat sensitive to the assumed degree of correlation between the stacked count uncertainties; if we neglect this correlation entirely, as in B11, the constraint obtained is $0.38\pm0.03$, which agrees well with the B11 result. As above, we consider allowing power-law redshift evolution of $\sigma_{\rm LF}$ and find again a mild (1.2$\sigma$) preference for a decrease with increasing redshift.

Including the deep GOODS-N counts from B12 (which were not included in the baseline fit -- see discussion in Section 3.9), assuming a 50 per cent correlation between the errors on the GOODS counts in each redshift bin, as for the stacked COSMOS counts, yields tighter constraints of $\alpha_{\rm LF}=1.17\pm0.12$ and $\sigma_{\rm LF}=0.29\pm0.02$. Being able to robustly extract deep number counts is clearly important for future faint-end LF constraints at high redshift.

Figure 4 shows the evolution of $L_c$ with redshift, including $1\sigma$ uncertainty contours. We also plot $L_c(z)$ when additional freedom in the $L_c$ and $\Phi_c$ evolution is allowed (with a third order term in $\ln(1+z)$ for each -- see equations 3 and 4). A preference for an increase in $L_c$ beyond $z\sim2.5$, and deviations from pure power-law evolution in $1+z$, are apparent in both cases. To further demonstrate the current data's high-redshift constraining power, we repeated our fit with $L_c(z)$ fixed to flatten into a plateau but otherwise be completely consistent with our best-fit model. Even allowing for higher-order terms in the $\Phi_c$ evolution to compensate the lack of $L_c$ freedom, this evolution is disfavoured at the 8, 6, and 3$\sigma$ levels for plateaus at redshift 2.0, 2.5 and 3.0, respectively.

It is possible that our $L_c(z)$ results are being biased by our simplistic SED treatment, or failing to explicitly account for multiple CIB source populations. Investigation of stacked count uncertainty correlations and different LF shape parametrizations is also clearly merited. We therefore do not attempt further interpretation, for instance inferring the star formation rate density, in the present work, however, our findings strongly suggest that joint fits to number counts and power spectra can provide meaningful constraints on the evolution of the dust luminosity function and, ultimately, star formation rates at high redshift.

Our approach is completely independent from existing analyses using luminosity function measurements of resolved sources \citep[for recent examples using \emph{Spitzer} data, see, e.g.,][]{magnelli/etal:2011,patel/etal:2013}. With future FIR, submm and microwave data improvements (see Section 5.7), we may expect analyses such as ours to yield constraints on high-redshift dust-obscured star formation that are both tighter and, thanks to directly probing the peak or tail of the thermal dust emission, more robust, than those probing the rest-frame MIR.

\begin{figure}
	\centering
	\includegraphics{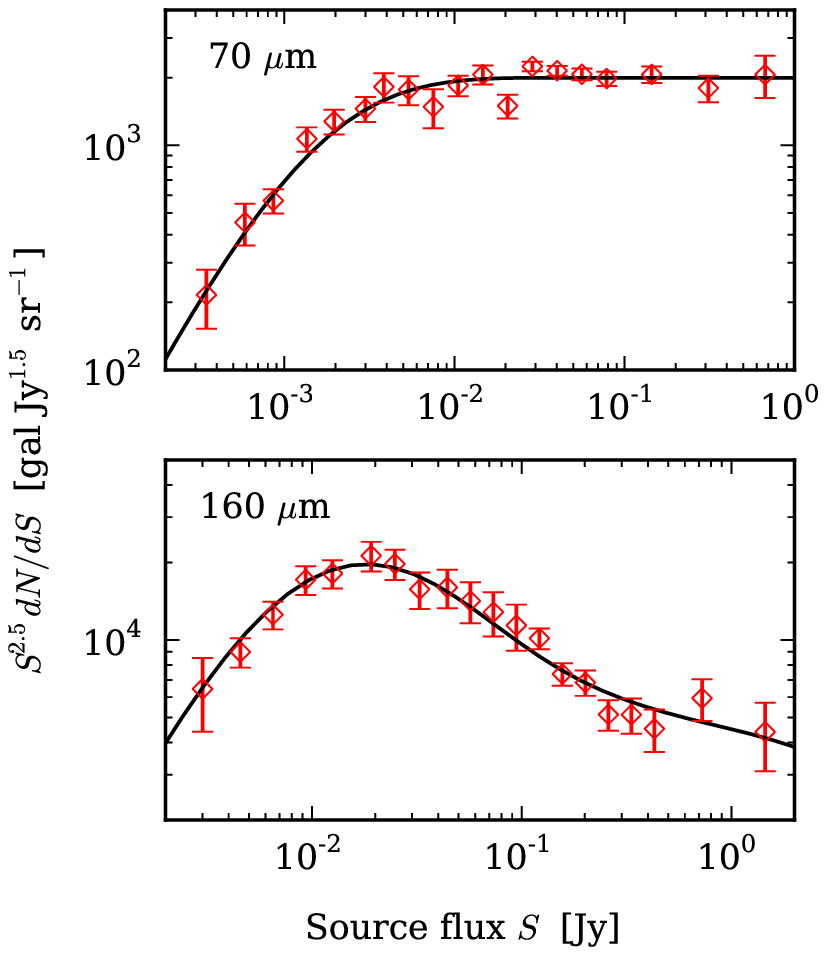}
	\caption{Far-infrared differential number counts from \emph{Spitzer}-MIPS \citep{bethermin/etal:2010}, with best-fit model overplotted. The counts are corrected for best-fit calibration, source SED and bandpass filter, as in Figure 2 (see Section 3).}
\end{figure}

\begin{figure}
	\centering
	\includegraphics{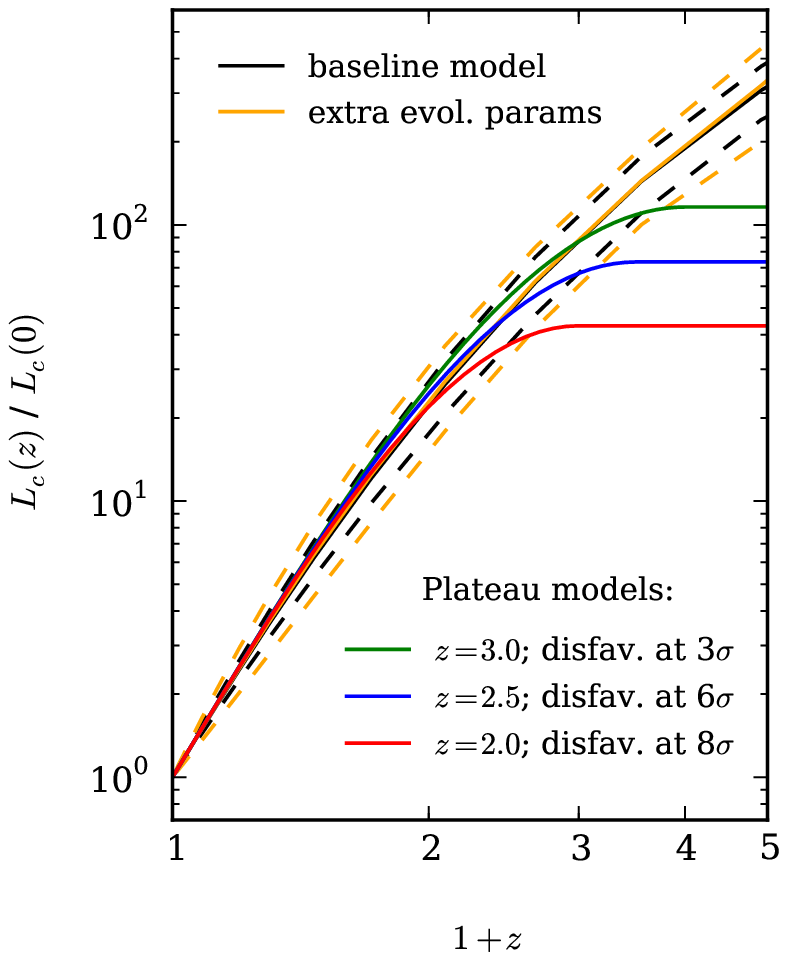}
	\caption{Redshift evolution of characteristic thermal dust luminosity, $L_c$. Model predictions are shown for the baseline model and when extra evolution parameters are added to $L_c$ and $\Phi_c$. Three evolution curves that are consistent with the baseline predictions at low redshift, but feature a plateau in $L_c$ at redshift 2.0, 2.5, and 3.0, are also shown. We repeated our fitting with the $L_c$ evolution fixed to each of these plateau models but all other parameters allowed to vary as before and found that the plateau models are disfavoured at the 8, 6 and 3$\sigma$ levels even if additional freedom is allowed in the $\Phi_c$ evolution. This demonstrates the ability of the counts and spectra to constrain high-redshift evolution.}
\end{figure}

\subsection{Spectral energy distribution}

We have been able to place tight constraints on our adopted thermal dust SED parameters, with two main results. Firstly, evolution of the temperature of the `$a$' dust component, which follows $T_a=T_0\left(\frac{1+z}{1+z_T}\right)^{\epsilon_T}$ for $z>z_T$, with $T_0=24.3\pm1.4$ K, $z_T=1.27^{+0.14}_{-0.15}$, and $\epsilon_T=0.75\pm0.10$, is strongly required (indicated by non-zero $\epsilon_T$). Secondly, an additional dust component, with redshift-independent temperature $T_b=52\pm7$ K, is also necessary -- as shown by the significant preference for non-zero $f_b=0.46\pm0.10$ (see equation 8).

It seems possible that the evolution in $T_a$ is due not to an evolution in the actual dust population, but to the increase in $L_c(z)$; as discussed in Section 2.2, a positive correlation between temperature and luminosity has been observed in various studies. The apparent evolution in dust temperature may also be associated with evolution in dust opacity. A more general form of the graybody equation is given by \citep[e.g.][]{blain/etal:2003}
\be
\phi_{\nu}\propto[1-\exp(-(\nu/\nu_c)^{\beta})]B_{\nu}(T),
\ee
which asymptotes to the optically thin limit adopted in the baseline model for small $\nu/\nu_c$. There is a strong degeneracy between $T$, $\beta$ and $\nu_c$, and we find that the effect of introducing $\nu_c$ may be largely reproduced by changes in $T_a$ and $\beta$, at least for $\nu_c\sim3000$ GHz \citep[adopting the value from][]{draine:2006}. We may expect a similar degeneracy to exist between parameters describing any redshift evolution in $\nu_c$ and $T_a$.

Regardless of the physical origin, the fact that the data require a non-zero $\epsilon_T$ at a significance level of over 7$\sigma$ strongly suggests that \emph{some} kind of SED evolution is required. We find that for the baseline parametrization, the evolution in temperature cannot be replaced with additional evolution terms in $L_c$ or $\Phi_c$, or by introducing evolution in emissivity index, $\beta$, or allowing a different $\beta$ for the `$a$' and `$b$' components.

We find that fitting for a power-law modification to the Wien tail, with free index, $\alpha_{\rm MIR}$, as an alternative to the second dust component, yields similar results to the baseline model, provided evolution in the dust temperature is still allowed. The MIR SED index is constrained to be $\alpha_{\rm MIR}=2.50\pm0.23$, somewhat higher than the value of 2.0 adopted in recent work \citep[e.g.,][]{hall/etal:2010,shang/etal:2012,viero/etal:2013}. With this modified MIR model, the temperature of the dust at low redshift is higher than $T_0$ in the baseline model by $\sim1.5\sigma$, and the emissivity index, $\beta$, is $\sim1.5\sigma$ lower. Strikingly, the evolution parameters for the dust temperature are both in excellent agreement with those of the `$a$' component in the baseline model, with temperature evolution again strongly required.

The baseline and modified MIR SED models yield similar goodness-of-fit, although the baseline model is mildly favoured. Stronger discrimination between these models is not possible with the current data because of their similarity over the range of frequency probed (illustrated in Figure 5). Data from shorter wavelengths, and an expanded SED model (including contributions other than thermal dust emission), will provide tighter constraints on the behaviour of the dust emission in the MIR range. We find no evidence for evolution in $\alpha_{\rm MIR}$ when we adopt the modified MIR power law SED, and we likewise find no evidence for evolution in $T_b$ in the baseline model. Allowing a modified MIR index \emph{in addition} to explicitly parametrizing a second temperature component does not lead to improvements in the model likelihood.

\begin{figure}
	\centering
	\includegraphics{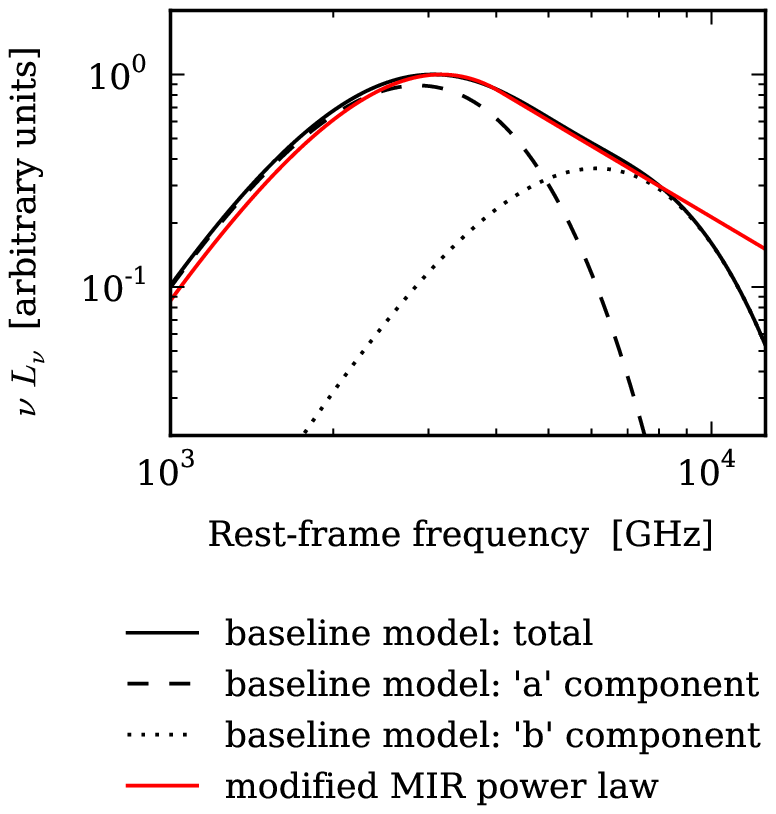}
	\caption{Comparison of the best-fit dust SED from the baseline model and an alternative model featuring a power-law modification to the MIR Wien tail instead of a second dust component, both shown at $z<z_T$, where the SED is not redshift-dependent, for the same integrated thermal dust luminosity. The shapes of the SEDs from the two models are extremely similar over a wide range of frequency, and the data considered do not strongly prefer one over the other.}
\end{figure}

While, as stated in Section 2, and discussed here, the extent to which our SED parameters can be associated with physical quantities is unclear, what our results demonstrate is that the counts and power spectra are capable of constraining the mean thermal dust SED at high redshift.

Our dust temperature and emissivity constraints show good general agreement with the $T\sim30-60$ K and $\beta\sim1.5-2$ obtained in SED fits to FIR-, submm- and mm-selected galaxies \citep[e.g.][]{dunne/eales:2001,chapman/etal:2005,coppin/etal:2006,michalowski/etal:2010a,chapin/etal:2011,greve/etal:2012}. We can also compare our model predictions with recent measurements relating to SED evolution. \cite{kirkpatrick/etal:2012} fit SEDs of 24 $\mu$m-selected sources using a two-temperature modified blackbody similar to our baseline SED parametrization. They found cold and hot dust temperatures of $25\pm2$ K and $55\pm6$ K for galaxies at $z\sim1$, and $28\pm2$ K and $59\pm5$ K at $z\sim2$. Our baseline model yields values of $24\pm1$ K and $52\pm7$ K at $z=1$, and $29\pm2$ K and $52\pm7$ at $z=2$, which are in good agreement with these measurements. \cite{magdis/etal:2012} stacked SEDs from multi-wavelength observations of galaxies and inferred an increase in effective dust temperature from $32\pm2$ to $38\pm2$ over $1<z<2$, assuming a single-temperature modified blackbody with $\beta$ fixed to 1.5. We can compare these values to our alternative single-temperature model featuring the modified MIR power law SED. Fixing $\beta=1.5$, we obtain effective temperatures of $29\pm1$ and $35\pm2$ at $z=1$ and $z=2$, again in agreement with the more direct observations.

The SED constraints obtained in our model hinge on the use of data covering a wide range of frequency; in this work, the submillimetre and microwave-band power spectra provide considerable improvement over the SPIRE and MIPS counts alone (see Section 5.5). Deep counts at lower frequencies than SPIRE (for instance, from the Submillimetre Common-User Bolometer Array-2 -- SCUBA-2 -- Cosmology Legacy Survey\footnote{http://www.jach.hawaii.edu/JCMT/surveys/Cosmology.html}) could likely fulfil this role to some extent as well, provided the effect of strong lensing on the bright-end counts can be robustly treated.

\subsection{Clustering properties}

The data strongly require scale-dependent dusty source biasing relative to the linear matter power spectrum, parametrized through $A_{\rm bias}=1.61\pm0.19$ and $k_c=4.8^{+0.7}_{-0.8}$ Mpc$^{-1}$. A preference for scale-dependent bias remains at high significance even when the non-linear matter power spectrum (calculated using HALOFIT from the CAMB distribution) is used in our clustered power calculation.

Figure 6 shows the impact of this scale dependence as a function of both $\ell$ and $k$; by $\ell\sim750$, the scale dependence results in a factor of two difference compared to the linear matter power spectrum shape. The fact that the scale dependence has a significant impact even on these angular scales -- where the one-halo term of the halo model is subdominant \citep[e.g.][]{amblard/etal:2011,xia/etal:2012,viero/etal:2013} -- suggests that it may not be valid to neglect the scale dependence of halo bias when calculating the two-halo halo model term, as discussed in Section 1.

Deviations from scale-independent bias are apparent around $k\sim0.1$ Mpc$^{-1}$ in our model, consistent with measurements from galaxy surveys and $N$-body simulations \citep[e.g.,][]{eisenstein/etal:2005,cole/etal:2005,yoo/etal:2009,fernandez/etal:2010}. At highly non-linear scales ($k\gtrsim1$ Mpc$^{-1}$), the clustered power is suppressed by the $\exp(-(k/k_c)^2)$ factor in equation (7). As stated in Section 2.3, physically this suppression is likely mimicking the roll-over the one-halo term, which occurs on scales roughly corresponding to the physical extent of a halo, in existing CIB clustering models \citep[e.g.][]{amblard/etal:2011,shang/etal:2012}.

\begin{figure}
	\centering
	\includegraphics{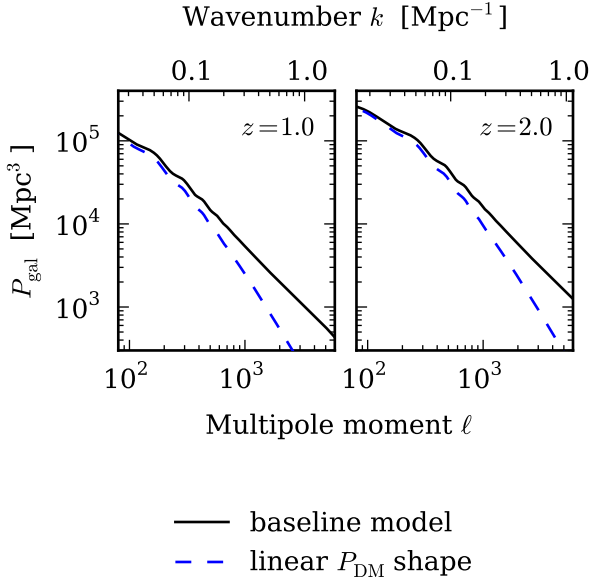}
	\caption{Dusty galaxy power spectrum, $P_{\rm gal}$, as a function of multipole moment and wavenumber (related by $k=(\ell+1/2)/\chi$ for small angles). We show the linear matter power spectra, scaled to match the large-scale $P_{\rm gal}$ behaviour, to illustrate the impact of the scale-dependent source bias, which is important for $k\gtrsim0.1$ Mpc$^{-1}$. At $z\sim1$, where the contribution to the clustered power peaks, the scale dependence results in a doubling of $P_{\rm gal}$ by $\ell\sim750$.}
\end{figure}

Figure 7 shows our constraint on the evolution of the large-scale CIB source bias. Also shown is the bias of dark matter haloes at several fixed masses as a function of redshift, using the parametric fit to $N$-body simulations presented by \cite{tinker/etal:2010}. Our baseline results are consistent with dusty sources inhabiting haloes of mass $M_{\rm halo}\sim10^{13}M_{\odot}$, which is higher than the values of $\sim10^{12}M_{\odot}$ associated with optimal star formation from recent studies \citep[e.g.][]{wang/etal:2013,bethermin/etal:2012a,behroozi/etal:2013b,behroozi/etal:2013}. Large number of fainter sources occupying massive groups and clusters may lead to an increase in the effective bias we have measured. We also find that lower values of $b_0$, corresponding to a lower characteristic halo mass, are allowed if we introduce a correlation between CIB source bias and dust luminosity (see Section 2.4). The data only provide weak constraints on such a correlation; we consider as a simple example a monotonic relation of the form
\be
b'_{\rm gal}(k,z,L)\propto b_{\rm gal}(k,z)\left(1+\gamma_{\rm corr}\frac{L}{L_c(z)}\right),
\ee
where $\gamma_{\rm corr}$ is a free parameter. In order to compare directly with the large-scale bias constraints in the absence of a $b_{\rm gal}-L$ correlation, the proportionality constant is fixed at each redshift by requiring the mean large-scale bias, averaged over all sources, to equal $b_0(1+z)^{\epsilon_{\rm bias}}$. We find a mild preference for a positive value of $\gamma_{\rm corr}$, which leads to a lower inferred large-scale bias, shown in Figure 7. As stated in Section 2.4, the data show only a mild (2$\sigma$) preference for a $b_{\rm gal}-L$ correlation of this form in terms of goodness-of-fit, and the introduction of the correlation does not impact significantly on LF or SED constraints.

Further work, involving more detailed modelling of the CIB source clustering, perhaps within a halo model framework, is required to explore the relation between dust luminosity and halo mass in more detail.

\begin{figure}
	\centering
	\includegraphics{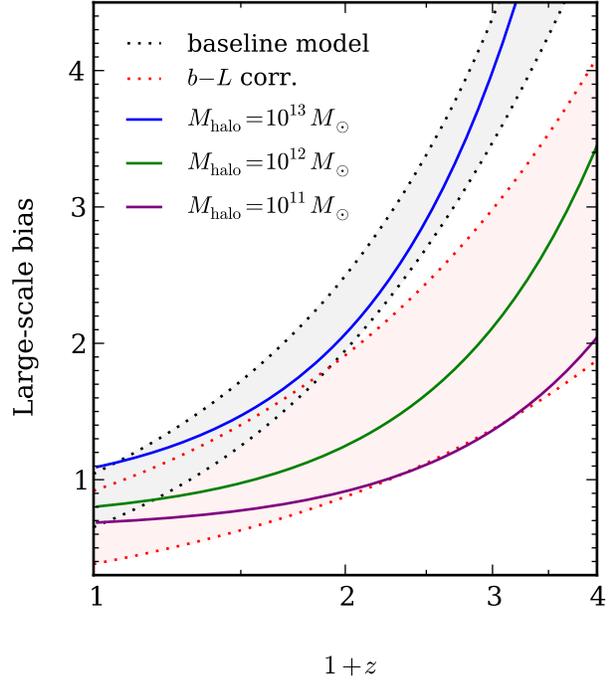}
	\caption{Evolution of large-scale CIB source bias: we show $1\sigma$ contours from both our baseline model, and with the addition of the $b_{\rm gal}-L$ correlation from equation 24. A positive $b_{\rm gal}-L$ correlation is mildly preferred, leading to lower bias values. The bias of dark matter haloes of several fixed masses are also shown as a function of redshift, using the fitting functions provided by \protect\cite{tinker/etal:2010}.}
\end{figure}

\subsection{Calibration}
Constraints on the photometric calibration parameters, which were included as additional free parameters in our fitting (Section 3.5) are shown in Table 4. The fact that the marginalised uncertainties from our fit are smaller than the nominal uncertainties is a consequence of fitting to a large number of data sets with substantial frequency overlap. As discussed in Section 3, what we are calling calibration parameters may also be absorbing uncertainties relating to bandpass filter transmission profiles and the filter response to the source SEDs. We therefore do not expect the $f_{\rm cal}$ constraints to be consistent with unity in every case.

Table 4 shows the constraints on the additional SPIRE beam area correction factors, $f_{\rm beam}$, introduced as described in Section 3.7. Also shown are the constraints obtained when neither these factors, nor the additional \emph{Planck} beam uncertainty, are allowed (``No beam corrections" column). Removing the additional beam-related freedom does not significantly affect the calibration values, except that the SPIRE spectrum calibrations absorb the effect of the beam area correction and lie further from unity. The marginalised CIB parameter constraints also change by less than 0.5$\sigma$ in all cases. 

We allowed additional freedom in the SPIRE and \emph{Planck} beam treatment in the baseline model, motivated by discussions in \cite{planckcib:2011} and \cite{viero/etal:2013}, but have demonstrated that this is not significantly affecting or biasing our results. We further note that removing either the SPIRE or \emph{Planck} spectra from the fit entirely does not qualitatively modify the shape of the luminosity function evolution or the requirement for SED evolution and the second temperature component discussed earlier in this section. Additionally, the preference for the low \emph{Planck} 545 GHz calibration remains even when the SPIRE spectra are not included.

\begin{table}
 \centering
 \caption{Marginalised constraints on calibration parameters}
 \begin{tabular}{lccc}
Data set&Baseline&No beam&Nominal\\
&model&corrections\\
\hline
\emph{Planck} 857 GHz spectrum&$0.95\pm0.04$&$0.94\pm0.04$&0.09$^{\dagger}$\\
\emph{Planck} 545 GHz spectrum&$0.88\pm0.03$&$0.86\pm0.03$&0.09$^{\dagger}$\\
\emph{Planck} 353 GHz spectrum&$1.01\pm0.03$&$0.99\pm0.03$&0.04$^{\dagger}$\\
\emph{Planck} 217 GHz spectrum&$1.06\pm0.03$&$1.04\pm0.03$&0.04$^{\dagger}$\\
\\
SPIRE 1200 GHz spectrum&$1.07\pm0.10$&$1.15\pm0.05$&0.15\\
SPIRE 1200 GHz beam&$1.07\pm0.08$&&0.10\\
SPIRE 857 GHz spectrum&$0.99\pm0.09$&$1.13\pm0.07$&0.15\\
SPIRE 857 GHz beam&$1.07\pm0.08$&&0.10\\
SPIRE 600 GHz spectrum&$1.01\pm0.09$&$1.05\pm0.04$&0.15\\
SPIRE 600 GHz beam&$1.01\pm0.09$&&0.10\\
\\
ACT 218 GHz spectrum&$1.02\pm0.03$&$1.04\pm0.03$&0.07\\
\\
SPT 220 GHz spectrum$^{\ddagger}$&$0.98\pm0.04$&$0.99\pm0.04$&0.048$^{\ddagger}$\\
\\
SPIRE 1200 GHz counts&$1.04\pm0.05$&$1.04\pm0.05$&0.07$^{*}$\\
SPIRE 857 GHz counts&$0.98\pm0.05$&$0.99\pm0.05$&0.07$^{*}$\\
SPIRE 600 GHz counts&$0.99\pm0.02$&$1.00\pm0.05$&0.07$^{*}$\\
\\
MIPS 4300 GHz counts&$0.94\pm0.06$&$0.94\pm0.06$&0.07\\
MIPS 1900 GHz counts&$0.98\pm0.03$&$0.98\pm0.03$&0.12\\
\end{tabular}
\begin{center}
$^{\dagger}$ an additional 2 per cent has been added to the \emph{Planck} calibration uncertainties (Section 3.6.1)\\
$^{\ddagger}$ SPT calibration is in units of power, otherwise units of flux or CMB temperature\\
$^{*}$ we enforce correlation between these calibration values, with band-to-band covariance of 0.05$^2$ (Section 3.5)
\end{center}
\end{table}

\section{Discussion}

In this section we compare our model's predictions with several measurements outside those used to constrain it, and discuss some of its implications for future work.

\subsection{Integrated CIB intensity}

The integrated CIB intensity, clustered anisotropy power and Poisson anisotropy power each receive different contributions from different populations of sources, with the brightest sources making a larger relative contribution to the Poisson power, for instance (see discussions in e.g. \citealt{hajian/etal:2012}, A12, \citealt{addison/etal:prep}). Our model correctly reproduces the anisotropy power spectra over a range of frequency; an obvious question is whether it also reproduces the integrated CIB intensity. We show our mean model predictions, with 1$\sigma$ uncertainty contours, in Figure 8, along with measurements of the integrated CIB SED from FIRAS \citep{lagache/etal:1999}. We also show measurements of the mean CIB intensity at 160 and 100 $\mu$m from IRAS and MIPS \citep{penin/etal:2012}.

Our model is in fairly good agreement with the CIB intensity measurements, with an underestimation of  $\sim1\sigma$ around the peak of the CIB emission. In principle, we could have used the FIRAS data to constrain the model, along with the spectra and number counts; we did not consider this because the \emph{Planck} 857 and 545 GHz maps were calibrated from the FIRAS measurements, and thus any systematic present in the FIRAS data could have effectively entered our modelling twice.

\begin{figure}
	\centering
	\includegraphics{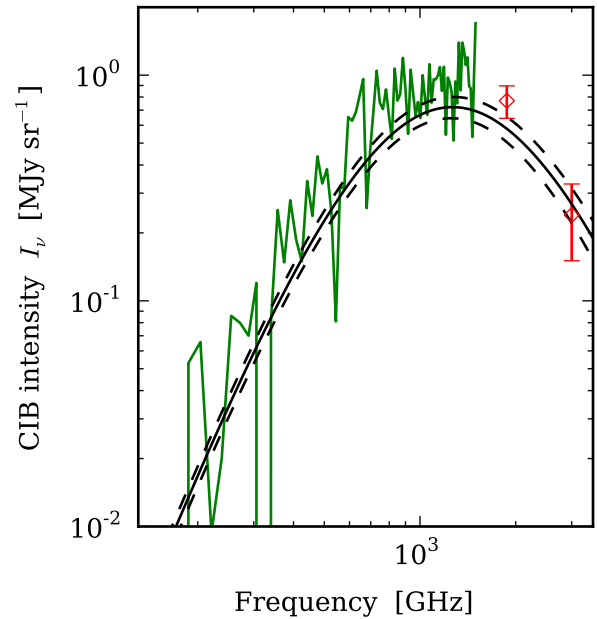}
	\caption{Mean CIB intensity spectrum, measured using FIRAS \citep[green line]{lagache/etal:1999} and from an analysis of IRAS and MIPS data at 100 and 160 $\mu$m \citep[red diamonds]{penin/etal:2012}. The solid black line shows our mean model prediction, and the dashed lines the 1$\sigma$ confidence contours.}
\end{figure}

\subsection{Power spectra at lower frequencies}

As stated in Section 3, we limited our analysis to spectra containing negligible contribution from the Sunyaev Zel'dovich effect or its cross-correlation with CIB sources. A good test of our model assumptions is to extrapolate the power spectrum predictions to lower frequencies. We compare our predictions with the most recent CIB power spectrum constraints from SPT \citep{reichardt/etal:2012} in Table 5. We compare to the SPT measurements obtained when a correlation between the tSZ effect and CIB sources is allowed, since \cite{addison/etal:prep} found that such a correlation may exist at the 10--20 per cent level in units of power.

There is fairly good consistency between our predictions and the SPT measurements, although our predictions for Poisson power are somewhat higher than the SPT results. Our constraints, at least on the clustered power amplitude, are competitive with the direct measurements. Extending our model to include lower frequency ($\nu<220$ GHz) ACT, SPT, and \emph{Planck} data may therefore improve constraints on both our dust emission model, and secondary anisotropies like the Sunyaev Zel'dovich effect.

\begin{table}
  \centering
  \caption{Power spectrum predictions for lower frequencies (units of $\ell(\ell+1)C_{\ell}/2\pi|_{\ell=3000}$, in $\mu$K$^2$ CMB)}
  \begin{tabular}{llcc}
&&Prediction&Measured$^{\dagger}$\\
\hline
150 GHz&Poisson&$9.6\pm0.9$&$8.1\pm 0.5$\\
&Clustered&$7.0\pm0.8$&$6.7\pm 0.7$\\
95 GHz&Poisson&$1.37\pm 0.21$&$1.04\pm 0.15$\\
&Clustered&$0.97\pm 0.17$&$0.87\pm 0.17$\\
\end{tabular}
\begin{center}
$^{\dagger}$ SPT constraints from \cite{reichardt/etal:2012} with tSZ--CIB correlation allowed
\end{center}
\end{table}

\subsection{Kinematic Sunyaev Zel'dovich constraints}

The kSZ constraint obtained in our baseline fit, $A_{\rm kSZ}=5.3^{+2.2}_{-2.4}$ $\mu$K$^2$, represents a preference for a non-zero kSZ contribution at slightly more than the 2$\sigma$ significance level, which remains even if no upper limit is imposed on the kSZ amplitude. The amplitude is consistent with recent predictions of $\sim2-4$ $\mu$K$^2$ \citep{shaw/etal:2012} and $\sim3.5-5.5$ $\mu$K$^2$ \citep[]{mesinger/etal:2012}, however the current uncertainty is too large for our result to be useful for constraining kSZ models. We find that the kSZ amplitude is not highly correlated with any of the dust model parameters, meaning that there is no single quantity which, if better known, would lead to a substantial improvement in the kSZ constraint. While constraining the kSZ power spectrum from data free from the tSZ is an attractive idea, it would appear that, at least currently, this is not feasible, given the small size of the kSZ power relative to the dust at 220 GHz.

\subsection{Comparison to AzTEC number counts}

Figure 9 shows differential number counts obtained at 1.1~mm using the AzTEC camera \citep{scott/etal:2012} with the 95\% confidence limit predictions from our baseline model. It is clear that the model significantly underpredicts the counts at $\sim3-6~$mJy. There are a number of possible explanations for this discrepancy. Our LF or SED modeling may be too simplistic to correctly predict the bright millimeter counts based on extrapolation from the FIR and submm number count constraints from MIPS and SPIRE. It is also possible that strong gravitational lensing significantly enhances the bright-end counts. Several recent source models \citep[e.g.][]{lapi/etal:2011,bethermin/etal:2012c} that include treatment of strong lensing predict that it has a significant effect only for $S>10~$mJy at $1.1~$mm, however these models do not include the effect of lensing magnification outside the strong lensing regime -- magnification factor $\mu<2$ -- and do not include the effect of baryons on lensing potentials. Finally, we note that the $24~\mu m$ MIPS counts used to extract the counts at 70 -- 500$~\mu$m by \cite{bethermin/etal:2010} and \cite{bethermin/etal:2012b} were obtained from an instrument with angular resolution of $\sim6\arcsec$, a factor or two or more better than AzTEC and SPIRE. An excess of measured bright sources could be at least partially an artefact of limited angular resolution, caused by blending of multiple faint sources. \cite{karim/etal:2013} found evidence for this phenomenon using high-resolution ALMA observations. The bright-end \cite{bethermin/etal:2012b} SPIRE counts are also systematically low compared to SPIRE counts obtained without using MIPS source positions (see their Figure 9). We expect future advances in lens modelling and understanding of resolution-dependent effects on source counts to help quantify the importance of these effects.

\begin{figure}
	\centering
	\includegraphics{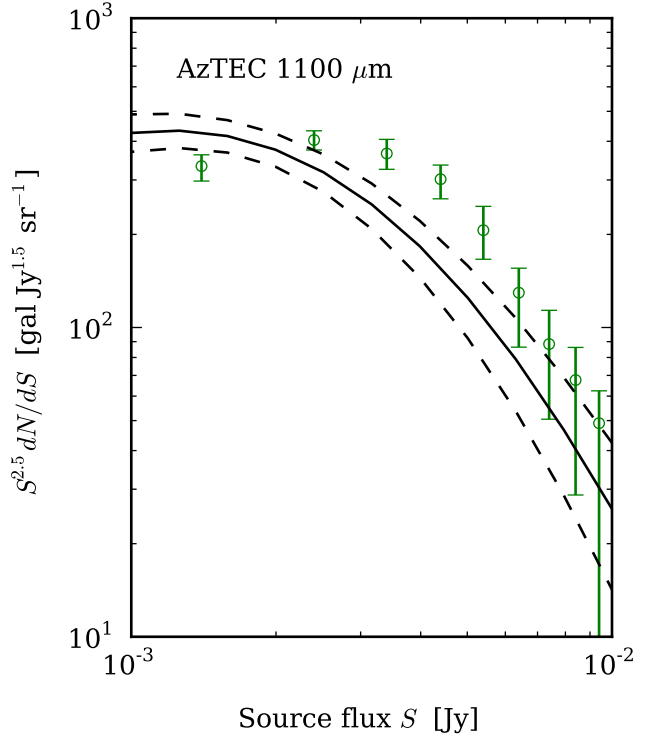}
	\caption{AzTEC counts at $1.1~$mm from \protect\cite{scott/etal:2012}, with 95\% confidence limits from our baseline model overplotted. Several possible causes of the discrepancy at intermediate flux values are discussed in the text (Section 5.4).}
\end{figure}

\subsection{The signal in the shot-noise}

An interesting result from our work is the constraining power that is carried by the Poisson component of the high-resolution spectra (mainly SPIRE, but also ACT and SPT). To illustrate this, Figure 10 shows two-dimensional parameter constraints, with contours enclosing 68 and 95 per cent of the MCMC samples, for several model parameters. We show constraints from three models: the baseline model, a model constrained from the SPIRE and MIPS number counts alone, and a model constrained by the counts plus clustering power alone. In this third model, the Poisson power in each of the nine power spectra is allowed to vary as an additional free parameter. The results of this comparison can be summarised as follows: including the clustered power leads to a significant improvement in SED parameter constraints over the counts alone, and inclusion of the Poisson power in the fit leads to moderate further improvements in the SED parameters constraints, and large improvements in the bias parameter constraints, in particular dramatically reducing the correlation between $b_0$ and $\epsilon_{\rm bias}$.

We also find that the model likelihood is not sufficiently improved by treating the Poisson levels independently to warrant a preference for this approach on goodness-of-fit grounds ($\Delta\chi^2=15$ for 9 fewer degrees of freedom, corresponding to a 1.3$\sigma$ preference). Furthermore, there is good consistency between the results from the three models for every parameter. This suggests that our results are not significantly biased by our treatment of bright source masking prior to power spectrum calculation (see Section 2.5).

\begin{figure}
	\centering
	\includegraphics{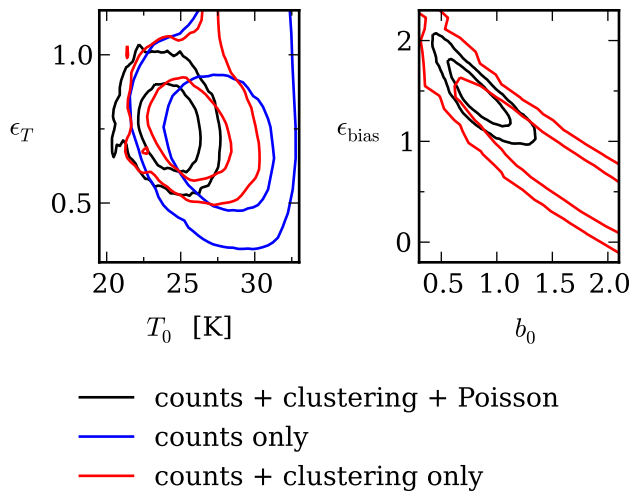}
	\caption{Two-dimensional parameter constraints with contours enclosing 68 and 95 per cent of the MCMC samples. Constraints on SED parameters (such as the low-redshift temperature of the `$a$' component, $T_0$, and the parameter describing the temperature evolution, $\epsilon_T$ -- see equation 7) are considerably improved by fitting the counts and clustered power over fitting the SPIRE and MIPS counts alone. Requiring the model to also fit the Poisson power spectrum further improves the SED constraints, and also significantly improves constraints on the large-scale bias of the CIB sources (given at $z=0$ by $b_0$) and the redshift evolution of the bias, parametrized by $\epsilon_{\rm bias}$.}
\end{figure}

\subsection{Redshift distribution of power and intensity}

Figure 11 shows the redshift distribution of clustered and Poisson anisotropy power, and CIB intensity, as a function of frequency, for our best-fit model parameters. Different levels of bright source removal is responsible for the behaviour of the Poisson power at low redshift, with notably fewer bright sources bring masked in the \emph{Planck} compared to SPIRE spectra. It should be noted that the redshift distributions in the submm and mm shift to higher redshift if only the brighter sources are considered. For example, considering only the sources with flux $S>15$ mJy at 353 GHz (850 $\mu$m) leads to a shift in the median of the intensity distribution to $z\sim2.5$, with virtually no contribution from $z\lesssim1$, in better agreement with the redshift-distribution of classical submm galaxies observed with SCUBA \citep[e.g.][]{chapman/etal:2005}.

Our model predicts that, as expected, lower frequencies receive greater relative contribution from higher redshifts \citep[e.g.][]{knox/etal:2001}, but, interestingly, there is only moderate evolution in the distribution over the submm and microwave bands. Physically, this corresponds to a high degree of coherence in the dust emission across this frequency range. Defining the degree of cross-correlation between frequency $\nu_1$ and $\nu_2$ as the ratio of the cross-spectrum power to the square root of the product of the auto-spectra, $C_{\ell,\nu_1\nu_2}/\sqrt{C_{\ell,\nu_1}C_{\ell,\nu_2}}$, we predict a correlation in units of power of around 90 per cent between the \emph{Planck} 857 and 217 GHz bands, for example.

This high degree of correlation would suggest that much of the dust that acts as a foreground contaminant for current CMB temperature cosmology (hampering detection of the kSZ power spectrum, for instance) may be removed either by direct cross-correlation of microwave maps with maps at higher frequencies, or understood through a joint fit with the high-frequency, dust-dominated data.

The most direct test of this prediction is comparison with measured power spectra from cross-correlating microwave and submm maps. We repeated our model fitting including cross-spectra presented by \cite{hajian/etal:2012}, calculated by correlating BLAST maps at 250, 350 and 500 $\mu$m with the ACT 218 GHz map, and found that our baseline model provides a good fit ($\Delta\chi^2=25$ for 27 additional bandpowers, treating the BLAST map calibration parameters as described in Section 3.5). This is consistent with the high degree of coherence preferred by most of the ACT / BLAST cross-spectra in the analyses of \cite{hajian/etal:2012} and A12, however, the BLAST / ACT cross-spectra bandpowers are noisy, and future cross-correlations (for instance from \emph{Planck}, or from cross-correlating SPIRE maps with ACT or SPT) will provide a stronger test of our prediction.

Since the evolution in $dC_{\ell}/dz$ with changing frequency is more significant for the Poisson than the clustered power, we further predict that the degree of correlation between different bands is $\ell$-dependent, with the correlation highest on large angular scales, where the Poisson is subdominant. Our predictions for the scale dependence of the SPIRE cross-correlations are in good agreement with the recent measurements of \cite{viero/etal:2013}. We predict cross-correlations of 0.92--0.96, 0.85--0.91 and 0.95--0.98 for the SPIRE 1200$\times$860, $1200\times600$ and $860\times600$ GHz spectra, respectively, with the range indicating the variation from $\ell\sim2\times10^4$ to $\ell\sim10^2$. \cite{viero/etal:2013} find cross-correlations of $0.95\pm0.04$, $0.86\pm0.04$, and $0.95\pm0.03$ for the same three cross-spectra (also see their Figure 10, which shows the trend of decreasing correlation with increasing $\ell$).

\begin{figure*}
	\centering
	\includegraphics[width=150mm]{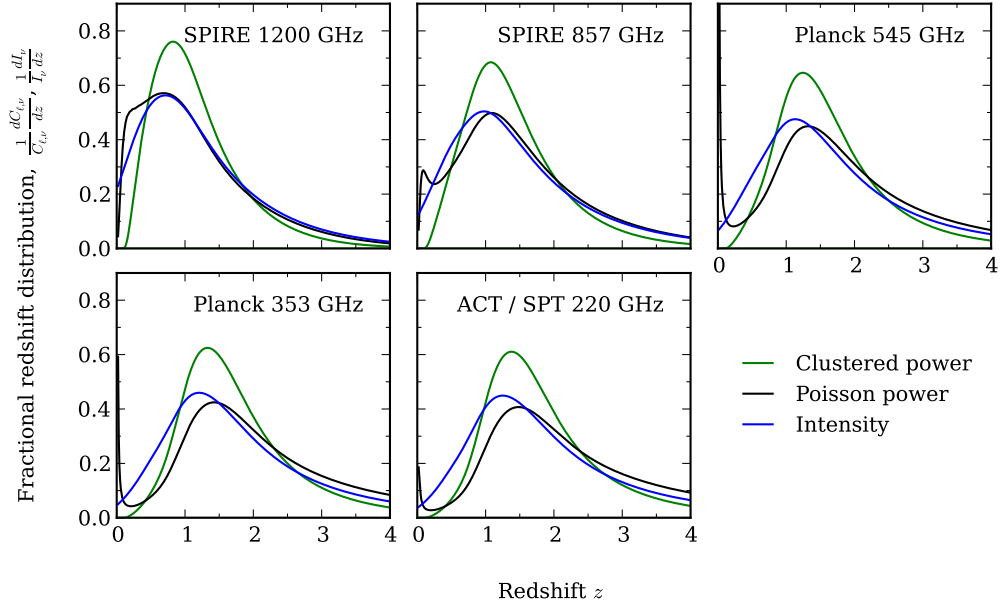}
	\caption{Redshift distribution of clustered and Poisson CIB anisotropy power, and CIB intensity, normalised such that the area under each curve is unity, for our best-fit model. The redshift distribution of the clustered power is shown at $\ell=3000$, although the redshift distribution does not vary strongly with scale over the range of scales where the clustered power contributes significantly to the total spectrum. A few per cent of the Poisson power (particularly for \emph{Planck}, where fewer bright sources are removed) is predicted to originate from low-redshift sources that are below the source removal limit, although in each case the power remains convergent. The high degree of overlap between the curves at different frequencies corresponds to a high degree of coherence in the dusty galaxy emission over the submillimeter and microwave bands. This is an important prediction of our model.}
\end{figure*}

\subsection{Model limitations and future work}

A large assumption in our modelling is that the dusty sources can essentially be treated as a homogenous population, whose average properties vary smoothly over time. Much of our constraining power is in the form of integrated quantities -- the power spectra -- meaning our results are probably not especially sensitive to rare, extreme objects (which could distort results obtained from a flux-limited study), but it is not clear that our model would correctly capture a sharp transition in either dust or clustering properties, which features in some existing work. In the model developed by \cite{granato/etal:2004}, \cite{lapi/etal:2011}, and \cite{xia/etal:2012}, for instance, highly-clustered, massive, proto-spheroidal galaxies dominate the dusty galaxy population at high redshift, but cease to exist at $z<1$, where merger-driven starbursts and (eventually) spiral galaxies take over. Allowing more explicitly for multiple source populations, with separate emission or clustering properties, is a clear improvement we can consider in future work.

Some possible limitations of our SED parametrization have already been discussed (Sections 2.2 and 4.3). The Atacama Large Millimeter/submillimeter Array (ALMA\footnote{http://www.almaobservatory.org/}) will greatly improve our understanding of the processes relating to dust-enshrouded star formation on galaxy scales, and may enable us move on from the graybody approximation to a more physical dust emission treatment. Additionally, extending our model to include higher frequencies, incorporating non-thermal dust contributions to the SED and bolometric IR luminosity, will facilitate comparison with infrared luminosity density and star formation rate constraints from other work.

The first release of \emph{Planck} maps, covering the entire sky (rather than just the $\sim$140 deg$^2$ used for the spectra we have included here), is expected in early 2013. Spectra from these maps, in addition to future \emph{Herschel}, ACT, SPT, ACTPol \citep{niemack/etal:2010} and SPTpol \citep{mcmahon/etal:2009,austermann/etal:2012} spectra, will significantly improve constraints over those considered here. Future number counts, from additional SPIRE fields, as well as, for example, SCUBA-2, will also give much-needed improved constraints on the properties of fainter high-redshift sources. The angular correlation function of flux-limited samples of CIB sources \citep[e.g.][]{cooray/etal:2010,maddox/etal:2010,guo/etal:2011}, from, for instance, the \emph{Herschel}-ATLAS \citep{eales/etal:2010} or SCUBA-2 will, when used in conjunction with the angular power spectrum, allow more detailed investigation of the $b_{\rm gal}-L$ correlation.

A host of CIB cross-correlations, using other tracers of large-scale structure, such as reconstructed CMB or galaxy weak lensing deflection maps, or optical data from the SDSS, are expected in the coming months. Being sensitive to different combinations of model parameters from the counts and CIB spectra, these statistics will be invaluable for furthering our understanding of extragalactic dust emission. The framework presented in Section 2 can be easily used to either predict these correlations, or extended to use them to improve constraints. Our model is particularly well-suited to predicting or interpreting the CMB lensing or integrated Sachs-Wolfe correlations with the CIB, where much of the measured signal will be on large angular scales, and relatively insensitive to the details of the small-scale clustering.

The primary motivation for studying extragalactic dust emission is to ultimately constrain the star formation history. Detectors onboard, for example, \emph{Spitzer} and WISE, probe re-processed starlight in the near- and mid-infrared, complementing the longer-wavelength data we have used. The near- and far-infrared emission are known to be correlated \citep[e.g.][]{bond/etal:2012}, and incorporating data at shorter wavelengths in our model will both improve understanding of this correlation, and lead to a more complete view of obscured star formation.

\section{Conclusions}

We have simultaneously constrained the luminosity function, spectral energy distribution and clustering properties of dusty CIB galaxies using a combined fit to number counts and angular power spectra from five current instruments -- \emph{Spitzer}, \emph{Herschel}, \emph{Planck}, ACT and SPT. Our analysis is based on fairly simple, phenomenological, parametrizations, but has yielded a number of important results:
\begin{enumerate}
\item evolution in the dust SED, parametrized here through evolution in graybody dust temperature, appears to be required,
\item the data show a strong preference for either a second graybody component or modified MIR tail behaviour,
\item joint fits to FIR, submm and microwave counts and spectra can constrain the evolution of the dusty source luminosity function at high redshift ($z\gtrsim2.5-3$),
\item pure power-law CIB source clustering is moderately disfavoured by combining current \emph{Planck} and SPIRE spectra,
\item we predict a high degree of submm / microwave band CIB correlation (about 90 per cent in units of power between \emph{Planck}'s 217 and 857 GHz bands, for example), and
\item the Poisson, shot-noise, component of the power spectrum can be used to improve CIB source constraints.
\end{enumerate}

None of the data sets considered in this work are individually capable of yielding constraints competitive with those from the combined fit. The basic approach we put forward -- combining data from a range of instruments, with a rigorous treatment of data and modelling uncertainties -- will thus be of great importance for realising the potential of future data sets.\\

GA acknowledges support from a Science and Technology Facilities Council studentship, and JD from a RCUK Fellowship and ERC grant 259505. The authors would like to thank Matthieu B\'ethermin for clarification regarding the B\'ethermin et al. stacking analysis, Alexandre Amblard for information relating to the SPIRE power spectra, and Amir Hajian, Sudeep Das and Thibaut Louis for discussions about the spectrum-to-spectrum correlations. We thank David Spergel for reading the manuscript and making useful suggestions, and are also grateful to Gaelen Marsden, Nick Battaglia, David Elbaz, Stephen Wilkins, Guilaine Lagache, George Efstathiou, Christian Reichardt, Jeremy Tinker, Beth Reid, Dan Eisenstein, Julien Devriendt, Douglas Watson, Erminia Calabrese and Mark Halpern for helpful discussions.

\footnotesize{
  \bibliographystyle{mn2e}

\begin{thebibliography}{178}
\expandafter\ifx\csname natexlab\endcsname\relax\def\natexlab#1{#1}\fi

\bibitem[{{Addison} {et~al}\mbox{.}(2012{\natexlab{a}}){Addison}, {Dunkley},
  {Hajian}, {Viero}, {Bond}, {Das}, {Devlin}, {Halpern}, {Hincks}, {Hlozek},
  {Marriage}, {Moodley}, {Page}, {Reese}, {Scott}, {Spergel}, {Staggs}, \&
  {Wollack}}]{addison/etal:2012}
{Addison} G.~E. {et~al.}, 2012{\natexlab{a}}, \apj, 752, 120

\bibitem[{{Addison} {et~al}\mbox{.}(2012{\natexlab{b}}){Addison}, {Dunkley}, \&
  {Spergel}}]{addison/etal:prep}
{Addison} G.~E., {Dunkley} J., {Spergel} D.~N., 2012{\natexlab{b}}, \mnras,
  427, 1741

\bibitem[{{Alexander} {et~al}\mbox{.}(2005){Alexander}, {Bauer}, {Chapman},
  {Smail}, {Blain}, {Brandt}, \& {Ivison}}]{alexander/etal:2005}
{Alexander} D.~M., {Bauer} F.~E., {Chapman} S.~C., {Smail} I., {Blain} A.~W.,
  {Brandt} W.~N., {Ivison} R.~J., 2005, \apj, 632, 736

\bibitem[{{Amblard} {et~al}\mbox{.}(2011){Amblard}, {Cooray}, {Serra},
  {Altieri}, {Arumugam}, {Aussel}, {Blain}, {Bock}, {Boselli}, {Buat},
  {Castro-Rodr{\'{\i}}guez}, {Cava}, {Chanial}, {Chapin}, {Clements}, {Conley},
  {Conversi}, {Dowell}, {Dwek}, {Eales}, {Elbaz}, {Farrah}, {Franceschini},
  {Gear}, {Glenn}, {Griffin}, {Halpern}, {Hatziminaoglou}, {Ibar}, {Isaak},
  {Ivison}, {Khostovan}, {Lagache}, {Levenson}, {Lu}, {Madden}, {Maffei},
  {Mainetti}, {Marchetti}, {Marsden}, {Mitchell-Wynne}, {Nguyen}, {O'Halloran},
  {Oliver}, {Omont}, {Page}, {Panuzzo}, {Papageorgiou}, {Pearson},
  {P{\'e}rez-Fournon}, {Pohlen}, {Rangwala}, {Roseboom}, {Rowan-Robinson},
  {Portal}, {Schulz}, {Scott}, {Seymour}, {Shupe}, {Smith}, {Stevens},
  {Symeonidis}, {Trichas}, {Tugwell}, {Vaccari}, {Valiante}, {Valtchanov},
  {Vieira}, {Vigroux}, {Wang}, {Ward}, {Wright}, {Xu}, \&
  {Zemcov}}]{amblard/etal:2011}
{Amblard} A. {et~al.}, 2011, \nat, 470, 510

\bibitem[{{Austermann} {et~al}\mbox{.}(2012){Austermann}, {Aird}, {Beall},
  {Becker}, {Bender}, {Benson}, {Bleem}, {Britton}, {Carlstrom}, {Chang},
  {Chiang}, {Cho}, {Crawford}, {Crites}, {Datesman}, {de Haan}, {Dobbs},
  {George}, {Halverson}, {Harrington}, {Henning}, {Hilton}, {Holder},
  {Holzapfel}, {Hoover}, {Huang}, {Hubmayr}, {Irwin}, {Keisler}, {Kennedy},
  {Knox}, {Lee}, {Leitch}, {Li}, {Lueker}, {Marrone}, {McMahon}, {Mehl},
  {Meyer}, {Montroy}, {Natoli}, {Nibarger}, {Niemack}, {Novosad}, {Padin},
  {Pryke}, {Reichardt}, {Ruhl}, {Saliwanchik}, {Sayre}, {Schaffer},
  {Shirokoff}, {Stark}, {Story}, {Vanderlinde}, {Vieira}, {Wang}, {Williamson},
  {Yefremenko}, {Yoon}, \& {Zahn}}]{austermann/etal:2012}
{Austermann} J.~E. {et~al.}, 2012, in Society of Photo-Optical Instrumentation
  Engineers (SPIE) Conference Series, Vol. 8452, Society of Photo-Optical
  Instrumentation Engineers (SPIE) Conference Series

\bibitem[{{Bai} {et~al}\mbox{.}(2007){Bai}, {Marcillac}, {Rieke}, {Rieke},
  {Tran}, {Hinz}, {Rudnick}, {Kelly}, \& {Blaylock}}]{bai/etal:2007}
{Bai} L. {et~al.}, 2007, \apj, 664, 181

\bibitem[{{Bai} {et~al}\mbox{.}(2006){Bai}, {Rieke}, {Rieke}, {Hinz}, {Kelly},
  \& {Blaylock}}]{bai/etal:2006}
{Bai} L., {Rieke} G.~H., {Rieke} M.~J., {Hinz} J.~L., {Kelly} D.~M., {Blaylock}
  M., 2006, \apj, 639, 827

\bibitem[{{Battaglia} {et~al}\mbox{.}(2012){Battaglia}, {Bond}, {Pfrommer}, \&
  {Sievers}}]{battaglia/etal:2012}
{Battaglia} N., {Bond} J.~R., {Pfrommer} C., {Sievers} J.~L., 2012, \apj, 758,
  75

\bibitem[{{Battaglia} {et~al}\mbox{.}(2010){Battaglia}, {Bond}, {Pfrommer},
  {Sievers}, \& {Sijacki}}]{battaglia/etal:2010}
{Battaglia} N., {Bond} J.~R., {Pfrommer} C., {Sievers} J.~L., {Sijacki} D.,
  2010, \apj, 725, 91

\bibitem[{{Behroozi} {et~al}\mbox{.}(2013{\natexlab{a}}){Behroozi}, {Wechsler},
  \& {Conroy}}]{behroozi/etal:2013}
{Behroozi} P.~S., {Wechsler} R.~H., {Conroy} C., 2013{\natexlab{a}}, \apjl,
  762, L31

\bibitem[{{Behroozi} {et~al}\mbox{.}(2013{\natexlab{b}}){Behroozi}, {Wechsler},
  \& {Conroy}}]{behroozi/etal:2013b}
{Behroozi} P.~S., {Wechsler} R.~H., {Conroy} C., 2013{\natexlab{b}}, \apj, 770,
  57

\bibitem[{{Benford} {et~al}\mbox{.}(1999){Benford}, {Cox}, {Omont}, {Phillips},
  \& {McMahon}}]{benford/etal:1999}
{Benford} D.~J., {Cox} P., {Omont} A., {Phillips} T.~G., {McMahon} R.~G., 1999,
  \apjl, 518, L65

\bibitem[{{Berlind} \& {Weinberg}(2002)}]{berlind/weinberg:2002}
{Berlind} A.~A., {Weinberg} D.~H., 2002, \apj, 575, 587

\bibitem[{{Berta} {et~al}\mbox{.}(2011){Berta}, {Magnelli}, {Nordon}, {Lutz},
  {Wuyts}, {Altieri}, {Andreani}, {Aussel}, {Casta{\~n}eda}, {Cepa}, {Cimatti},
  {Daddi}, {Elbaz}, {F{\"o}rster Schreiber}, {Genzel}, {Le Floc'h}, {Maiolino},
  {P{\'e}rez-Fournon}, {Poglitsch}, {Popesso}, {Pozzi}, {Riguccini},
  {Rodighiero}, {Sanchez-Portal}, {Sturm}, {Tacconi}, \&
  {Valtchanov}}]{berta/etal:2011}
{Berta} S. {et~al.}, 2011, \aap, 532, A49

\bibitem[{{B{\'e}thermin} {et~al}\mbox{.}(2012{\natexlab{a}}){B{\'e}thermin},
  {Daddi}, {Magdis}, {Sargent}, {Hezaveh}, {Elbaz}, {Le Borgne}, {Mullaney},
  {Pannella}, {Buat}, {Charmandaris}, {Lagache}, \&
  {Scott}}]{bethermin/etal:2012c}
{B{\'e}thermin} M. {et~al.}, 2012{\natexlab{a}}, \apjl, 757, L23

\bibitem[{{B{\'e}thermin} {et~al}\mbox{.}(2010){B{\'e}thermin}, {Dole},
  {Beelen}, \& {Aussel}}]{bethermin/etal:2010}
{B{\'e}thermin} M., {Dole} H., {Beelen} A., {Aussel} H., 2010, \aap, 512, A78

\bibitem[{{B{\'e}thermin} {et~al}\mbox{.}(2011){B{\'e}thermin}, {Dole},
  {Lagache}, {Le Borgne}, \& {Penin}}]{bethermin/etal:2011}
{B{\'e}thermin} M., {Dole} H., {Lagache} G., {Le Borgne} D., {Penin} A., 2011,
  \aap, 529, A4+

\bibitem[{{B{\'e}thermin} {et~al}\mbox{.}(2012{\natexlab{b}}){B{\'e}thermin},
  {Dor{\'e}}, \& {Lagache}}]{bethermin/etal:2012a}
{B{\'e}thermin} M., {Dor{\'e}} O., {Lagache} G., 2012{\natexlab{b}}, \aap, 537,
  L5

\bibitem[{{B{\'e}thermin} {et~al}\mbox{.}(2012{\natexlab{c}}){B{\'e}thermin},
  {Le Floc'h}, {Ilbert}, {Conley}, {Lagache}, {Amblard}, {Arumugam}, {Aussel},
  {Berta}, {Bock}, {Boselli}, {Buat}, {Casey}, {Castro-Rodr{\'{\i}}guez},
  {Cava}, {Clements}, {Cooray}, {Dowell}, {Eales}, {Farrah}, {Franceschini},
  {Glenn}, {Griffin}, {Hatziminaoglou}, {Heinis}, {Ibar}, {Ivison},
  {Kartaltepe}, {Levenson}, {Magdis}, {Marchetti}, {Marsden}, {Nguyen},
  {O'Halloran}, {Oliver}, {Omont}, {Page}, {Panuzzo}, {Papageorgiou},
  {Pearson}, {P{\'e}rez-Fournon}, {Pohlen}, {Rigopoulou}, {Roseboom},
  {Rowan-Robinson}, {Salvato}, {Schulz}, {Scott}, {Seymour}, {Shupe}, {Smith},
  {Symeonidis}, {Trichas}, {Tugwell}, {Vaccari}, {Valtchanov}, {Vieira},
  {Viero}, {Wang}, {Xu}, \& {Zemcov}}]{bethermin/etal:2012b}
{B{\'e}thermin} M. {et~al.}, 2012{\natexlab{c}}, \aap, 542, A58

\bibitem[{{Blain}(1996)}]{blain:1996}
{Blain} A.~W., 1996, \mnras, 283, 1340

\bibitem[{{Blain}(1999{\natexlab{a}})}]{blain:1999b}
{Blain} A.~W., 1999{\natexlab{a}}, \mnras, 309, 955

\bibitem[{{Blain}(1999{\natexlab{b}})}]{blain:1999a}
{Blain} A.~W., 1999{\natexlab{b}}, \mnras, 304, 669

\bibitem[{{Blain} {et~al}\mbox{.}(2003){Blain}, {Barnard}, \&
  {Chapman}}]{blain/etal:2003}
{Blain} A.~W., {Barnard} V.~E., {Chapman} S.~C., 2003, \mnras, 338, 733

\bibitem[{{Blain} {et~al}\mbox{.}(1998){Blain}, {Ivison}, \&
  {Smail}}]{blain/etal:1998}
{Blain} A.~W., {Ivison} R.~J., {Smail} I., 1998, \mnras, 296, L29

\bibitem[{{Blake} {et~al}\mbox{.}(2007){Blake}, {Collister}, {Bridle}, \&
  {Lahav}}]{blake/etal:2007}
{Blake} C., {Collister} A., {Bridle} S., {Lahav} O., 2007, \mnras, 374, 1527

\bibitem[{Bond(1996)}]{bond:1996}
Bond J.~R., 1996, in Cosmology and Large Scale Structure, Les Houches Session
  LX, Schaeffer R., ed., Elsevier, London, UK, p. 496

\bibitem[{{Bond} {et~al}\mbox{.}(1991{\natexlab{a}}){Bond}, {Carr}, \&
  {Hogan}}]{bond/etal:1991c}
{Bond} J.~R., {Carr} B.~J., {Hogan} C.~J., 1991{\natexlab{a}}, \apj, 367, 420

\bibitem[{{Bond} {et~al}\mbox{.}(1991{\natexlab{b}}){Bond}, {Cole},
  {Efstathiou}, \& {Kaiser}}]{bond/etal:1991}
{Bond} J.~R., {Cole} S., {Efstathiou} G., {Kaiser} N., 1991{\natexlab{b}},
  \apj, 379, 440

\bibitem[{{Bond} {et~al}\mbox{.}(2012){Bond}, {Benford}, {Gardner}, {Amblard},
  {Fleuren}, {Blain}, {Dunne}, {Smith}, {Maddox}, {Hoyos}, {Baes}, {Bonfield},
  {Bourne}, {Bridge}, {Buttiglione}, {Cava}, {Clements}, {Cooray}, {Dariush},
  {de Zotti}, {Driver}, {Dye}, {Eales}, {Eisenhardt}, {Hopwood}, {Ibar},
  {Ivison}, {Jarvis}, {Kelvin}, {Robotham}, {Temi}, {Thompson}, {Tsai}, {van
  der Werf}, {Wright}, {Wu}, \& {Yan}}]{bond/etal:2012}
{Bond} N.~A. {et~al.}, 2012, \apjl, 750, L18

\bibitem[{{Boselli} \& {Gavazzi}(2006)}]{boselli/gavazzi:2006}
{Boselli} A., {Gavazzi} G., 2006, \pasp, 118, 517

\bibitem[{{Bracco} {et~al}\mbox{.}(2011){Bracco}, {Cooray}, {Veneziani},
  {Amblard}, {Serra}, {Wardlow}, {Thompson}, {White}, {Auld}, {Baes},
  {Bertoldi}, {Buttiglione}, {Cava}, {Clements}, {Dariush}, {de Zotti},
  {Dunne}, {Dye}, {Eales}, {Fritz}, {Gomez}, {Hopwood}, {Ibar}, {Ivison},
  {Jarvis}, {Lagache}, {Lee}, {Leeuw}, {Maddox}, {Micha{\l}owski}, {Pearson},
  {Pohlen}, {Rigby}, {Rodighiero}, {Smith}, {Temi}, {Vaccari}, \& {van der
  Werf}}]{bracco/etal:2011}
{Bracco} A. {et~al.}, 2011, \mnras, 412, 1151

\bibitem[{{Caputi} {et~al}\mbox{.}(2007){Caputi}, {Lagache}, {Yan}, {Dole},
  {Bavouzet}, {Le Floc'h}, {Choi}, {Helou}, \& {Reddy}}]{caputi/etal:2007}
{Caputi} K.~I. {et~al.}, 2007, \apj, 660, 97

\bibitem[{{Carlstrom} {et~al}\mbox{.}(2011){Carlstrom}, {Ade}, {Aird},
  {Benson}, {Bleem}, {Busetti}, {Chang}, {Chauvin}, {Cho}, {Crawford},
  {Crites}, {Dobbs}, {Halverson}, {Heimsath}, {Holzapfel}, {Hrubes}, {Joy},
  {Keisler}, {Lanting}, {Lee}, {Leitch}, {Leong}, {Lu}, {Lueker}, {Luong-van},
  {McMahon}, {Mehl}, {Meyer}, {Mohr}, {Montroy}, {Padin}, {Plagge}, {Pryke},
  {Ruhl}, {Schaffer}, {Schwan}, {Shirokoff}, {Spieler}, {Staniszewski},
  {Stark}, {Tucker}, {Vanderlinde}, {Vieira}, \&
  {Williamson}}]{carlstrom/etal:2011}
{Carlstrom} J.~E. {et~al.}, 2011, \pasp, 123, 568

\bibitem[{{Chapin} {et~al}\mbox{.}(2011){Chapin}, {Chapman}, {Coppin},
  {Devlin}, {Dunlop}, {Greve}, {Halpern}, {Hasselfield}, {Hughes}, {Ivison},
  {Marsden}, {Moncelsi}, {Netterfield}, {Pascale}, {Scott}, {Smail}, {Viero},
  {Walter}, {Weiss}, \& {van der Werf}}]{chapin/etal:2011}
{Chapin} E.~L. {et~al.}, 2011, \mnras, 411, 505

\bibitem[{{Chapman} {et~al}\mbox{.}(2005){Chapman}, {Blain}, {Smail}, \&
  {Ivison}}]{chapman/etal:2005}
{Chapman} S.~C., {Blain} A.~W., {Smail} I., {Ivison} R.~J., 2005, \apj, 622,
  772

\bibitem[{{Cole} {et~al}\mbox{.}(2005){Cole}, {Percival}, {Peacock}, {Norberg},
  {Baugh}, {Frenk}, {Baldry}, {Bland-Hawthorn}, {Bridges}, {Cannon}, {Colless},
  {Collins}, {Couch}, {Cross}, {Dalton}, {Eke}, {De Propris}, {Driver},
  {Efstathiou}, {Ellis}, {Glazebrook}, {Jackson}, {Jenkins}, {Lahav}, {Lewis},
  {Lumsden}, {Maddox}, {Madgwick}, {Peterson}, {Sutherland}, \&
  {Taylor}}]{cole/etal:2005}
{Cole} S. {et~al.}, 2005, \mnras, 362, 505

\bibitem[{{Cooray} {et~al}\mbox{.}(2010){Cooray}, {Amblard}, {Wang},
  {Arumugam}, {Auld}, {Aussel}, {Babbedge}, {Blain}, {Bock}, {Boselli}, {Buat},
  {Burgarella}, {Castro-Rodriguez}, {Cava}, {Chanial}, {Clements}, {Conley},
  {Conversi}, {Dowell}, {Dwek}, {Eales}, {Elbaz}, {Farrah}, {Fox},
  {Franceschini}, {Gear}, {Glenn}, {Griffin}, {Halpern}, {Hatziminaoglou},
  {Ibar}, {Isaak}, {Ivison}, {Khostovan}, {Lagache}, {Levenson}, {Lu},
  {Madden}, {Maffei}, {Mainetti}, {Marchetti}, {Marsden}, {Mitchell-Wynne},
  {Mortier}, {Nguyen}, {O'Halloran}, {Oliver}, {Omont}, {Page}, {Panuzzo},
  {Papageorgiou}, {Pearson}, {Perez Fournon}, {Pohlen}, {Rawlings}, {Raymond},
  {Rigopoulou}, {Rizzo}, {Roseboom}, {Rowan-Robinson}, {Schulz}, {Scott},
  {Serra}, {Seymour}, {Shupe}, {Smith}, {Stevens}, {Symeonidis}, {Trichas},
  {Tugwell}, {Vaccari}, {Valtchanov}, {Vieira}, {Vigroux}, {Ward}, {Wright},
  {Xu}, \& {Zemcov}}]{cooray/etal:2010}
{Cooray} A. {et~al.}, 2010, \aap, 518, L22+

\bibitem[{{Cooray} \& {Sheth}(2002)}]{cooray/sheth:2002}
{Cooray} A., {Sheth} R., 2002, \physrep, 372, 1

\bibitem[{{Coppin} {et~al}\mbox{.}(2006){Coppin}, {Chapin}, {Mortier}, {Scott},
  {Borys}, {Dunlop}, {Halpern}, {Hughes}, {Pope}, {Scott}, {Serjeant}, {Wagg},
  {Alexander}, {Almaini}, {Aretxaga}, {Babbedge}, {Best}, {Blain}, {Chapman},
  {Clements}, {Crawford}, {Dunne}, {Eales}, {Edge}, {Farrah}, {Gazta{\~n}aga},
  {Gear}, {Granato}, {Greve}, {Fox}, {Ivison}, {Jarvis}, {Jenness}, {Lacey},
  {Lepage}, {Mann}, {Marsden}, {Martinez-Sansigre}, {Oliver}, {Page},
  {Peacock}, {Pearson}, {Percival}, {Priddey}, {Rawlings}, {Rowan-Robinson},
  {Savage}, {Seigar}, {Sekiguchi}, {Silva}, {Simpson}, {Smail}, {Stevens},
  {Takagi}, {Vaccari}, {van Kampen}, \& {Willott}}]{coppin/etal:2006}
{Coppin} K. {et~al.}, 2006, \mnras, 372, 1621

\bibitem[{{Daddi} {et~al}\mbox{.}(2005){Daddi}, {Dickinson}, {Chary}, {Pope},
  {Morrison}, {Alexander}, {Bauer}, {Brandt}, {Giavalisco}, {Ferguson}, {Lee},
  {Lehmer}, {Papovich}, \& {Renzini}}]{daddi/etal:2005}
{Daddi} E. {et~al.}, 2005, \apjl, 631, L13

\bibitem[{{Das} {et~al}\mbox{.}(2011){Das}, {Marriage}, {Ade}, {Aguirre},
  {Amiri}, {Appel}, {Barrientos}, {Battistelli}, {Bond}, {Brown}, {Burger},
  {Chervenak}, {Devlin}, {Dicker}, {Bertrand Doriese}, {Dunkley}, {D{\"u}nner},
  {Essinger-Hileman}, {Fisher}, {Fowler}, {Hajian}, {Halpern}, {Hasselfield},
  {Hern{\'a}ndez-Monteagudo}, {Hilton}, {Hilton}, {Hincks}, {Hlozek},
  {Huffenberger}, {Hughes}, {Hughes}, {Infante}, {Irwin}, {Baptiste Juin},
  {Kaul}, {Klein}, {Kosowsky}, {Lau}, {Limon}, {Lin}, {Lupton}, {Marsden},
  {Martocci}, {Mauskopf}, {Menanteau}, {Moodley}, {Moseley}, {Netterfield},
  {Niemack}, {Nolta}, {Page}, {Parker}, {Partridge}, {Reid}, {Sehgal},
  {Sherwin}, {Sievers}, {Spergel}, {Staggs}, {Swetz}, {Switzer}, {Thornton},
  {Trac}, {Tucker}, {Warne}, {Wollack}, \& {Zhao}}]{das/etal:2011}
{Das} S. {et~al.}, 2011, \apj, 729, 62

\bibitem[{{De Bernardis} \& {Cooray}(2012)}]{debernardis/cooray:2012}
{De Bernardis} F., {Cooray} A., 2012, \apj, 760, 14

\bibitem[{{Desert} {et~al}\mbox{.}(1990){Desert}, {Boulanger}, \&
  {Puget}}]{desert/etal:1990}
{Desert} F.-X., {Boulanger} F., {Puget} J.~L., 1990, \aap, 237, 215

\bibitem[{{Dole} {et~al}\mbox{.}(2003){Dole}, {Lagache}, \&
  {Puget}}]{dole/etal:2003}
{Dole} H., {Lagache} G., {Puget} J.-L., 2003, \apj, 585, 617

\bibitem[{{Dole} {et~al}\mbox{.}(2006){Dole}, {Lagache}, {Puget}, {Caputi},
  {Fern{\'a}ndez-Conde}, {Le Floc'h}, {Papovich}, {P{\'e}rez-Gonz{\'a}lez},
  {Rieke}, \& {Blaylock}}]{dole/etal:2006}
{Dole} H. {et~al.}, 2006, \aap, 451, 417

\bibitem[{{Draine}(2006)}]{draine:2006}
{Draine} B.~T., 2006, \apj, 636, 1114

\bibitem[{{Dunkley} {et~al}\mbox{.}(2005){Dunkley}, {Bucher}, {Ferreira},
  {Moodley}, \& {Skordis}}]{dunkley/etal:2005}
{Dunkley} J., {Bucher} M., {Ferreira} P.~G., {Moodley} K., {Skordis} C., 2005,
  \mnras, 356, 925

\bibitem[{{Dunkley} {et~al}\mbox{.}(2011){Dunkley}, {Hlozek}, {Sievers},
  {Acquaviva}, {Ade}, {Aguirre}, {Amiri}, {Appel}, {Barrientos}, {Battistelli},
  {Bond}, {Brown}, {Burger}, {Chervenak}, {Das}, {Devlin}, {Dicker}, {Bertrand
  Doriese}, {D{\"u}nner}, {Essinger-Hileman}, {Fisher}, {Fowler}, {Hajian},
  {Halpern}, {Hasselfield}, {Hern{\'a}ndez-Monteagudo}, {Hilton}, {Hilton},
  {Hincks}, {Huffenberger}, {Hughes}, {Hughes}, {Infante}, {Irwin}, {Juin},
  {Kaul}, {Klein}, {Kosowsky}, {Lau}, {Limon}, {Lin}, {Lupton}, {Marriage},
  {Marsden}, {Mauskopf}, {Menanteau}, {Moodley}, {Moseley}, {Netterfield},
  {Niemack}, {Nolta}, {Page}, {Parker}, {Partridge}, {Reid}, {Sehgal},
  {Sherwin}, {Spergel}, {Staggs}, {Swetz}, {Switzer}, {Thornton}, {Trac},
  {Tucker}, {Warne}, {Wollack}, \& {Zhao}}]{dunkley/etal:2011}
{Dunkley} J. {et~al.}, 2011, \apj, 739, 52

\bibitem[{{Dunne} {et~al}\mbox{.}(2000){Dunne}, {Eales}, {Edmunds}, {Ivison},
  {Alexander}, \& {Clements}}]{dunne/etal:2000}
{Dunne} L., {Eales} S., {Edmunds} M., {Ivison} R., {Alexander} P., {Clements}
  D.~L., 2000, \mnras, 315, 115

\bibitem[{{Dunne} \& {Eales}(2001)}]{dunne/eales:2001}
{Dunne} L., {Eales} S.~A., 2001, \mnras, 327, 697

\bibitem[{{D{\"u}nner} {et~al}\mbox{.}(2013){D{\"u}nner}, {Hasselfield},
  {Marriage}, {Sievers}, {Acquaviva}, {Addison}, {Ade}, {Aguirre}, {Amiri},
  {Appel}, {Barrientos}, {Battistelli}, {Bond}, {Brown}, {Burger}, {Calabrese},
  {Chervenak}, {Das}, {Devlin}, {Dicker}, {Bertrand Doriese}, {Dunkley},
  {Essinger-Hileman}, {Fisher}, {Gralla}, {Fowler}, {Hajian}, {Halpern},
  {Hern{\'a}ndez-Monteagudo}, {Hilton}, {Hilton}, {Hincks}, {Hlozek},
  {Huffenberger}, {Hughes}, {Hughes}, {Infante}, {Irwin}, {Baptiste Juin},
  {Kaul}, {Klein}, {Kosowsky}, {Lau}, {Limon}, {Lin}, {Louis}, {Lupton},
  {Marsden}, {Martocci}, {Mauskopf}, {Menanteau}, {Moodley}, {Moseley},
  {Netterfield}, {Niemack}, {Nolta}, {Page}, {Parker}, {Partridge}, {Quintana},
  {Reid}, {Sehgal}, {Sherwin}, {Spergel}, {Staggs}, {Swetz}, {Switzer},
  {Thornton}, {Trac}, {Tucker}, {Warne}, {Wilson}, {Wollack}, \&
  {Zhao}}]{dunner/etal:2013}
{D{\"u}nner} R. {et~al.}, 2013, \apj, 762, 10

\bibitem[{{Eales} {et~al}\mbox{.}(2010){Eales}, {Dunne}, {Clements}, {Cooray},
  {de Zotti}, {Dye}, {Ivison}, {Jarvis}, {Lagache}, {Maddox}, {Negrello},
  {Serjeant}, {Thompson}, {van Kampen}, {Amblard}, {Andreani}, {Baes},
  {Beelen}, {Bendo}, {Benford}, {Bertoldi}, {Bock}, {Bonfield}, {Boselli},
  {Bridge}, {Buat}, {Burgarella}, {Carlberg}, {Cava}, {Chanial}, {Charlot},
  {Christopher}, {Coles}, {Cortese}, {Dariush}, {da Cunha}, {Dalton}, {Danese},
  {Dannerbauer}, {Driver}, {Dunlop}, {Fan}, {Farrah}, {Frayer}, {Frenk},
  {Geach}, {Gardner}, {Gomez}, {Gonz{\'a}lez-Nuevo}, {Gonz{\'a}lez-Solares},
  {Griffin}, {Hardcastle}, {Hatziminaoglou}, {Herranz}, {Hughes}, {Ibar},
  {Jeong}, {Lacey}, {Lapi}, {Lawrence}, {Lee}, {Leeuw}, {Liske},
  {L{\'o}pez-Caniego}, {M{\"u}ller}, {Nandra}, {Panuzzo}, {Papageorgiou},
  {Patanchon}, {Peacock}, {Pearson}, {Phillipps}, {Pohlen}, {Popescu},
  {Rawlings}, {Rigby}, {Rigopoulou}, {Robotham}, {Rodighiero}, {Sansom},
  {Schulz}, {Scott}, {Smith}, {Sibthorpe}, {Smail}, {Stevens}, {Sutherland},
  {Takeuchi}, {Tedds}, {Temi}, {Tuffs}, {Trichas}, {Vaccari}, {Valtchanov},
  {van der Werf}, {Verma}, {Vieria}, {Vlahakis}, \& {White}}]{eales/etal:2010}
{Eales} S. {et~al.}, 2010, \pasp, 122, 499

\bibitem[{{Eales} {et~al}\mbox{.}(1999){Eales}, {Lilly}, {Gear}, {Dunne},
  {Bond}, {Hammer}, {Le F{\`e}vre}, \& {Crampton}}]{eales/etal:1999}
{Eales} S., {Lilly} S., {Gear} W., {Dunne} L., {Bond} J.~R., {Hammer} F., {Le
  F{\`e}vre} O., {Crampton} D., 1999, \apj, 515, 518

\bibitem[{{Eisenstein} {et~al}\mbox{.}(2005){Eisenstein}, {Zehavi}, {Hogg},
  {Scoccimarro}, {Blanton}, {Nichol}, {Scranton}, {Seo}, {Tegmark}, {Zheng},
  {Anderson}, {Annis}, {Bahcall}, {Brinkmann}, {Burles}, {Castander},
  {Connolly}, {Csabai}, {Doi}, {Fukugita}, {Frieman}, {Glazebrook}, {Gunn},
  {Hendry}, {Hennessy}, {Ivezi{\'c}}, {Kent}, {Knapp}, {Lin}, {Loh}, {Lupton},
  {Margon}, {McKay}, {Meiksin}, {Munn}, {Pope}, {Richmond}, {Schlegel},
  {Schneider}, {Shimasaku}, {Stoughton}, {Strauss}, {SubbaRao}, {Szalay},
  {Szapudi}, {Tucker}, {Yanny}, \& {York}}]{eisenstein/etal:2005}
{Eisenstein} D.~J. {et~al.}, 2005, \apj, 633, 560

\bibitem[{{Elbaz} {et~al}\mbox{.}(2011){Elbaz}, {Dickinson}, {Hwang},
  {D{\'{\i}}az-Santos}, {Magdis}, {Magnelli}, {Le Borgne}, {Galliano},
  {Pannella}, {Chanial}, {Armus}, {Charmandaris}, {Daddi}, {Aussel}, {Popesso},
  {Kartaltepe}, {Altieri}, {Valtchanov}, {Coia}, {Dannerbauer}, {Dasyra},
  {Leiton}, {Mazzarella}, {Alexander}, {Buat}, {Burgarella}, {Chary}, {Gilli},
  {Ivison}, {Juneau}, {Le Floc'h}, {Lutz}, {Morrison}, {Mullaney}, {Murphy},
  {Pope}, {Scott}, {Brodwin}, {Calzetti}, {Cesarsky}, {Charlot}, {Dole},
  {Eisenhardt}, {Ferguson}, {F{\"o}rster Schreiber}, {Frayer}, {Giavalisco},
  {Huynh}, {Koekemoer}, {Papovich}, {Reddy}, {Surace}, {Teplitz}, {Yun}, \&
  {Wilson}}]{elbaz/etal:2011}
{Elbaz} D. {et~al.}, 2011, \aap, 533, A119

\bibitem[{{Fernandez} {et~al}\mbox{.}(2010){Fernandez}, {Komatsu}, {Iliev}, \&
  {Shapiro}}]{fernandez/etal:2010}
{Fernandez} E.~R., {Komatsu} E., {Iliev} I.~T., {Shapiro} P.~R., 2010, \apj,
  710, 1089

\bibitem[{{Fixsen} {et~al}\mbox{.}(1998){Fixsen}, {Dwek}, {Mather}, {Bennett},
  \& {Shafer}}]{fixsen/etal:1998}
{Fixsen} D.~J., {Dwek} E., {Mather} J.~C., {Bennett} C.~L., {Shafer} R.~A.,
  1998, \apj, 508, 123

\bibitem[{{Fowler} {et~al}\mbox{.}(2010){Fowler}, {Acquaviva}, {Ade},
  {Aguirre}, {Amiri}, {Appel}, {Barrientos}, {Battistelli}, {Bond}, {Brown},
  {Burger}, {Chervenak}, {Das}, {Devlin}, {Dicker}, {Doriese}, {Dunkley},
  {D{\"u}nner}, {Essinger-Hileman}, {Fisher}, {Hajian}, {Halpern},
  {Hasselfield}, {Hern{\'a}ndez-Monteagudo}, {Hilton}, {Hilton}, {Hincks},
  {Hlozek}, {Huffenberger}, {Hughes}, {Hughes}, {Infante}, {Irwin}, {Jimenez},
  {Juin}, {Kaul}, {Klein}, {Kosowsky}, {Lau}, {Limon}, {Lin}, {Lupton},
  {Marriage}, {Marsden}, {Martocci}, {Mauskopf}, {Menanteau}, {Moodley},
  {Moseley}, {Netterfield}, {Niemack}, {Nolta}, {Page}, {Parker}, {Partridge},
  {Quintana}, {Reid}, {Sehgal}, {Sievers}, {Spergel}, {Staggs}, {Swetz},
  {Switzer}, {Thornton}, {Trac}, {Tucker}, {Verde}, {Warne}, {Wilson},
  {Wollack}, \& {Zhao}}]{fowler/etal:2010}
{Fowler} J.~W. {et~al.}, 2010, \apj, 722, 1148

\bibitem[{Fowler {et~al}\mbox{.}(2007)Fowler, Niemack, Dicker, Aboobaker, Ade,
  Battistelli, Devlin, Fisher, Halpern, Hargrave, Hincks, Kaul, Klein, Lau,
  Limon, Marriage, Mauskopf, Page, Staggs, Swetz, Switzer, Thornton, \&
  Tucker}]{fowler/etal:2007}
Fowler J.~W. {et~al.}, 2007, Appl. Opt., 46, 3444

\bibitem[{{Gautier} {et~al}\mbox{.}(1992){Gautier}, {Boulanger}, {Perault}, \&
  {Puget}}]{gautier/etal:1992}
{Gautier}, III T.~N., {Boulanger} F., {Perault} M., {Puget} J.~L., 1992, \aj,
  103, 1313

\bibitem[{Gelman {et~al}\mbox{.}(1996)Gelman, Roberts, \&
  Gilks}]{gelman/roberts/gilks:1996}
Gelman A., Roberts G.~O., Gilks W.~R., 1996, in Bayesian Statistics 5, et~al.
  J.~B., ed., Oxford University Press, pp. 599--607

\bibitem[{{Gispert} {et~al}\mbox{.}(2000){Gispert}, {Lagache}, \&
  {Puget}}]{gispert/etal:2000}
{Gispert} R., {Lagache} G., {Puget} J.~L., 2000, \aap, 360, 1

\bibitem[{{Glenn} {et~al}\mbox{.}(2010){Glenn}, {Conley}, {B{\'e}thermin},
  {Altieri}, {Amblard}, {Arumugam}, {Aussel}, {Babbedge}, {Blain}, {Bock},
  {Boselli}, {Buat}, {Castro-Rodr{\'{\i}}guez}, {Cava}, {Chanial}, {Clements},
  {Conversi}, {Cooray}, {Dowell}, {Dwek}, {Eales}, {Elbaz}, {Ellsworth-Bowers},
  {Fox}, {Franceschini}, {Gear}, {Griffin}, {Halpern}, {Hatziminaoglou},
  {Ibar}, {Isaak}, {Ivison}, {Lagache}, {Laurent}, {Levenson}, {Lu}, {Madden},
  {Maffei}, {Mainetti}, {Marchetti}, {Marsden}, {Nguyen}, {O'Halloran},
  {Oliver}, {Omont}, {Page}, {Panuzzo}, {Papageorgiou}, {Pearson},
  {P{\'e}rez-Fournon}, {Pohlen}, {Rigopoulou}, {Rizzo}, {Roseboom},
  {Rowan-Robinson}, {Portal}, {Schulz}, {Scott}, {Seymour}, {Shupe}, {Smith},
  {Stevens}, {Symeonidis}, {Trichas}, {Tugwell}, {Vaccari}, {Valtchanov},
  {Vieira}, {Vigroux}, {Wang}, {Ward}, {Wright}, {Xu}, \&
  {Zemcov}}]{glenn/etal:2010}
{Glenn} J. {et~al.}, 2010, \mnras, 409, 109

\bibitem[{{Gordon} {et~al}\mbox{.}(2007){Gordon}, {Engelbracht}, {Fadda},
  {Stansberry}, {Wachter}, {Frayer}, {Rieke}, {Noriega-Crespo}, {Latter},
  {Young}, {Neugebauer}, {Balog}, {Beeman}, {Dole}, {Egami}, {Haller}, {Hines},
  {Kelly}, {Marleau}, {Misselt}, {Morrison}, {P{\'e}rez-Gonz{\'a}lez}, {Rho},
  \& {Wheaton}}]{gordon/etal:2007}
{Gordon} K.~D. {et~al.}, 2007, \pasp, 119, 1019

\bibitem[{{Granato} {et~al}\mbox{.}(2004){Granato}, {De Zotti}, {Silva},
  {Bressan}, \& {Danese}}]{granato/etal:2004}
{Granato} G.~L., {De Zotti} G., {Silva} L., {Bressan} A., {Danese} L., 2004,
  \apj, 600, 580

\bibitem[{{Greve} {et~al}\mbox{.}(2012){Greve}, {Vieira}, {Wei{\ss}},
  {Aguirre}, {Aird}, {Ashby}, {Benson}, {Bleem}, {Bradford}, {Brodwin},
  {Carlstrom}, {Chang}, {Chapman}, {Crawford}, {de Breuck}, {de Haan}, {Dobbs},
  {Downes}, {Fassnacht}, {Fazio}, {George}, {Gladders}, {Gonzalez},
  {Halverson}, {Hezaveh}, {High}, {Holder}, {Holzapfel}, {Hoover}, {Hrubes},
  {Johnson}, {Keisler}, {Knox}, {Lee}, {Leitch}, {Lueker}, {Luong-Van},
  {Malkan}, {Marrone}, {McIntyre}, {McMahon}, {Mehl}, {Menten}, {Meyer},
  {Montroy}, {Murphy}, {Natoli}, {Padin}, {Plagge}, {Pryke}, {Reichardt},
  {Rest}, {Rosenman}, {Ruel}, {Ruhl}, {Schaffer}, {Sharon}, {Shaw},
  {Shirokoff}, {Stalder}, {Stanford}, {Staniszewski}, {Stark}, {Story},
  {Vanderlinde}, {Walsh}, {Welikala}, \& {Williamson}}]{greve/etal:2012}
{Greve} T.~R. {et~al.}, 2012, \apj, 756, 101

\bibitem[{{Griffin} {et~al}\mbox{.}(2010){Griffin}, {Abergel}, {Abreu}, {Ade},
  {Andr{\'e}}, {Augueres}, {Babbedge}, {Bae}, {Baillie}, {Baluteau}, {Barlow},
  {Bendo}, {Benielli}, {Bock}, {Bonhomme}, {Brisbin}, {Brockley-Blatt},
  {Caldwell}, {Cara}, {Castro-Rodriguez}, {Cerulli}, {Chanial}, {Chen},
  {Clark}, {Clements}, {Clerc}, {Coker}, {Communal}, {Conversi}, {Cox},
  {Crumb}, {Cunningham}, {Daly}, {Davis}, {de Antoni}, {Delderfield}, {Devin},
  {di Giorgio}, {Didschuns}, {Dohlen}, {Donati}, {Dowell}, {Dowell}, {Duband},
  {Dumaye}, {Emery}, {Ferlet}, {Ferrand}, {Fontignie}, {Fox}, {Franceschini},
  {Frerking}, {Fulton}, {Garcia}, {Gastaud}, {Gear}, {Glenn}, {Goizel},
  {Griffin}, {Grundy}, {Guest}, {Guillemet}, {Hargrave}, {Harwit}, {Hastings},
  {Hatziminaoglou}, {Herman}, {Hinde}, {Hristov}, {Huang}, {Imhof}, {Isaak},
  {Israelsson}, {Ivison}, {Jennings}, {Kiernan}, {King}, {Lange}, {Latter},
  {Laurent}, {Laurent}, {Leeks}, {Lellouch}, {Levenson}, {Li}, {Li},
  {Lilienthal}, {Lim}, {Liu}, {Lu}, {Madden}, {Mainetti}, {Marliani}, {McKay},
  {Mercier}, {Molinari}, {Morris}, {Moseley}, {Mulder}, {Mur}, {Naylor},
  {Nguyen}, {O'Halloran}, {Oliver}, {Olofsson}, {Olofsson}, {Orfei}, {Page},
  {Pain}, {Panuzzo}, {Papageorgiou}, {Parks}, {Parr-Burman}, {Pearce},
  {Pearson}, {P{\'e}rez-Fournon}, {Pinsard}, {Pisano}, {Podosek}, {Pohlen},
  {Polehampton}, {Pouliquen}, {Rigopoulou}, {Rizzo}, {Roseboom}, {Roussel},
  {Rowan-Robinson}, {Rownd}, {Saraceno}, {Sauvage}, {Savage}, {Savini},
  {Sawyer}, {Scharmberg}, {Schmitt}, {Schneider}, {Schulz}, {Schwartz},
  {Shafer}, {Shupe}, {Sibthorpe}, {Sidher}, {Smith}, {Smith}, {Smith},
  {Spencer}, {Stobie}, {Sudiwala}, {Sukhatme}, {Surace}, {Stevens}, {Swinyard},
  {Trichas}, {Tourette}, {Triou}, {Tseng}, {Tucker}, {Turner}, {Vaccari},
  {Valtchanov}, {Vigroux}, {Virique}, {Voellmer}, {Walker}, {Ward}, {Waskett},
  {Weilert}, {Wesson}, {White}, {Whitehouse}, {Wilson}, {Winter}, {Woodcraft},
  {Wright}, {Xu}, {Zavagno}, {Zemcov}, {Zhang}, \& {Zonca}}]{griffin/etal:2010}
{Griffin} M.~J. {et~al.}, 2010, \aap, 518, L3

\bibitem[{{Guo} {et~al}\mbox{.}(2011){Guo}, {Cole}, {Lacey}, {Baugh}, {Frenk},
  {Norberg}, {Auld}, {Baldry}, {Bamford}, {Bourne}, {Buttiglione}, {Cava},
  {Cooray}, {Croom}, {Dariush}, {de Zotti}, {Driver}, {Dunne}, {Dye}, {Eales},
  {Fritz}, {Hopkins}, {Hopwood}, {Ibar}, {Ivison}, {Jarvis}, {Jones}, {Kelvin},
  {Liske}, {Loveday}, {Maddox}, {Parkinson}, {Pascale}, {Peacock}, {Pohlen},
  {Prescott}, {Rigby}, {Robotham}, {Rodighiero}, {Sharp}, {Smith}, {Temi}, \&
  {van Kampen}}]{guo/etal:2011}
{Guo} Q. {et~al.}, 2011, \mnras, 412, 2277

\bibitem[{{Haiman} \& {Knox}(2000)}]{haiman/knox:2000}
{Haiman} Z., {Knox} L., 2000, \apj, 530, 124

\bibitem[{{Hajian} {et~al}\mbox{.}(2011){Hajian}, {Acquaviva}, {Ade},
  {Aguirre}, {Amiri}, {Appel}, {Barrientos}, {Battistelli}, {Bond}, {Brown},
  {Burger}, {Chervenak}, {Das}, {Devlin}, {Dicker}, {Bertrand Doriese},
  {Dunkley}, {D{\"u}nner}, {Essinger-Hileman}, {Fisher}, {Fowler}, {Halpern},
  {Hasselfield}, {Hern{\'a}ndez-Monteagudo}, {Hilton}, {Hilton}, {Hincks},
  {Hlozek}, {Huffenberger}, {Hughes}, {Hughes}, {Infante}, {Irwin}, {Baptiste
  Juin}, {Kaul}, {Klein}, {Kosowsky}, {Lau}, {Limon}, {Lin}, {Lupton},
  {Marriage}, {Marsden}, {Mauskopf}, {Menanteau}, {Moodley}, {Moseley},
  {Netterfield}, {Niemack}, {Nolta}, {Page}, {Parker}, {Partridge}, {Reid},
  {Sehgal}, {Sherwin}, {Sievers}, {Spergel}, {Staggs}, {Swetz}, {Switzer},
  {Thornton}, {Trac}, {Tucker}, {Warne}, {Wollack}, \&
  {Zhao}}]{hajian/etal:2011}
{Hajian} A. {et~al.}, 2011, \apj, 740, 86

\bibitem[{{Hajian} {et~al}\mbox{.}(2012){Hajian}, {Viero}, {Addison},
  {Aguirre}, {Appel}, {Battaglia}, {Bock}, {Bond}, {Das}, {Devlin}, {Dicker},
  {Dunkley}, {D{\"u}nner}, {Essinger-Hileman}, {Hughes}, {Fowler}, {Halpern},
  {Hasselfield}, {Hilton}, {Hincks}, {Hlozek}, {Irwin}, {Klein}, {Kosowsky},
  {Lin}, {Marriage}, {Marsden}, {Marsden}, {Menanteau}, {Moncelsi}, {Moodley},
  {Netterfield}, {Niemack}, {Nolta}, {Page}, {Parker}, {Patanchon}, {Scott},
  {Sehgal}, {Sievers}, {Spergel}, {Staggs}, {Swetz}, {Switzer}, {Thornton}, \&
  {Wollack}}]{hajian/etal:2012}
{Hajian} A. {et~al.}, 2012, \apj, 744, 40

\bibitem[{{Hall} {et~al}\mbox{.}(2010){Hall}, {Keisler}, {Knox}, {Reichardt},
  {Ade}, {Aird}, {Benson}, {Bleem}, {Carlstrom}, {Chang}, {Cho}, {Crawford},
  {Crites}, {de Haan}, {Dobbs}, {George}, {Halverson}, {Holder}, {Holzapfel},
  {Hrubes}, {Joy}, {Lee}, {Leitch}, {Lueker}, {McMahon}, {Mehl}, {Meyer},
  {Mohr}, {Montroy}, {Padin}, {Plagge}, {Pryke}, {Ruhl}, {Schaffer}, {Shaw},
  {Shirokoff}, {Spieler}, {Stalder}, {Staniszewski}, {Stark}, {Switzer},
  {Vanderlinde}, {Vieira}, {Williamson}, \& {Zahn}}]{hall/etal:2010}
{Hall} N.~R. {et~al.}, 2010, \apj, 718, 632

\bibitem[{{Hashimoto} {et~al}\mbox{.}(1998){Hashimoto}, {Oemler}, {Lin}, \&
  {Tucker}}]{hashimoto/etal:1998}
{Hashimoto} Y., {Oemler}, Jr. A., {Lin} H., {Tucker} D.~L., 1998, \apj, 499,
  589

\bibitem[{{Hauser} \& {Dwek}(2001)}]{hauser/dwek:2001}
{Hauser} M.~G., {Dwek} E., 2001, \araa, 39, 249

\bibitem[{{Hayward} {et~al}\mbox{.}(2012){Hayward}, {Jonsson}, {Kere{\v s}},
  {Magnelli}, {Hernquist}, \& {Cox}}]{hayward/etal:2012}
{Hayward} C.~C., {Jonsson} P., {Kere{\v s}} D., {Magnelli} B., {Hernquist} L.,
  {Cox} T.~J., 2012, \mnras, 424, 951

\bibitem[{{Hezaveh} \& {Holder}(2011)}]{hezaveh/holder:2011}
{Hezaveh} Y.~D., {Holder} G.~P., 2011, \apj, 734, 52

\bibitem[{{Hezaveh} {et~al}\mbox{.}(2012){Hezaveh}, {Marrone}, \&
  {Holder}}]{hezaveh/etal:2012}
{Hezaveh} Y.~D., {Marrone} D.~P., {Holder} G.~P., 2012, \apj, 761, 20

\bibitem[{{Hildebrand}(1983)}]{hildebrand:1983}
{Hildebrand} R.~H., 1983, \qjras, 24, 267

\bibitem[{{Holland} {et~al}\mbox{.}(1999){Holland}, {Robson}, {Gear},
  {Cunningham}, {Lightfoot}, {Jenness}, {Ivison}, {Stevens}, {Ade}, {Griffin},
  {Duncan}, {Murphy}, \& {Naylor}}]{holland/etal:1999}
{Holland} W.~S. {et~al.}, 1999, \mnras, 303, 659

\bibitem[{{Hughes} {et~al}\mbox{.}(1998){Hughes}, {Serjeant}, {Dunlop},
  {Rowan-Robinson}, {Blain}, {Mann}, {Ivison}, {Peacock}, {Efstathiou}, {Gear},
  {Oliver}, {Lawrence}, {Longair}, {Goldschmidt}, \&
  {Jenness}}]{hughes/etal:1998}
{Hughes} D.~H. {et~al.}, 1998, \nat, 394, 241

\bibitem[{{Kaiser}(1992)}]{kaiser:1992}
{Kaiser} N., 1992, \apj, 388, 272

\bibitem[{{Karim} {et~al}\mbox{.}(2013){Karim}, {Swinbank}, {Hodge}, {Smail},
  {Walter}, {Biggs}, {Simpson}, {Danielson}, {Alexander}, {Bertoldi}, {de
  Breuck}, {Chapman}, {Coppin}, {Dannerbauer}, {Edge}, {Greve}, {Ivison},
  {Knudsen}, {Menten}, {Schinnerer}, {Wardlow}, {Wei{\ss}}, \& {van der
  Werf}}]{karim/etal:2013}
{Karim} A. {et~al.}, 2013, \mnras, 432, 2

\bibitem[{{Kashlinsky} \& {Odenwald}(2000)}]{kashlinsky/odenwald:2000}
{Kashlinsky} A., {Odenwald} S., 2000, \apj, 528, 74

\bibitem[{{Keisler} {et~al}\mbox{.}(2011){Keisler}, {Reichardt}, {Aird},
  {Benson}, {Bleem}, {Carlstrom}, {Chang}, {Cho}, {Crawford}, {Crites}, {de
  Haan}, {Dobbs}, {Dudley}, {George}, {Halverson}, {Holder}, {Holzapfel},
  {Hoover}, {Hou}, {Hrubes}, {Joy}, {Knox}, {Lee}, {Leitch}, {Lueker},
  {Luong-Van}, {McMahon}, {Mehl}, {Meyer}, {Millea}, {Mohr}, {Montroy},
  {Natoli}, {Padin}, {Plagge}, {Pryke}, {Ruhl}, {Schaffer}, {Shaw},
  {Shirokoff}, {Spieler}, {Staniszewski}, {Stark}, {Story}, {van Engelen},
  {Vanderlinde}, {Vieira}, {Williamson}, \& {Zahn}}]{keisler/etal:2011}
{Keisler} R. {et~al.}, 2011, \apj, 743, 28

\bibitem[{{Kennicutt}(1983)}]{kennicutt:1983}
{Kennicutt}, Jr. R.~C., 1983, \aj, 88, 483

\bibitem[{{Kennicutt}(1998)}]{kennicutt:1998}
{Kennicutt}, Jr. R.~C., 1998, \apj, 498, 541

\bibitem[{{Kirkpatrick} {et~al}\mbox{.}(2012){Kirkpatrick}, {Pope},
  {Alexander}, {Charmandaris}, {Daddi}, {Dickinson}, {Elbaz}, {Gabor}, {Hwang},
  {Ivison}, {Mullaney}, {Pannella}, {Scott}, {Altieri}, {Aussel}, {Bournaud},
  {Buat}, {Coia}, {Dannerbauer}, {Dasyra}, {Kartaltepe}, {Leiton}, {Lin},
  {Magdis}, {Magnelli}, {Morrison}, {Popesso}, \&
  {Valtchanov}}]{kirkpatrick/etal:2012}
{Kirkpatrick} A. {et~al.}, 2012, \apj, 759, 139

\bibitem[{{Knox} {et~al}\mbox{.}(2001){Knox}, {Cooray}, {Eisenstein}, \&
  {Haiman}}]{knox/etal:2001}
{Knox} L., {Cooray} A., {Eisenstein} D., {Haiman} Z., 2001, \apj, 550, 7

\bibitem[{{Komatsu} {et~al}\mbox{.}(2011){Komatsu}, {Smith}, {Dunkley},
  {Bennett}, {Gold}, {Hinshaw}, {Jarosik}, {Larson}, {Nolta}, {Page},
  {Spergel}, {Halpern}, {Hill}, {Kogut}, {Limon}, {Meyer}, {Odegard}, {Tucker},
  {Weiland}, {Wollack}, \& {Wright}}]{komatsu/etal:2011}
{Komatsu} E. {et~al.}, 2011, \apjs, 192, 18

\bibitem[{{Kravtsov} {et~al}\mbox{.}(2004){Kravtsov}, {Berlind}, {Wechsler},
  {Klypin}, {Gottl{\"o}ber}, {Allgood}, \& {Primack}}]{kravtsov/etal:2004}
{Kravtsov} A.~V., {Berlind} A.~A., {Wechsler} R.~H., {Klypin} A.~A.,
  {Gottl{\"o}ber} S., {Allgood} B., {Primack} J.~R., 2004, \apj, 609, 35

\bibitem[{{Lagache} {et~al}\mbox{.}(1999){Lagache}, {Abergel}, {Boulanger},
  {D{\'e}sert}, \& {Puget}}]{lagache/etal:1999}
{Lagache} G., {Abergel} A., {Boulanger} F., {D{\'e}sert} F.~X., {Puget} J.-L.,
  1999, \aap, 344, 322

\bibitem[{{Lagache} {et~al}\mbox{.}(2007){Lagache}, {Bavouzet},
  {Fernandez-Conde}, {Ponthieu}, {Rodet}, {Dole}, {Miville-Desch{\^e}nes}, \&
  {Puget}}]{lagache/etal:2007}
{Lagache} G., {Bavouzet} N., {Fernandez-Conde} N., {Ponthieu} N., {Rodet} T.,
  {Dole} H., {Miville-Desch{\^e}nes} M.-A., {Puget} J.-L., 2007, \apjl, 665,
  L89

\bibitem[{{Lagache} {et~al}\mbox{.}(2003){Lagache}, {Dole}, \&
  {Puget}}]{lagache/etal:2003}
{Lagache} G., {Dole} H., {Puget} J.-L., 2003, \mnras, 338, 555

\bibitem[{{Lagache} {et~al}\mbox{.}(2004){Lagache}, {Dole}, {Puget},
  {P{\'e}rez-Gonz{\'a}lez}, {Le Floc'h}, {Rieke}, {Papovich}, {Egami},
  {Alonso-Herrero}, {Engelbracht}, {Gordon}, {Misselt}, \&
  {Morrison}}]{lagache/etal:2004}
{Lagache} G. {et~al.}, 2004, \apjs, 154, 112

\bibitem[{{Lamarre} {et~al}\mbox{.}(2010){Lamarre}, {Puget}, {Ade}, {Bouchet},
  {Guyot}, {Lange}, {Pajot}, {Arondel}, {Benabed}, {Beney}, {Beno{\^i}t},
  {Bernard}, {Bhatia}, {Blanc}, {Bock}, {Br{\'e}elle}, {Bradshaw}, {Camus},
  {Catalano}, {Charra}, {Charra}, {Church}, {Couchot}, {Coulais}, {Crill},
  {Crook}, {Dassas}, {de Bernardis}, {Delabrouille}, {de Marcillac}, {Delouis},
  {D{\'e}sert}, {Dumesnil}, {Dupac}, {Efstathiou}, {Eng}, {Evesque},
  {Fourmond}, {Ganga}, {Giard}, {Gispert}, {Guglielmi}, {Haissinski},
  {Henrot-Versill{\'e}}, {Hivon}, {Holmes}, {Jones}, {Koch}, {Lagard{\`e}re},
  {Lami}, {Land{\'e}}, {Leriche}, {Leroy}, {Longval},
  {Mac{\'{\i}}as-P{\'e}rez}, {Maciaszek}, {Maffei}, {Mansoux}, {Marty}, {Masi},
  {Mercier}, {Miville-Desch{\^e}nes}, {Moneti}, {Montier}, {Murphy},
  {Narbonne}, {Nexon}, {Paine}, {Pahn}, {Perdereau}, {Piacentini}, {Piat},
  {Plaszczynski}, {Pointecouteau}, {Pons}, {Ponthieu}, {Prunet}, {Rambaud},
  {Recouvreur}, {Renault}, {Ristorcelli}, {Rosset}, {Santos}, {Savini},
  {Serra}, {Stassi}, {Sudiwala}, {Sygnet}, {Tauber}, {Torre}, {Tristram},
  {Vibert}, {Woodcraft}, {Yurchenko}, \& {Yvon}}]{lamarre/etal:2010}
{Lamarre} J.-M. {et~al.}, 2010, \aap, 520, A9

\bibitem[{{Lapi} {et~al}\mbox{.}(2011){Lapi}, {Gonz{\'a}lez-Nuevo}, {Fan},
  {Bressan}, {De Zotti}, {Danese}, {Negrello}, {Dunne}, {Eales}, {Maddox},
  {Auld}, {Baes}, {Bonfield}, {Buttiglione}, {Cava}, {Clements}, {Cooray},
  {Dariush}, {Dye}, {Fritz}, {Herranz}, {Hopwood}, {Ibar}, {Ivison}, {Jarvis},
  {Kaviraj}, {L{\'o}pez-Caniego}, {Massardi}, {Micha{\l}owski}, {Pascale},
  {Pohlen}, {Rigby}, {Rodighiero}, {Serjeant}, {Smith}, {Temi}, {Wardlow}, \&
  {van der Werf}}]{lapi/etal:2011}
{Lapi} A. {et~al.}, 2011, \apj, 742, 24

\bibitem[{{Le Floc'h} {et~al}\mbox{.}(2005){Le Floc'h}, {Papovich}, {Dole},
  {Bell}, {Lagache}, {Rieke}, {Egami}, {P{\'e}rez-Gonz{\'a}lez},
  {Alonso-Herrero}, {Rieke}, {Blaylock}, {Engelbracht}, {Gordon}, {Hines},
  {Misselt}, {Morrison}, \& {Mould}}]{lefloc'h/etal:2005}
{Le Floc'h} E. {et~al.}, 2005, \apj, 632, 169

\bibitem[{{Le Floc'h} {et~al}\mbox{.}(2007){Le Floc'h}, {Willmer}, {Noeske},
  {Konidaris}, {Laird}, {Koo}, {Nandra}, {Bundy}, {Salim}, {Maiolino},
  {Conselice}, {Lotz}, {Papovich}, {Smith}, {Bai}, {Coil}, {Barmby}, {Ashby},
  {Huang}, {Blaylock}, {Rieke}, {Newman}, {Ivison}, {Chapman}, {Dole}, {Egami},
  \& {Elbaz}}]{lefloch/etal:2007}
{Le Floc'h} E. {et~al.}, 2007, \apjl, 660, L65

\bibitem[{{Leger} \& {Puget}(1984)}]{leger/puget:1984}
{Leger} A., {Puget} J.~L., 1984, \aap, 137, L5

\bibitem[{{Lewis} {et~al}\mbox{.}(2000){Lewis}, {Challinor}, \&
  {Lasenby}}]{lewis/etal:2000}
{Lewis} A., {Challinor} A., {Lasenby} A., 2000, \apj, 538, 473

\bibitem[{{Lima} {et~al}\mbox{.}(2010{\natexlab{a}}){Lima}, {Jain}, \&
  {Devlin}}]{lima/etal:2010b}
{Lima} M., {Jain} B., {Devlin} M., 2010{\natexlab{a}}, \mnras, 406, 2352

\bibitem[{{Lima} {et~al}\mbox{.}(2010{\natexlab{b}}){Lima}, {Jain}, {Devlin},
  \& {Aguirre}}]{lima/etal:2010}
{Lima} M., {Jain} B., {Devlin} M., {Aguirre} J., 2010{\natexlab{b}}, \apjl,
  717, L31

\bibitem[{{Limber}(1953)}]{limber:1953}
{Limber} D.~N., 1953, \apj, 117, 134

\bibitem[{{Lueker} {et~al}\mbox{.}(2010){Lueker}, {Reichardt}, {Schaffer},
  {Zahn}, {Ade}, {Aird}, {Benson}, {Bleem}, {Carlstrom}, {Chang}, {Cho},
  {Crawford}, {Crites}, {de Haan}, {Dobbs}, {George}, {Hall}, {Halverson},
  {Holder}, {Holzapfel}, {Hrubes}, {Joy}, {Keisler}, {Knox}, {Lee}, {Leitch},
  {McMahon}, {Mehl}, {Meyer}, {Mohr}, {Montroy}, {Padin}, {Plagge}, {Pryke},
  {Ruhl}, {Shaw}, {Shirokoff}, {Spieler}, {Stalder}, {Staniszewski}, {Stark},
  {Vanderlinde}, {Vieira}, \& {Williamson}}]{lueker/etal:2010}
{Lueker} M. {et~al.}, 2010, \apj, 719, 1045

\bibitem[{{Lutz} {et~al}\mbox{.}(2005){Lutz}, {Valiante}, {Sturm}, {Genzel},
  {Tacconi}, {Lehnert}, {Sternberg}, \& {Baker}}]{lutz/etal:2005}
{Lutz} D., {Valiante} E., {Sturm} E., {Genzel} R., {Tacconi} L.~J., {Lehnert}
  M.~D., {Sternberg} A., {Baker} A.~J., 2005, \apjl, 625, L83

\bibitem[{{Maddox} {et~al}\mbox{.}(2010){Maddox}, {Dunne}, {Rigby}, {Eales},
  {Cooray}, {Scott}, {Peacock}, {Negrello}, {Smith}, {Benford}, {Amblard},
  {Auld}, {Baes}, {Bonfield}, {Burgarella}, {Buttiglione}, {Cava}, {Clements},
  {Dariush}, {de Zotti}, {Dye}, {Frayer}, {Fritz}, {Gonzalez-Nuevo}, {Herranz},
  {Ibar}, {Ivison}, {Jarvis}, {Lagache}, {Leeuw}, {Lopez-Caniego}, {Pascale},
  {Pohlen}, {Rodighiero}, {Samui}, {Serjeant}, {Temi}, {Thompson}, \&
  {Verma}}]{maddox/etal:2010}
{Maddox} S.~J. {et~al.}, 2010, \aap, 518, L11

\bibitem[{{Magdis} {et~al}\mbox{.}(2012){Magdis}, {Daddi}, {B{\'e}thermin},
  {Sargent}, {Elbaz}, {Pannella}, {Dickinson}, {Dannerbauer}, {da Cunha},
  {Walter}, {Rigopoulou}, {Charmandaris}, {Hwang}, \&
  {Kartaltepe}}]{magdis/etal:2012}
{Magdis} G.~E. {et~al.}, 2012, \apj, 760, 6

\bibitem[{{Magnelli} {et~al}\mbox{.}(2011){Magnelli}, {Elbaz}, {Chary},
  {Dickinson}, {Le Borgne}, {Frayer}, \& {Willmer}}]{magnelli/etal:2011}
{Magnelli} B., {Elbaz} D., {Chary} R.~R., {Dickinson} M., {Le Borgne} D.,
  {Frayer} D.~T., {Willmer} C.~N.~A., 2011, \aap, 528, A35+

\bibitem[{{Mandelbaum} {et~al}\mbox{.}(2013){Mandelbaum}, {Slosar}, {Baldauf},
  {Seljak}, {Hirata}, {Nakajima}, {Reyes}, \& {Smith}}]{mandelbaum/etal:2013}
{Mandelbaum} R., {Slosar} A., {Baldauf} T., {Seljak} U., {Hirata} C.~M.,
  {Nakajima} R., {Reyes} R., {Smith} R.~E., 2013, \mnras

\bibitem[{{Marsden} {et~al}\mbox{.}(2009){Marsden}, {Ade}, {Bock}, {Chapin},
  {Devlin}, {Dicker}, {Griffin}, {Gundersen}, {Halpern}, {Hargrave}, {Hughes},
  {Klein}, {Mauskopf}, {Magnelli}, {Moncelsi}, {Netterfield}, {Ngo}, {Olmi},
  {Pascale}, {Patanchon}, {Rex}, {Scott}, {Semisch}, {Thomas}, {Truch},
  {Tucker}, {Tucker}, {Viero}, \& {Wiebe}}]{marsden/etal:2009}
{Marsden} G. {et~al.}, 2009, \apj, 707, 1729

\bibitem[{{McMahon} {et~al}\mbox{.}(2009){McMahon}, {Aird}, {Benson}, {Bleem},
  {Britton}, {Carlstrom}, {Chang}, {Cho}, {de Haan}, {Crawford}, {Crites},
  {Datesman}, {Dobbs}, {Everett}, {Halverson}, {Holder}, {Holzapfel}, {Hrubes},
  {Irwin}, {Joy}, {Keisler}, {Lanting}, {Lee}, {Leitch}, {Loehr}, {Lueker},
  {Mehl}, {Meyer}, {Mohr}, {Montroy}, {Niemack}, {Ngeow}, {Novosad}, {Padin},
  {Plagge}, {Pryke}, {Reichardt}, {Ruhl}, {Schaffer}, {Shaw}, {Shirokoff},
  {Spieler}, {Stadler}, {Stark}, {Staniszewski}, {Vanderlinde}, {Vieira},
  {Wang}, {Williamson}, {Yefremenko}, {Yoon}, {Zhan}, \&
  {Zenteno}}]{mcmahon/etal:2009}
{McMahon} J.~J. {et~al.}, 2009, in American Institute of Physics Conference
  Series, Vol. 1185, American Institute of Physics Conference Series, {Young}
  B., {Cabrera} B., {Miller} A., eds., pp. 511--514

\bibitem[{{Mead} {et~al}\mbox{.}(2010){Mead}, {King}, {Sijacki}, {Leonard},
  {Puchwein}, \& {McCarthy}}]{mead/etal:2010}
{Mead} J.~M.~G., {King} L.~J., {Sijacki} D., {Leonard} A., {Puchwein} E.,
  {McCarthy} I.~G., 2010, \mnras, 406, 434

\bibitem[{{Mesinger} {et~al}\mbox{.}(2012){Mesinger}, {McQuinn}, \&
  {Spergel}}]{mesinger/etal:2012}
{Mesinger} A., {McQuinn} M., {Spergel} D.~N., 2012, \mnras, 422, 1403

\bibitem[{{Metropolis} {et~al}\mbox{.}(1953){Metropolis}, {Rosenbluth}, \&
  {Rosenbluth}}]{metropolis/etal:1953}
{Metropolis} N., {Rosenbluth} A.~W., {Rosenbluth}, M.~N.~and~{Teller} A.~H.,
  1953, J.~Chem.~Phys., 21, 1087

\bibitem[{{Micha{\l}owski} {et~al}\mbox{.}(2010{\natexlab{a}}){Micha{\l}owski},
  {Hjorth}, \& {Watson}}]{michalowski/etal:2010a}
{Micha{\l}owski} M., {Hjorth} J., {Watson} D., 2010{\natexlab{a}}, \aap, 514,
  A67

\bibitem[{{Micha{\l}owski} {et~al}\mbox{.}(2010{\natexlab{b}}){Micha{\l}owski},
  {Watson}, \& {Hjorth}}]{michalowski/etal:2010b}
{Micha{\l}owski} M.~J., {Watson} D., {Hjorth} J., 2010{\natexlab{b}}, \apj,
  712, 942

\bibitem[{{Miville-Desch{\^e}nes} {et~al}\mbox{.}(2007){Miville-Desch{\^e}nes},
  {Lagache}, {Boulanger}, \& {Puget}}]{miville-deschenes/etal:2007}
{Miville-Desch{\^e}nes} M.-A., {Lagache} G., {Boulanger} F., {Puget} J.-L.,
  2007, \aap, 469, 595

\bibitem[{{Navarro} {et~al}\mbox{.}(1996){Navarro}, {Frenk}, \&
  {White}}]{navarro/etal:1996}
{Navarro} J.~F., {Frenk} C.~S., {White} S.~D.~M., 1996, \apj, 462, 563

\bibitem[{{Negrello} {et~al}\mbox{.}(2010){Negrello}, {Hopwood}, {De Zotti},
  {Cooray}, {Verma}, {Bock}, {Frayer}, {Gurwell}, {Omont}, {Neri},
  {Dannerbauer}, {Leeuw}, {Barton}, {Cooke}, {Kim}, {da Cunha}, {Rodighiero},
  {Cox}, {Bonfield}, {Jarvis}, {Serjeant}, {Ivison}, {Dye}, {Aretxaga},
  {Hughes}, {Ibar}, {Bertoldi}, {Valtchanov}, {Eales}, {Dunne}, {Driver},
  {Auld}, {Buttiglione}, {Cava}, {Grady}, {Clements}, {Dariush}, {Fritz},
  {Hill}, {Hornbeck}, {Kelvin}, {Lagache}, {Lopez-Caniego}, {Gonzalez-Nuevo},
  {Maddox}, {Pascale}, {Pohlen}, {Rigby}, {Robotham}, {Simpson}, {Smith},
  {Temi}, {Thompson}, {Woodgate}, {York}, {Aguirre}, {Beelen}, {Blain},
  {Baker}, {Birkinshaw}, {Blundell}, {Bradford}, {Burgarella}, {Danese},
  {Dunlop}, {Fleuren}, {Glenn}, {Harris}, {Kamenetzky}, {Lupu}, {Maddalena},
  {Madore}, {Maloney}, {Matsuhara}, {Michaowski}, {Murphy}, {Naylor}, {Nguyen},
  {Popescu}, {Rawlings}, {Rigopoulou}, {Scott}, {Scott}, {Seibert}, {Smail},
  {Tuffs}, {Vieira}, {van der Werf}, \& {Zmuidzinas}}]{negrello/etal:2010}
{Negrello} M. {et~al.}, 2010, Science, 330, 800

\bibitem[{{Negrello} {et~al}\mbox{.}(2007){Negrello}, {Perrotta},
  {Gonz{\'a}lez-Nuevo}, {Silva}, {de Zotti}, {Granato}, {Baccigalupi}, \&
  {Danese}}]{negrello/etal:2007}
{Negrello} M., {Perrotta} F., {Gonz{\'a}lez-Nuevo} J., {Silva} L., {de Zotti}
  G., {Granato} G.~L., {Baccigalupi} C., {Danese} L., 2007, \mnras, 377, 1557

\bibitem[{{Neugebauer} {et~al}\mbox{.}(1984){Neugebauer}, {Habing}, {van
  Duinen}, {Aumann}, {Baud}, {Beichman}, {Beintema}, {Boggess}, {Clegg}, {de
  Jong}, {Emerson}, {Gautier}, {Gillett}, {Harris}, {Hauser}, {Houck},
  {Jennings}, {Low}, {Marsden}, {Miley}, {Olnon}, {Pottasch}, {Raimond},
  {Rowan-Robinson}, {Soifer}, {Walker}, {Wesselius}, \&
  {Young}}]{neugebauer/etal:1984}
{Neugebauer} G. {et~al.}, 1984, \apjl, 278, L1

\bibitem[{{Niemack} {et~al}\mbox{.}(2010){Niemack}, {Ade}, {Aguirre},
  {Barrientos}, {Beall}, {Bond}, {Britton}, {Cho}, {Das}, {Devlin}, {Dicker},
  {Dunkley}, {D{\"u}nner}, {Fowler}, {Hajian}, {Halpern}, {Hasselfield},
  {Hilton}, {Hilton}, {Hubmayr}, {Hughes}, {Infante}, {Irwin}, {Jarosik},
  {Klein}, {Kosowsky}, {Marriage}, {McMahon}, {Menanteau}, {Moodley},
  {Nibarger}, {Nolta}, {Page}, {Partridge}, {Reese}, {Sievers}, {Spergel},
  {Staggs}, {Thornton}, {Tucker}, {Wollack}, \& {Yoon}}]{niemack/etal:2010}
{Niemack} M.~D. {et~al.}, 2010, in Society of Photo-Optical Instrumentation
  Engineers (SPIE) Conference Series, Vol. 7741, Society of Photo-Optical
  Instrumentation Engineers (SPIE) Conference Series

\bibitem[{{Oliver} {et~al}\mbox{.}(2010){Oliver}, {Wang}, {Smith}, {Altieri},
  {Amblard}, {Arumugam}, {Auld}, {Aussel}, {Babbedge}, {Blain}, {Bock},
  {Boselli}, {Buat}, {Burgarella}, {Castro-Rodr{\'{\i}}guez}, {Cava},
  {Chanial}, {Clements}, {Conley}, {Conversi}, {Cooray}, {Dowell}, {Dwek},
  {Eales}, {Elbaz}, {Fox}, {Franceschini}, {Gear}, {Glenn}, {Griffin},
  {Halpern}, {Hatziminaoglou}, {Ibar}, {Isaak}, {Ivison}, {Lagache},
  {Levenson}, {Lu}, {Madden}, {Maffei}, {Mainetti}, {Marchetti},
  {Mitchell-Wynne}, {Mortier}, {Nguyen}, {O'Halloran}, {Omont}, {Page},
  {Panuzzo}, {Papageorgiou}, {Pearson}, {P{\'e}rez-Fournon}, {Pohlen},
  {Rawlings}, {Raymond}, {Rigopoulou}, {Rizzo}, {Roseboom}, {Rowan-Robinson},
  {S{\'a}nchez Portal}, {Savage}, {Schulz}, {Scott}, {Seymour}, {Shupe},
  {Stevens}, {Symeonidis}, {Trichas}, {Tugwell}, {Vaccari}, {Valiante},
  {Valtchanov}, {Vieira}, {Vigroux}, {Ward}, {Wright}, {Xu}, \&
  {Zemcov}}]{oliver/etal:2010}
{Oliver} S.~J. {et~al.}, 2010, \aap, 518, L21

\bibitem[{{Pascale} {et~al}\mbox{.}(2009){Pascale}, {Ade}, {Bock}, {Chapin},
  {Devlin}, {Dye}, {Eales}, {Griffin}, {Gundersen}, {Halpern}, {Hargrave},
  {Hughes}, {Klein}, {Marsden}, {Mauskopf}, {Moncelsi}, {Ngo}, {Netterfield},
  {Olmi}, {Patanchon}, {Rex}, {Scott}, {Semisch}, {Thomas}, {Truch}, {Tucker},
  {Tucker}, {Viero}, \& {Wiebe}}]{pascale/etal:2009}
{Pascale} E. {et~al.}, 2009, \apj, 707, 1740

\bibitem[{{Patanchon} {et~al}\mbox{.}(2009){Patanchon}, {Ade}, {Bock},
  {Chapin}, {Devlin}, {Dicker}, {Griffin}, {Gundersen}, {Halpern}, {Hargrave},
  {Hughes}, {Klein}, {Marsden}, {Mauskopf}, {Moncelsi}, {Netterfield}, {Olmi},
  {Pascale}, {Rex}, {Scott}, {Semisch}, {Thomas}, {Truch}, {Tucker}, {Tucker},
  {Viero}, \& {Wiebe}}]{patanchon/etal:2009}
{Patanchon} G. {et~al.}, 2009, \apj, 707, 1750

\bibitem[{{Patel} {et~al}\mbox{.}(2013){Patel}, {Clements}, {Vaccari},
  {Mortlock}, {Rowan-Robinson}, {P{\'e}rez-Fournon}, \&
  {Afonso-Luis}}]{patel/etal:2013}
{Patel} H., {Clements} D.~L., {Vaccari} M., {Mortlock} D.~J., {Rowan-Robinson}
  M., {P{\'e}rez-Fournon} I., {Afonso-Luis} A., 2013, \mnras, 428, 291

\bibitem[{{Peacock} \& {Smith}(2000)}]{peacock/smith:2000}
{Peacock} J.~A., {Smith} R.~E., 2000, \mnras, 318, 1144

\bibitem[{{Peebles}(1980)}]{peebles:1980}
{Peebles} P.~J.~E., 1980, {The large-scale structure of the universe}. Research
  supported by the National Science Foundation.~Princeton, N.J., Princeton
  University Press, 1980.~435 p.

\bibitem[{{P{\'e}nin} {et~al}\mbox{.}(2012{\natexlab{a}}){P{\'e}nin},
  {Dor{\'e}}, {Lagache}, \& {B{\'e}thermin}}]{penin/etal:2012b}
{P{\'e}nin} A., {Dor{\'e}} O., {Lagache} G., {B{\'e}thermin} M.,
  2012{\natexlab{a}}, \aap, 537, A137

\bibitem[{{P{\'e}nin} {et~al}\mbox{.}(2012{\natexlab{b}}){P{\'e}nin},
  {Lagache}, {Noriega-Crespo}, {Grain}, {Miville-Desch{\^e}nes}, {Ponthieu},
  {Martin}, {Blagrave}, \& {Lockman}}]{penin/etal:2012}
{P{\'e}nin} A. {et~al.}, 2012{\natexlab{b}}, \aap, 543, A123

\bibitem[{{P{\'e}rez-Gonz{\'a}lez}
  {et~al}\mbox{.}(2005){P{\'e}rez-Gonz{\'a}lez}, {Rieke}, {Egami},
  {Alonso-Herrero}, {Dole}, {Papovich}, {Blaylock}, {Jones}, {Rieke}, {Rigby},
  {Barmby}, {Fazio}, {Huang}, \& {Martin}}]{perez-gonzales/etal:2005}
{P{\'e}rez-Gonz{\'a}lez} P.~G. {et~al.}, 2005, \apj, 630, 82

\bibitem[{{Perrotta} {et~al}\mbox{.}(2002){Perrotta}, {Baccigalupi},
  {Bartelmann}, {De Zotti}, \& {Granato}}]{perrotta/etal:2002}
{Perrotta} F., {Baccigalupi} C., {Bartelmann} M., {De Zotti} G., {Granato}
  G.~L., 2002, \mnras, 329, 445

\bibitem[{{Pilbratt} {et~al}\mbox{.}(2010){Pilbratt}, {Riedinger}, {Passvogel},
  {Crone}, {Doyle}, {Gageur}, {Heras}, {Jewell}, {Metcalfe}, {Ott}, \&
  {Schmidt}}]{pilbratt/etal:2010}
{Pilbratt} G.~L. {et~al.}, 2010, \aap, 518, L1

\bibitem[{{Planck Collaboration} {et~al}\mbox{.}(2011){Planck Collaboration},
  {Ade}, {Aghanim}, {Arnaud}, {Ashdown}, {Aumont}, {Baccigalupi}, {Balbi},
  {Banday}, {Barreiro}, \& et~al.}]{planckcib:2011}
{Planck Collaboration} {et~al.}, 2011, \aap, 536, A18

\bibitem[{{Planck HFI Core Team} {et~al}\mbox{.}(2011{\natexlab{a}}){Planck HFI
  Core Team}, {Ade}, {Aghanim}, {Ansari}, {Arnaud}, {Ashdown}, {Aumont},
  {Banday}, {Bartelmann}, {Bartlett}, {Battaner}, {Benabed}, {Beno{\^i}t},
  {Bernard}, {Bersanelli}, {Bhatia}, {Bock}, {Bond}, {Borrill}, {Bouchet},
  {Boulanger}, {Bradshaw}, {Br{\'e}elle}, {Bucher}, {Camus}, {Cardoso},
  {Catalano}, {Challinor}, {Chamballu}, {Charra}, {Charra}, {Chary}, {Chiang},
  {Church}, {Clements}, {Colombi}, {Couchot}, {Coulais}, {Cressiot}, {Crill},
  {Crook}, {de Bernardis}, {Delabrouille}, {Delouis}, {D{\'e}sert}, {Dolag},
  {Dole}, {Dor{\'e}}, {Douspis}, {Efstathiou}, {Eng}, {Filliard}, {Forni},
  {Fosalba}, {Fourmond}, {Ganga}, {Giard}, {Girard}, {Giraud-H{\'e}raud},
  {Gispert}, {G{\'o}rski}, {Gratton}, {Griffin}, {Guyot}, {Haissinski},
  {Harrison}, {Helou}, {Henrot-Versill{\'e}}, {Hern{\'a}ndez-Monteagudo},
  {Hildebrandt}, {Hills}, {Hivon}, {Hobson}, {Holmes}, {Huffenberger}, {Jaffe},
  {Jones}, {Kaplan}, {Kneissl}, {Knox}, {Lagache}, {Lamarre}, {Lami}, {Lange},
  {Lasenby}, {Lavabre}, {Lawrence}, {Leriche}, {Leroy}, {Longval},
  {Mac{\'{\i}}as-P{\'e}rez}, {Maciaszek}, {MacTavish}, {Maffei}, {Mandolesi},
  {Mann}, {Mansoux}, {Masi}, {Matsumura}, {McGehee}, {Melin}, {Mercier},
  {Miville-Desch{\^e}nes}, {Moneti}, {Montier}, {Mortlock}, {Murphy}, {Nati},
  {Netterfield}, {N{\o}rgaard-Nielsen}, {North}, {Noviello}, {Novikov},
  {Osborne}, {Paine}, {Pajot}, {Patanchon}, {Peacocke}, {Pearson}, {Perdereau},
  {Perotto}, {Piacentini}, {Piat}, {Plaszczynski}, {Pointecouteau}, {Pons},
  {Ponthieu}, {Pr{\'e}zeau}, {Prunet}, {Puget}, {Reach}, {Renault},
  {Ristorcelli}, {Rocha}, {Rosset}, {Roudier}, {Rowan-Robinson}, {Rusholme},
  {Santos}, {Savini}, {Schaefer}, {Shellard}, {Spencer}, {Starck}, {Stassi},
  {Stolyarov}, {Stompor}, {Sudiwala}, {Sunyaev}, {Sygnet}, {Tauber}, {Thum},
  {Torre}, {Touze}, {Tristram}, {van Leeuwen}, {Vibert}, {Vibert}, {Wade},
  {Wandelt}, {White}, {Wiesemeyer}, {Woodcraft}, {Yurchenko}, {Yvon}, \&
  {Zacchei}}]{planckHFIperformance:2011}
{Planck HFI Core Team} {et~al.}, 2011{\natexlab{a}}, \aap, 536, A4

\bibitem[{{Planck HFI Core Team} {et~al}\mbox{.}(2011{\natexlab{b}}){Planck HFI
  Core Team}, {Ade}, {Aghanim}, {Ansari}, {Arnaud}, {Ashdown}, {Aumont},
  {Banday}, {Bartelmann}, {Bartlett}, {Battaner}, {Benabed}, {Beno{\^i}t},
  {Bernard}, {Bersanelli}, {Bock}, {Bond}, {Borrill}, {Bouchet}, {Boulanger},
  {Bradshaw}, {Bucher}, {Cardoso}, {Castex}, {Catalano}, {Challinor},
  {Chamballu}, {Chary}, {Chen}, {Chiang}, {Church}, {Clements}, {Colley},
  {Colombi}, {Couchot}, {Coulais}, {Cressiot}, {Crill}, {Crook}, {de
  Bernardis}, {Delabrouille}, {Delouis}, {D{\'e}sert}, {Dolag}, {Dole},
  {Dor{\'e}}, {Douspis}, {Dunkley}, {Efstathiou}, {Filliard}, {Forni},
  {Fosalba}, {Ganga}, {Giard}, {Girard}, {Giraud-H{\'e}raud}, {Gispert},
  {G{\'o}rski}, {Gratton}, {Griffin}, {Guyot}, {Haissinski}, {Harrison},
  {Helou}, {Henrot-Versill{\'e}}, {Hern{\'a}ndez-Monteagudo}, {Hildebrandt},
  {Hills}, {Hivon}, {Hobson}, {Holmes}, {Huffenberger}, {Jaffe}, {Jones},
  {Kaplan}, {Kneissl}, {Knox}, {Kunz}, {Lagache}, {Lamarre}, {Lange},
  {Lasenby}, {Lavabre}, {Lawrence}, {Le Jeune}, {Leroy}, {Lesgourgues},
  {Mac{\'{\i}}as-P{\'e}rez}, {MacTavish}, {Maffei}, {Mandolesi}, {Mann},
  {Marleau}, {Marshall}, {Masi}, {Matsumura}, {McAuley}, {McGehee}, {Melin},
  {Mercier}, {Mitra}, {Miville-Desch{\^e}nes}, {Moneti}, {Montier}, {Mortlock},
  {Murphy}, {Nati}, {Netterfield}, {N{\o}rgaard-Nielsen}, {North}, {Noviello},
  {Novikov}, {Osborne}, {Pajot}, {Patanchon}, {Peacocke}, {Pearson},
  {Perdereau}, {Perotto}, {Piacentini}, {Piat}, {Plaszczynski},
  {Pointecouteau}, {Ponthieu}, {Pr{\'e}zeau}, {Prunet}, {Puget}, {Reach},
  {Remazeilles}, {Renault}, {Riazuelo}, {Ristorcelli}, {Rocha}, {Rosset},
  {Roudier}, {Rowan-Robinson}, {Rusholme}, {Saha}, {Santos}, {Savini},
  {Schaefer}, {Shellard}, {Spencer}, {Starck}, {Stolyarov}, {Stompor},
  {Sudiwala}, {Sunyaev}, {Sutton}, {Sygnet}, {Tauber}, {Thum}, {Torre},
  {Touze}, {Tristram}, {van Leeuwen}, {Vibert}, {Vibert}, {Wade}, {Wandelt},
  {White}, {Wiesemeyer}, {Woodcraft}, {Yurchenko}, {Yvon}, \&
  {Zacchei}}]{planckHFI:2011}
{Planck HFI Core Team} {et~al.}, 2011{\natexlab{b}}, \aap, 536, A6

\bibitem[{{Pozzi} {et~al}\mbox{.}(2004){Pozzi}, {Gruppioni}, {Oliver},
  {Matute}, {La Franca}, {Lari}, {Zamorani}, {Serjeant}, {Franceschini}, \&
  {Rowan-Robinson}}]{pozzi/etal:2004}
{Pozzi} F. {et~al.}, 2004, \apj, 609, 122

\bibitem[{{Puget} {et~al}\mbox{.}(1996){Puget}, {Abergel}, {Bernard},
  {Boulanger}, {Burton}, {Desert}, \& {Hartmann}}]{puget/etal:1996}
{Puget} J.-L., {Abergel} A., {Bernard} J.-P., {Boulanger} F., {Burton} W.~B.,
  {Desert} F.-X., {Hartmann} D., 1996, \aap, 308, L5+

\bibitem[{{Reichardt} {et~al}\mbox{.}(2012){Reichardt}, {Shaw}, {Zahn}, {Aird},
  {Benson}, {Bleem}, {Carlstrom}, {Chang}, {Cho}, {Crawford}, {Crites}, {de
  Haan}, {Dobbs}, {Dudley}, {George}, {Halverson}, {Holder}, {Holzapfel},
  {Hoover}, {Hou}, {Hrubes}, {Joy}, {Keisler}, {Knox}, {Lee}, {Leitch},
  {Lueker}, {Luong-Van}, {McMahon}, {Mehl}, {Meyer}, {Millea}, {Mohr},
  {Montroy}, {Natoli}, {Padin}, {Plagge}, {Pryke}, {Ruhl}, {Schaffer},
  {Shirokoff}, {Spieler}, {Staniszewski}, {Stark}, {Story}, {van Engelen},
  {Vanderlinde}, {Vieira}, \& {Williamson}}]{reichardt/etal:2012}
{Reichardt} C.~L. {et~al.}, 2012, \apj, 755, 70

\bibitem[{{Rieke} {et~al}\mbox{.}(2004){Rieke}, {Young}, {Engelbracht},
  {Kelly}, {Low}, {Haller}, {Beeman}, {Gordon}, {Stansberry}, {Misselt},
  {Cadien}, {Morrison}, {Rivlis}, {Latter}, {Noriega-Crespo}, {Padgett},
  {Stapelfeldt}, {Hines}, {Egami}, {Muzerolle}, {Alonso-Herrero}, {Blaylock},
  {Dole}, {Hinz}, {Le Floc'h}, {Papovich}, {P{\'e}rez-Gonz{\'a}lez}, {Smith},
  {Su}, {Bennett}, {Frayer}, {Henderson}, {Lu}, {Masci}, {Pesenson}, {Rebull},
  {Rho}, {Keene}, {Stolovy}, {Wachter}, {Wheaton}, {Werner}, \&
  {Richards}}]{rieke/etal:2004}
{Rieke} G.~H. {et~al.}, 2004, \apjs, 154, 25

\bibitem[{{Sajina} {et~al}\mbox{.}(2006){Sajina}, {Scott}, {Dennefeld}, {Dole},
  {Lacy}, \& {Lagache}}]{sajina/etal:2006}
{Sajina} A., {Scott} D., {Dennefeld} M., {Dole} H., {Lacy} M., {Lagache} G.,
  2006, \mnras, 369, 939

\bibitem[{{Sajina} {et~al}\mbox{.}(2012){Sajina}, {Yan}, {Fadda}, {Dasyra}, \&
  {Huynh}}]{sajina/etal:2012}
{Sajina} A., {Yan} L., {Fadda} D., {Dasyra} K., {Huynh} M., 2012, \apj, 757, 13

\bibitem[{{Saunders} {et~al}\mbox{.}(1990){Saunders}, {Rowan-Robinson},
  {Lawrence}, {Efstathiou}, {Kaiser}, {Ellis}, \& {Frenk}}]{saunders/etal:1990}
{Saunders} W., {Rowan-Robinson} M., {Lawrence} A., {Efstathiou} G., {Kaiser}
  N., {Ellis} R.~S., {Frenk} C.~S., 1990, \mnras, 242, 318

\bibitem[{{Scoccimarro} {et~al}\mbox{.}(2001){Scoccimarro}, {Feldman}, {Fry},
  \& {Frieman}}]{scoccimarro/etal:2001}
{Scoccimarro} R., {Feldman} H.~A., {Fry} J.~N., {Frieman} J.~A., 2001, \apj,
  546, 652

\bibitem[{{Scott} {et~al}\mbox{.}(2012){Scott}, {Wilson}, {Aretxaga},
  {Austermann}, {Chapin}, {Dunlop}, {Ezawa}, {Halpern}, {Hatsukade}, {Hughes},
  {Kawabe}, {Kim}, {Kohno}, {Lowenthal}, {Monta{\~n}a}, {Nakanishi}, {Oshima},
  {Sanders}, {Scott}, {Scoville}, {Tamura}, {Welch}, {Yun}, \&
  {Zeballos}}]{scott/etal:2012}
{Scott} K.~S. {et~al.}, 2012, \mnras, 423, 575

\bibitem[{{Seljak}(2000)}]{seljak:2000}
{Seljak} U., 2000, \mnras, 318, 203

\bibitem[{{Shang} {et~al}\mbox{.}(2012){Shang}, {Haiman}, {Knox}, \&
  {Oh}}]{shang/etal:2012}
{Shang} C., {Haiman} Z., {Knox} L., {Oh} S.~P., 2012, \mnras, 2559

\bibitem[{{Shaw} {et~al}\mbox{.}(2012){Shaw}, {Rudd}, \&
  {Nagai}}]{shaw/etal:2012}
{Shaw} L.~D., {Rudd} D.~H., {Nagai} D., 2012, \apj, 756, 15

\bibitem[{{Sheth} \& {Tormen}(1999)}]{sheth/tormen:1999}
{Sheth} R.~K., {Tormen} G., 1999, \mnras, 308, 119

\bibitem[{{Shirokoff} {et~al}\mbox{.}(2011){Shirokoff}, {Reichardt}, {Shaw},
  {Millea}, {Ade}, {Aird}, {Benson}, {Bleem}, {Carlstrom}, {Chang}, {Cho},
  {Crawford}, {Crites}, {de Haan}, {Dobbs}, {Dudley}, {George}, {Halverson},
  {Holder}, {Holzapfel}, {Hrubes}, {Joy}, {Keisler}, {Knox}, {Lee}, {Leitch},
  {Lueker}, {Luong-Van}, {McMahon}, {Mehl}, {Meyer}, {Mohr}, {Montroy},
  {Padin}, {Plagge}, {Pryke}, {Ruhl}, {Schaffer}, {Spieler}, {Staniszewski},
  {Stark}, {Story}, {Vanderlinde}, {Vieira}, {Williamson}, \&
  {Zahn}}]{shirokoff/etal:2011}
{Shirokoff} E. {et~al.}, 2011, \apj, 736, 61

\bibitem[{{Smail} {et~al}\mbox{.}(1997){Smail}, {Ivison}, \&
  {Blain}}]{smail/etal:1997}
{Smail} I., {Ivison} R.~J., {Blain} A.~W., 1997, \apjl, 490, L5

\bibitem[{{Spergel} {et~al}\mbox{.}(2003){Spergel}, {Verde}, {Peiris},
  {Komatsu}, {Nolta}, {Bennett}, {Halpern}, {Hinshaw}, {Jarosik}, {Kogut},
  {Limon}, {Meyer}, {Page}, {Tucker}, {Weiland}, {Wollack}, \&
  {Wright}}]{spergel/etal:2003}
{Spergel} D.~N. {et~al.}, 2003, \apjs, 148, 175

\bibitem[{{Stansberry} {et~al}\mbox{.}(2007){Stansberry}, {Gordon},
  {Bhattacharya}, {Engelbracht}, {Rieke}, {Marleau}, {Fadda}, {Frayer},
  {Noriega-Crespo}, {Wachter}, {Young}, {M{\"u}ller}, {Kelly}, {Blaylock},
  {Henderson}, {Neugebauer}, {Beeman}, \& {Haller}}]{stansberry/etal:2007}
{Stansberry} J.~A. {et~al.}, 2007, \pasp, 119, 1038

\bibitem[{{Sunyaev} \& {Zeldovich}(1980)}]{sunyaev/zeldovich:1980}
{Sunyaev} R.~A., {Zeldovich} I.~B., 1980, \mnras, 190, 413

\bibitem[{{Sunyaev} \& {Zel'dovich}(1970)}]{sunyaev/zeldovich:1970}
{Sunyaev} R.~A., {Zel'dovich} Y.~B., 1970, Ap\&SS, 7, 3

\bibitem[{{Swetz} {et~al}\mbox{.}(2011){Swetz}, {Ade}, {Amiri}, {Appel},
  {Battistelli}, {Burger}, {Chervenak}, {Devlin}, {Dicker}, {Doriese},
  {D{\"u}nner}, {Essinger-Hileman}, {Fisher}, {Fowler}, {Halpern},
  {Hasselfield}, {Hilton}, {Hincks}, {Irwin}, {Jarosik}, {Kaul}, {Klein},
  {Lau}, {Limon}, {Marriage}, {Marsden}, {Martocci}, {Mauskopf}, {Moseley},
  {Netterfield}, {Niemack}, {Nolta}, {Page}, {Parker}, {Staggs}, {Stryzak},
  {Switzer}, {Thornton}, {Tucker}, {Wollack}, \& {Zhao}}]{swetz/etal:2011}
{Swetz} D.~S. {et~al.}, 2011, \apjs, 194, 41

\bibitem[{{Swinyard} {et~al}\mbox{.}(2010){Swinyard}, {Ade}, {Baluteau},
  {Aussel}, {Barlow}, {Bendo}, {Benielli}, {Bock}, {Brisbin}, {Conley},
  {Conversi}, {Dowell}, {Dowell}, {Ferlet}, {Fulton}, {Glenn}, {Glauser},
  {Griffin}, {Griffin}, {Guest}, {Imhof}, {Isaak}, {Jones}, {King}, {Leeks},
  {Levenson}, {Lim}, {Lu}, {Makiwa}, {Naylor}, {Nguyen}, {Oliver}, {Panuzzo},
  {Papageorgiou}, {Pearson}, {Pohlen}, {Polehampton}, {Pouliquen},
  {Rigopoulou}, {Ronayette}, {Roussel}, {Rykala}, {Savini}, {Schulz},
  {Schwartz}, {Shupe}, {Sibthorpe}, {Sidher}, {Smith}, {Spencer}, {Trichas},
  {Triou}, {Valtchanov}, {Wesson}, {Woodcraft}, {Xu}, {Zemcov}, \&
  {Zhang}}]{swinyard/etal:2010}
{Swinyard} B.~M. {et~al.}, 2010, \aap, 518, L4

\bibitem[{{Tegmark} {et~al}\mbox{.}(2004){Tegmark}, {Blanton}, {Strauss},
  {Hoyle}, {Schlegel}, {Scoccimarro}, {Vogeley}, {Weinberg}, {Zehavi},
  {Berlind}, {Budavari}, {Connolly}, {Eisenstein}, {Finkbeiner}, {Frieman},
  {Gunn}, {Hamilton}, {Hui}, {Jain}, {Johnston}, {Kent}, {Lin}, {Nakajima},
  {Nichol}, {Ostriker}, {Pope}, {Scranton}, {Seljak}, {Sheth}, {Stebbins},
  {Szalay}, {Szapudi}, {Verde}, {Xu}, {Annis}, {Bahcall}, {Brinkmann},
  {Burles}, {Castander}, {Csabai}, {Loveday}, {Doi}, {Fukugita}, {Gott},
  {Hennessy}, {Hogg}, {Ivezi{\' c}}, {Knapp}, {Lamb}, {Lee}, {Lupton}, {McKay},
  {Kunszt}, {Munn}, {O'Connell}, {Peoples}, {Pier}, {Richmond}, {Rockosi},
  {Schneider}, {Stoughton}, {Tucker}, {Vanden Berk}, {Yanny}, \&
  {York}}]{tegmark/etal:2004}
{Tegmark} M. {et~al.}, 2004, \apj, 606, 702

\bibitem[{{Tinker} {et~al}\mbox{.}(2010){Tinker}, {Robertson}, {Kravtsov},
  {Klypin}, {Warren}, {Yepes}, \& {Gottl{\"o}ber}}]{tinker/etal:2010}
{Tinker} J.~L., {Robertson} B.~E., {Kravtsov} A.~V., {Klypin} A., {Warren}
  M.~S., {Yepes} G., {Gottl{\"o}ber} S., 2010, \apj, 724, 878

\bibitem[{{Tinker} {et~al}\mbox{.}(2012){Tinker}, {Sheldon}, {Wechsler},
  {Becker}, {Rozo}, {Zu}, {Weinberg}, {Zehavi}, {Blanton}, {Busha}, \&
  {Koester}}]{tinker/etal:2012}
{Tinker} J.~L. {et~al.}, 2012, \apj, 745, 16

\bibitem[{{Tinker} {et~al}\mbox{.}(2005){Tinker}, {Weinberg}, {Zheng}, \&
  {Zehavi}}]{tinker/etal:2005}
{Tinker} J.~L., {Weinberg} D.~H., {Zheng} Z., {Zehavi} I., 2005, \apj, 631, 41

\bibitem[{{Viero} {et~al}\mbox{.}(2009){Viero}, {Ade}, {Bock}, {Chapin},
  {Devlin}, {Griffin}, {Gundersen}, {Halpern}, {Hargrave}, {Hughes}, {Klein},
  {MacTavish}, {Marsden}, {Martin}, {Mauskopf}, {Moncelsi}, {Negrello},
  {Netterfield}, {Olmi}, {Pascale}, {Patanchon}, {Rex}, {Scott}, {Semisch},
  {Thomas}, {Truch}, {Tucker}, {Tucker}, \& {Wiebe}}]{viero/etal:2009}
{Viero} M.~P. {et~al.}, 2009, \apj, 707, 1766

\bibitem[{{Viero} {et~al}\mbox{.}(2012){Viero}, {Moncelsi}, {Mentuch},
  {Buitrago}, {Bauer}, {Chapin}, {Conselice}, {Devlin}, {Halpern}, {Marsden},
  {Netterfield}, {Pascale}, {P{\'e}rez-Gonz{\'a}lez}, {Rex}, {Scott}, {Smith},
  {Truch}, {Trujillo}, \& {Wiebe}}]{viero/etal:2012}
{Viero} M.~P. {et~al.}, 2012, \mnras, 421, 2161

\bibitem[{{Viero} {et~al}\mbox{.}(2013){Viero}, {Wang}, {Zemcov}, {Addison},
  {Amblard}, {Arumugam}, {Aussel}, {B{\'e}thermin}, {Bock}, {Boselli}, {Buat},
  {Burgarella}, {Casey}, {Clements}, {Conley}, {Conversi}, {Cooray}, {De
  Zotti}, {Dowell}, {Farrah}, {Franceschini}, {Glenn}, {Griffin},
  {Hatziminaoglou}, {Heinis}, {Ibar}, {Ivison}, {Lagache}, {Levenson},
  {Marchetti}, {Marsden}, {Nguyen}, {O'Halloran}, {Oliver}, {Omont}, {Page},
  {Papageorgiou}, {Pearson}, {P{\'e}rez-Fournon}, {Pohlen}, {Rigopoulou},
  {Roseboom}, {Rowan-Robinson}, {Schulz}, {Scott}, {Seymour}, {Shupe}, {Smith},
  {Symeonidis}, {Vaccari}, {Valtchanov}, {Vieira}, {Wardlow}, \&
  {Xu}}]{viero/etal:2013}
{Viero} M.~P. {et~al.}, 2013, \apj, 772, 77

\bibitem[{{Wang} {et~al}\mbox{.}(2013){Wang}, {Farrah}, {Oliver}, {Amblard},
  {B{\'e}thermin}, {Bock}, {Conley}, {Cooray}, {Halpern}, {Heinis}, {Ibar},
  {Ilbert}, {Ivison}, {Marsden}, {Roseboom}, {Rowan-Robinson}, {Schulz},
  {Smith}, {Viero}, \& {Zemcov}}]{wang/etal:2013}
{Wang} L. {et~al.}, 2013, \mnras, 431, 648

\bibitem[{{Wardlow} {et~al}\mbox{.}(2013){Wardlow}, {Cooray}, {De Bernardis},
  {Amblard}, {Arumugam}, {Aussel}, {Baker}, {B{\'e}thermin}, {Blundell},
  {Bock}, {Boselli}, {Bridge}, {Buat}, {Burgarella}, {Bussmann},
  {Cabrera-Lavers}, {Calanog}, {Carpenter}, {Casey}, {Castro-Rodr{\'{\i}}guez},
  {Cava}, {Chanial}, {Chapin}, {Chapman}, {Clements}, {Conley}, {Cox},
  {Dowell}, {Dye}, {Eales}, {Farrah}, {Ferrero}, {Franceschini}, {Frayer},
  {Frazer}, {Fu}, {Gavazzi}, {Glenn}, {Gonz{\'a}lez Solares}, {Griffin},
  {Gurwell}, {Harris}, {Hatziminaoglou}, {Hopwood}, {Hyde}, {Ibar}, {Ivison},
  {Kim}, {Lagache}, {Levenson}, {Marchetti}, {Marsden}, {Martinez-Navajas},
  {Negrello}, {Neri}, {Nguyen}, {O'Halloran}, {Oliver}, {Omont}, {Page},
  {Panuzzo}, {Papageorgiou}, {Pearson}, {P{\'e}rez-Fournon}, {Pohlen},
  {Riechers}, {Rigopoulou}, {Roseboom}, {Rowan-Robinson}, {Schulz}, {Scott},
  {Scoville}, {Seymour}, {Shupe}, {Smith}, {Streblyanska}, {Strom},
  {Symeonidis}, {Trichas}, {Vaccari}, {Vieira}, {Viero}, {Wang}, {Xu}, {Yan},
  \& {Zemcov}}]{wardlow/etal:2013}
{Wardlow} J.~L. {et~al.}, 2013, \apj, 762, 59

\bibitem[{{Watson} {et~al}\mbox{.}(2011){Watson}, {Berlind}, \&
  {Zentner}}]{watson/etal:2011}
{Watson} D.~F., {Berlind} A.~A., {Zentner} A.~R., 2011, \apj, 738, 22

\bibitem[{{Werner} {et~al}\mbox{.}(2004){Werner}, {Roellig}, {Low}, {Rieke},
  {Rieke}, {Hoffmann}, {Young}, {Houck}, {Brandl}, {Fazio}, {Hora}, {Gehrz},
  {Helou}, {Soifer}, {Stauffer}, {Keene}, {Eisenhardt}, {Gallagher}, {Gautier},
  {Irace}, {Lawrence}, {Simmons}, {Van Cleve}, {Jura}, {Wright}, \&
  {Cruikshank}}]{werner/etal:2004}
{Werner} M.~W. {et~al.}, 2004, \apjs, 154, 1

\bibitem[{{White} {et~al}\mbox{.}(2011){White}, {Blanton}, {Bolton},
  {Schlegel}, {Tinker}, {Berlind}, {da Costa}, {Kazin}, {Lin}, {Maia},
  {McBride}, {Padmanabhan}, {Parejko}, {Percival}, {Prada}, {Ramos}, {Sheldon},
  {de Simoni}, {Skibba}, {Thomas}, {Wake}, {Zehavi}, {Zheng}, {Nichol},
  {Schneider}, {Strauss}, {Weaver}, \& {Weinberg}}]{white/etal:2011}
{White} M. {et~al.}, 2011, \apj, 728, 126

\bibitem[{{Wright} {et~al}\mbox{.}(2010){Wright}, {Eisenhardt}, {Mainzer},
  {Ressler}, {Cutri}, {Jarrett}, {Kirkpatrick}, {Padgett}, {McMillan},
  {Skrutskie}, {Stanford}, {Cohen}, {Walker}, {Mather}, {Leisawitz}, {Gautier},
  {McLean}, {Benford}, {Lonsdale}, {Blain}, {Mendez}, {Irace}, {Duval}, {Liu},
  {Royer}, {Heinrichsen}, {Howard}, {Shannon}, {Kendall}, {Walsh}, {Larsen},
  {Cardon}, {Schick}, {Schwalm}, {Abid}, {Fabinsky}, {Naes}, \&
  {Tsai}}]{wright/etal:2010}
{Wright} E.~L. {et~al.}, 2010, \aj, 140, 1868

\bibitem[{{Xia} {et~al}\mbox{.}(2012){Xia}, {Negrello}, {Lapi}, {de Zotti},
  {Danese}, \& {Viel}}]{xia/etal:2012}
{Xia} J.-Q., {Negrello} M., {Lapi} A., {de Zotti} G., {Danese} L., {Viel} M.,
  2012, \mnras, 2554

\bibitem[{{Yan} {et~al}\mbox{.}(2005){Yan}, {Chary}, {Armus}, {Teplitz},
  {Helou}, {Frayer}, {Fadda}, {Surace}, \& {Choi}}]{yan/etal:2005}
{Yan} L. {et~al.}, 2005, \apj, 628, 604

\bibitem[{{Yoo} {et~al}\mbox{.}(2009){Yoo}, {Weinberg}, {Tinker}, {Zheng}, \&
  {Warren}}]{yoo/etal:2009}
{Yoo} J., {Weinberg} D.~H., {Tinker} J.~L., {Zheng} Z., {Warren} M.~S., 2009,
  \apj, 698, 967

\bibitem[{{Zahn} {et~al}\mbox{.}(2012){Zahn}, {Reichardt}, {Shaw}, {Lidz},
  {Aird}, {Benson}, {Bleem}, {Carlstrom}, {Chang}, {Cho}, {Crawford}, {Crites},
  {de Haan}, {Dobbs}, {Dor{\'e}}, {Dudley}, {George}, {Halverson}, {Holder},
  {Holzapfel}, {Hoover}, {Hou}, {Hrubes}, {Joy}, {Keisler}, {Knox}, {Lee},
  {Leitch}, {Lueker}, {Luong-Van}, {McMahon}, {Mehl}, {Meyer}, {Millea},
  {Mohr}, {Montroy}, {Natoli}, {Padin}, {Plagge}, {Pryke}, {Ruhl}, {Schaffer},
  {Shirokoff}, {Spieler}, {Staniszewski}, {Stark}, {Story}, {van Engelen},
  {Vanderlinde}, {Vieira}, \& {Williamson}}]{zahn/etal:2012}
{Zahn} O. {et~al.}, 2012, \apj, 756, 65

\bibitem[{{Zehavi} {et~al}\mbox{.}(2002){Zehavi}, {Blanton}, {Frieman},
  {Weinberg}, {Mo}, {Strauss}, {Anderson}, {Annis}, {Bahcall}, {Bernardi},
  {Briggs}, {Brinkmann}, {Burles}, {Carey}, {Castander}, {Connolly}, {Csabai},
  {Dalcanton}, {Dodelson}, {Doi}, {Eisenstein}, {Evans}, {Finkbeiner},
  {Friedman}, {Fukugita}, {Gunn}, {Hennessy}, {Hindsley}, {Ivezi{\'c}}, {Kent},
  {Knapp}, {Kron}, {Kunszt}, {Lamb}, {Leger}, {Long}, {Loveday}, {Lupton},
  {McKay}, {Meiksin}, {Merrelli}, {Munn}, {Narayanan}, {Newcomb}, {Nichol},
  {Owen}, {Peoples}, {Pope}, {Rockosi}, {Schlegel}, {Schneider}, {Scoccimarro},
  {Sheth}, {Siegmund}, {Smee}, {Snir}, {Stebbins}, {Stoughton}, {SubbaRao},
  {Szalay}, {Szapudi}, {Tegmark}, {Tucker}, {Uomoto}, {Vanden Berk}, {Vogeley},
  {Waddell}, {Yanny}, \& {York}}]{zehavi/etal:2002}
{Zehavi} I. {et~al.}, 2002, \apj, 571, 172

\bibitem[{{Zehavi} {et~al}\mbox{.}(2011){Zehavi}, {Zheng}, {Weinberg},
  {Blanton}, {Bahcall}, {Berlind}, {Brinkmann}, {Frieman}, {Gunn}, {Lupton},
  {Nichol}, {Percival}, {Schneider}, {Skibba}, {Strauss}, {Tegmark}, \&
  {York}}]{zehavi/etal:2011}
{Zehavi} I. {et~al.}, 2011, \apj, 736, 59

\bibitem[{{Zehavi} {et~al}\mbox{.}(2005){Zehavi}, {Zheng}, {Weinberg},
  {Frieman}, {Berlind}, {Blanton}, {Scoccimarro}, {Sheth}, {Strauss}, {Kayo},
  {Suto}, {Fukugita}, {Nakamura}, {Bahcall}, {Brinkmann}, {Gunn}, {Hennessy},
  {Ivezi{\'c}}, {Knapp}, {Loveday}, {Meiksin}, {Schlegel}, {Schneider},
  {Szapudi}, {Tegmark}, {Vogeley}, \& {York}}]{zehavi/etal:2005}
{Zehavi} I. {et~al.}, 2005, \apj, 630, 1

\bibitem[{{Zheng} {et~al}\mbox{.}(2009){Zheng}, {Zehavi}, {Eisenstein},
  {Weinberg}, \& {Jing}}]{zheng/etal:2009}
{Zheng} Z., {Zehavi} I., {Eisenstein} D.~J., {Weinberg} D.~H., {Jing} Y.~P.,
  2009, \apj, 707, 554

\end{thebibliography}

}

\end{document}